\documentclass[a4paper,11pt]{article}
\pdfoutput=1 % if your are submitting a pdflatex (i.e. if you have images in pdf, png or jpg format)

\usepackage{jinstpub} % for details on the use of the package, please see the JINST-author-manual
\usepackage{lineno}

\newcommand*\degr{\ensuremath{^\circ}}
\newcommand*\arcmin{\ensuremath{^\prime}}

\newcommand{\quotes}[1]{``#1''}
\newcommand{\srsr}{{SRSR-D$^{\copyright}$}}

\title{\boldmath The LSPE--Strip feed horn array}

\author[a,b,1]{C.~Franceschet,\note{Corresponding author.}}
\author[a]{F.~Del Torto,}
\author[c]{F.~Villa,}
\author[a,b]{S.~Realini,}
\author[a]{R.~Bongiolatti,}
\author[d]{O.~A.~Peverini,}
\author[a,b]{F.~Pezzotta,}
\author[a,b]{D.~M.~Vigan\'o,}
\author[d]{G.~Addamo,}
\author[a,b]{M.~Bersanelli,}
\author[a,b]{F.~Cavaliere,}
\author[c]{F.~Cuttaia,}
\author[e,f]{M.~Gervasi,}
\author[a,b]{A.~Mennella,}
\author[c]{G.~Morgante,}
\author[g]{A.~C.~Taylor,}
\author[d]{G.~Virone}
\author[e,f]{and M.~Zannoni}

\affiliation[a]{Universit\'a degli Studi di Milano, Via Celoria 16, 20133 Milano, Italy}
\affiliation[b]{INFN Sezione di Milano, Via Celoria 16, 20133 Milano, Italy}
\affiliation[c]{INAF-OAS Bologna, Via Gobetti 101, 40129 Bologna, Italy}
\affiliation[d]{IEIIT-CNR, Politecnico di Torino, Corso Duca degli Abruzzi 24, 10129 Torino, Italy}
\affiliation[e]{Universit\'a degli Studi di Milano-Bicocca, Piazza della Scienza 3, 20126 Milano, Italy}
\affiliation[f]{INFN–Sezione di Milano Bicocca, Piazza della Scienza 3, 20126 Milano, Italy}
\affiliation[g]{University of Oxford, Denys Wilkinson Building, Keble Road, Oxford OX1 3RH, UK}
\emailAdd{cristian.franceschet@fisica.unimi.it}

\abstract{In this paper we discuss the design, manufacturing and characterization of the feed horn array of the Strip instrument of the Large Scale Polarization Explorer (LSPE) experiment. Strip is a microwave telescope, operating in the Q- and W-band, for the observation of the polarized emissions from the sky in a large fraction (about 37\%) of the Northern hemisphere with sub-degree angular resolution. The Strip focal plane is populated by forty-nine Q–band and six W-band corrugated horns, each feeding a cryogenically cooled polarimeter for the detection of the Stokes $Q$ and $U$ components of the polarized signal from the sky. The Q-band channel is designed to accurately monitor Galactic polarized synchrotron emission, while the combination of Q- and W-band will allow the study of atmospheric effects at the observation site, the Observatorio del Teide, in Tenerife. In this paper we focus on the development of the Strip corrugated feed horns, including design requirements, engineering and manufacturing, as well as detailed characterization and performance verification.}


\begin{document}
\maketitle
\flushbottom

%----------------------------------------
\section{Introduction}
\label{sec:intro}
Since its first detection in 2001 by the DASI instrument~\cite{DASI}, the polarized component of the Cosmic Microwave Background (CMB) has been measured with increasing precision and accuracy, culminating with the full-sky survey by the Planck satellite~\cite{Planck2020}. In recent years, the search for the the so-called B-mode pattern~\cite{Kamionkowski2016} has further motivated major experimental efforts in this area. If detected, the B-mode CMB polarization would highlight the imprint of primordial gravitational waves and would provide strong evidence for an inflation era in the very early universe.
However, measuring B-modes is a remarkably challenging task. Their signal is extremely weak, fractions of a $\mu$K, much smaller than the level of polarized emissions from our Galaxy, which can mimic the B-mode pattern in the sky~\cite{PhysRevLett.114.101301}. Therefore, separating the weakly polarized component of the CMB from the Galactic foregrounds calls for observations over a wide frequency range. Moreover, while the emission from Earth's atmosphere is essentially unpolarized in the microwave, its emission and fluctuations introduce significant noise components in ground-based experiments. Finally, strict control of the spurious polarization induced by the instrument itself, either observing from suborbital platforms or from space, is crucial to achieve high-precision measurements.

It is in this context that the Large Scale Polarization Explorer (LSPE) experiment~\cite{LSPE2020} will operate. LSPE is a combined ground-based and balloon-borne experiment, funded by the Italian Space Agency and the Istituto Nazionale di Fisica Nucleare (INFN), with the participation of numerous European and international partners. Aiming at large angular scales, where the peak of the intensity of the B-modes is expected, LSPE will guarantee a wide frequency coverage by including two independent instruments observing about 1/4 of the full sky in the same portion of the Northern Hemisphere: Strip, a HEMT-based polarimeter instrument operating at 43 and 95 GHz and SWIPE, a bolometer-based instrument observing at 140, 220 and 240 GHz from a stratospheric balloon during a long-duration flight from the Svalbard Islands.

LSPE will be sensitive to the B-modes of CMB polarization at a level corresponding to a tensor-to-scalar ratio $r=0.03$ with 99.7\% confidence level (CL) and an upper limit of $r=0.015$ at 95\% CL.
%--- begin rev. #1
The current upper limit on the tensor-to-scalar ratio is $r<0.044$ at 95\% CL, obtained combining data from the Planck satellite and the BICEP/Keck ground telescopes~\cite{Planck2020}~\cite{2021tristram}~\cite{bicep2018}. A considerable effort in detecting the B-modes or, at least, improving the upper limit of the tensor-to-scalar ratio comes from ground-based observations: among the several experiments aiming at the CMB polarization at large angular scales, the CLASS telescopes array will observe from the Chilean Atacama desert at
%38, 93 and 145/217 GHz, similarly to LSPE, 
the LSPE frequencies, placing upper limits on $r$ down to a level of 0.01 at 95\% CL~\cite{CLASS2016}. Broadband observations at low $\ell$ on both celestial hemispheres provide complementary information, while allowing for an in-depth understanding of the polarized CMB and foreground emissions.
%--- end rev. #1
Furthermore, the LSPE measurement of the CMB E-mode polarization at large angular scales will provide improved measurements of the optical depth $\tau$, as well as of the so called {\em low-$\ell$} anomaly~\cite{lowell2015}~\cite{low-ell}. LSPE will also deliver extended maps of the diffuse polarized emission produced in our Galaxy by synchrotron and interstellar dust. These data will be important in the study of the Galactic magnetic field and the properties of ionized gas and thermal dust in the Milky Way. Strip will also study the quality of the atmosphere at Teide Observatory in Tenerife for CMB polarization measurements.

In this paper, we focus on the the development and characterization of the array of corrugated horn antennas populating the Strip focal plane.

%----------------------------------------
\subsection{The Strip instrument}
\label{sec:strip}
Strip is an array of fifty-five coherent polarimeters coupled to a fully steerable crossed-Dragone telescope with 1.5 m projected aperture~\cite{STRIP2018}. The array is populated with forty-nine elements in the Q-band, to measure synchrotron emission at a frequency near the foreground minimum; and with six elements in W-band, mainly to provide simultaneous measurements of the atmospheric emission in the opacity window centered at 95~GHz. The Strip polarimeters, cryogenically cooled to 20~K, allow us to directly measure the Stokes $Q$ and $U$ parameters through a double-demodulation scheme already implemented into the QUIET experiment~\cite{Bischoff_2013} receivers\footnote{Strip actually includes most of the QUIET's original polarimeters, used in the two observation seasons between 2008 and 2010 at the Chajnantor plateau in the Atacama Desert in Chile.}, ensuring excellent rejection of $1/f$ noise from amplifier gain fluctuations as well as of temperature-to-polarization leakage, without the need of extra optical elements to modulate the CMB polarized signal~\cite{LSPE2020}.

%----------------------------------------
\subsubsection{The dual-reflector telescope}
\label{sec:optics}
The Strip dual-reflector telescope, originally developed for the Clover experiment~\cite{TAYLOR2006993}, includes a parabolic primary mirror with 1500~mm diameter and a hyperbolic secondary mirror with a 1660~mm $\times$ 1720~mm wide elliptical rim, arranged in a Dragonian cross-fed design. This configuration preserves polarization purity on the optical axis and gives low aberrations across a wide, flat focal plane. The equivalent focal length of the Strip optics is 2700~mm, resulting in a focal ratio of f$\sim$1.8 and providing an angular resolution of $\sim$20\arcmin\ in the Q-band and $\sim$10\arcmin\ in the W-band. 
Both reflectors are oversized, and a feed illumination angle of $\sim$15\degr\ is required to optimize the illumination of the telescope. Corrugated feed horns have been specifically designed to achieve this. The whole feed horn array is placed in the focal region of the telescope, ensuring no obstruction of the field of view. To reduce the contamination due to stray-light, the telescope is surrounded by a co-moving baffle made of aluminum plates coated with a millimetre-wave absorber.

\begin{figure}[htbp]
   \centering
   \includegraphics[width=.4\textwidth]{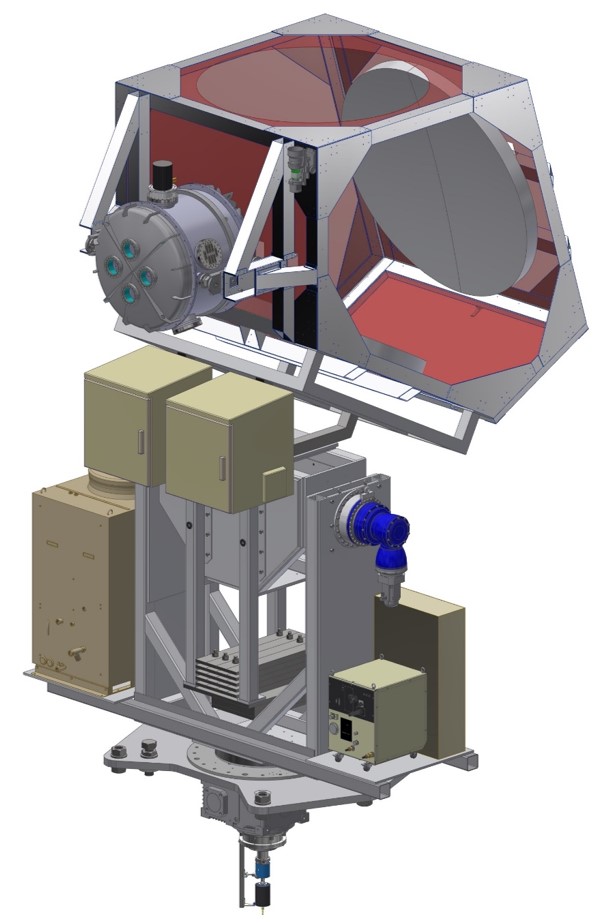}
   \caption{Overview of the Strip instrument. The telescope, including the two mirrors surrounded by the co-moving baffle and the cryostat (on the left), is held by the alt-azimuth mount. The warm electronics boxes are assembled on the mount structure. The rotary joint is shown at the bottom of the picture.}
   \label{fig:strip-overview}
\end{figure}

To meet the instrument nominal scanning strategy (i.e. continuous spinning around the azimuth axis at a fixed elevation 20\degr\ from the Zenith), the Strip  optical assembly is installed on top of an alt-azimuth mount equipped with a rotary joint, which ensures signal and power transmission while allowing for the continuous rotation of the entire telescope assembly. Figure~\ref{fig:strip-overview} shows a view of the Strip telescope~\cite{Villa2021}.

Our simulations of the Strip optical system result in a cross-polar discrimination better than $-30$~dB and a sidelobes rejection of order of $-55$~dB and $-65$~dB for near and far sidelobes, respectively~\cite{Realini2021}.

%----------------------------------------
\subsubsection{The focal plane arrays}
\label{sec:fpu}
Each array element in the focal plane~\cite{Mennella2021} includes a corrugated feed horn, followed by a dual-circular polarization chain (a groove polarizer~\cite{peverini2015}~\cite{Peverini2021} coupled to an orthomode transducer~\cite{Virone2014} in the Q–band; and a septum polarizer in the W–band), feeding the polarimetric module based on High Electron Mobility Transistor low noise amplifiers. The signal is then amplified and digitized in the warm electronics unit on board the telescope structure.

\begin{figure}[htbp]
   \centering
   \includegraphics[width=.4\textwidth]{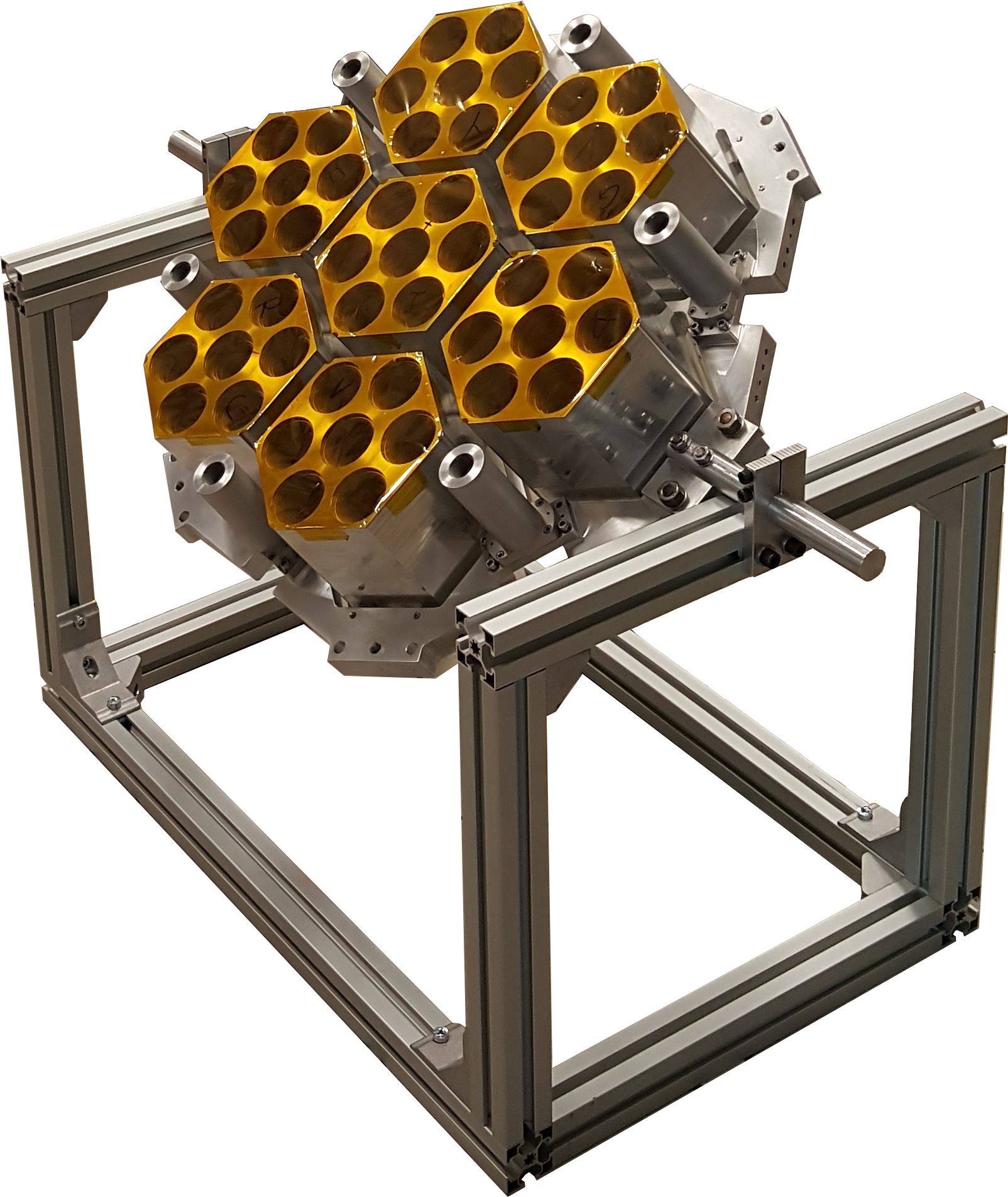}
   %\caption{{\em{Left}}: the simulated focal surface of the Strip telescope. {\em{Right}}: the focal plane unit including the Q-band and W-band feedhorns, as displaced according the shape of the telescope focal surface.}
   \caption{The Strip focal plane unit including the Q-band and W-band feed horns.}
   \label{fig:focal_surface}
\end{figure}

The forty-nine Q-band corrugated feed horns are arranged into a honeycomb lattice of seven hexagonal modules, each including seven elements. To optimize the telescope resolution, directivity and sidelobe level, the feed horn modules are arranged to closely match the telescope focal surface and oriented to illuminate the center on the primary mirror. The six W-band feed horns are placed and oriented on the telescope focal surface according to a similar criterion, paying attention to minimize the contribution of the stray-light, as they are placed in the outermost region of the instrument focal plane. %The right panel of
Figure~\ref{fig:focal_surface} shows the focal plane unit with the passive chains assembled into the mechanical support structure.

%----------------------------------------
\section{Development of the feed horn arrays}
\label{sec:fh_development}
The feed design, for both Q- and W-band, is based on profiled corrugated horns. These are the best performing antennas for such kind of applications, given their high cross-polarization discrimination, beam symmetry and impedance matching over wide frequency bands.
%--- begin rev. #2 #3
Table~\ref{tab:TopSpecs} shows the top level electromagnetic requirements which drive the feed horns design.

\begin{table}[htbp]
   \centering
   \caption[Top level electromagnetic requirements for the Q-band and W-band feed horns]{\label{tab:TopSpecs}Top level electromagnetic requirements for the Q-band and W-band feed horns.}
   \begin{tabular}{lcc}
    \hline
    %Feature                             & \multicolumn{2}{c}{Value}\\
                                        & Q-band               & W-band\\
    \hline
    Matching over bandwidth             & $<-30$ dB             & $<-30$ dB\\
    Cross-polarization over bandwidth   & $<-30$ dB             & $<-30$ dB\\
    Edge taper at center frequency      & $-30$ dB at $20\degr$   & $-20$ dB at $15\degr$\\
    Frequency range                     & $39-48$ GHz      & $85-105$ GHz\\
    \hline
   \end{tabular}
\end{table}
%--- end rev. #2 #3

The severe under-illumination of the Strip mirrors requires focal plane antennas with quite large apertures. An edge taper of about $-30$~dB at $\sim$20\degr, which meets the sidelobe requirements, leads to a feed aperture $\sim$50~mm and $\sim$37~mm at 43 and 95~GHz, respectively. These correspond almost exactly to the projected size of the radiometer's modules, and therefore it does not impact the inter-axis distance and the overall array design. A compact focal plane array guarantees minimal aberrations for the outer elements of the array of receivers, i.e., those further apart from the focus.

%----------------------------------------
\subsection{The corrugated horn electromagnetic design}
\label{sec:fh_design}
Several horn profiles have been studied to meet the Strip specifications and optimize its optical response. This has let to the design of a dual-profiled corrugated horn~\cite{gentili2000}, giving the best compromise between performance and compactness. A {\em{$sin^2-exp$}} profile has been chosen with the following analytical expression of the horn section radius, $R(z)$, as a function of the longitudinal coordinate $z$:
\begin{equation}
\label{eq:RzSin}
R(z)=R_{i}+\left(R_{s}-R_{i}\right)\left[(1-A) \frac{z}{L_{s}}+A \sin\left(\frac{\pi}{2} \frac{z}{L_{s}}\right)\right]^2
\end{equation}
in the sine squared section, $0 \leq z \leq L_{s}$, and
\begin{equation}
\label{eq:RzExp}
\begin{split}
R(z) = &\ R_{s}+e^{\alpha\left(z-L_{s}\right)}-1 ;
\\
\alpha = &\ \frac{1}{L_{e}} \ln \left(1+R_{a p}-R_{s}\right)
\end{split}
\end{equation}
in the exponential section, $L_{s} \leq z \leq L_{e}+L_{s}$.\\
$R_{i}$ is the throat radius, $R_{s}$ is the sine squared region end radius (or exponential region initial radius), $R_{ap}$ is the aperture radius, $L_{s}$ is the sine squared region length and $L_{e}$ is the exponential region length. The parameter $A$ ($0 \leq A \leq 1$) tapers the first region profile between linear and pure sine squared type. The parameters $L_{e} / \left(L_{e} + L_{s}\right)$, $A$ and $R_{s}$ can be used to optimize the position and frequency stability of the phase center and the compactness of the structure~\cite{Villa2002}.

%--- begin rev. #2 #3
With reference to equations~\ref{eq:RzSin} and~\ref{eq:RzExp}, Table~\ref{tab:DesignSpecs} summarizes the design parameters and some derived mechanical specifications, such as aperture diameter, circular waveguide diameter and total length of the feed horns. Note that the circular waveguide diameter is also driven by the coupling between horn and polarizer waveguides.

\begin{table}[htbp]
   \centering
   \caption[Q-band and W-band feed horn design requirements]{\label{tab:DesignSpecs}Q-band and W-band feed horn design specifications and derived mechanical parameters.}
   \begin{tabular}{lcc}
    \hline
    %Feature                             & \multicolumn{2}{c}{Value}\\
                                        & Q-band              & W-band\\
    \hline
    $R_{i}$ (mm)                        & 3.4                 & 1.31\\
    $R_{s}$ (mm)                        & 16.0                 & 7.0\\
    $R_{ap}$ (mm)                       & 25.0                & 11.08\\
    $L_{s}$ (mm)                        & 74.0                & 60.0\\
    $L_{e}$ (mm)                        & 46.0                & 31.8\\
    $A$                                 & 0.8                 & 0.8\\
    \hline
    Aperture diameter (mm)              & 50.0                & 22.16\\
    Circular waveguide diameter (mm)    & 6.8                 & 2.62\\
    Total feed length (mm)              & 137.0               & 101.8\\
    \hline
    \end{tabular}
\end{table}
%--- end rev. #2 #3

\begin{figure}[htbp]
   \centering
   \includegraphics[width=.8\textwidth]{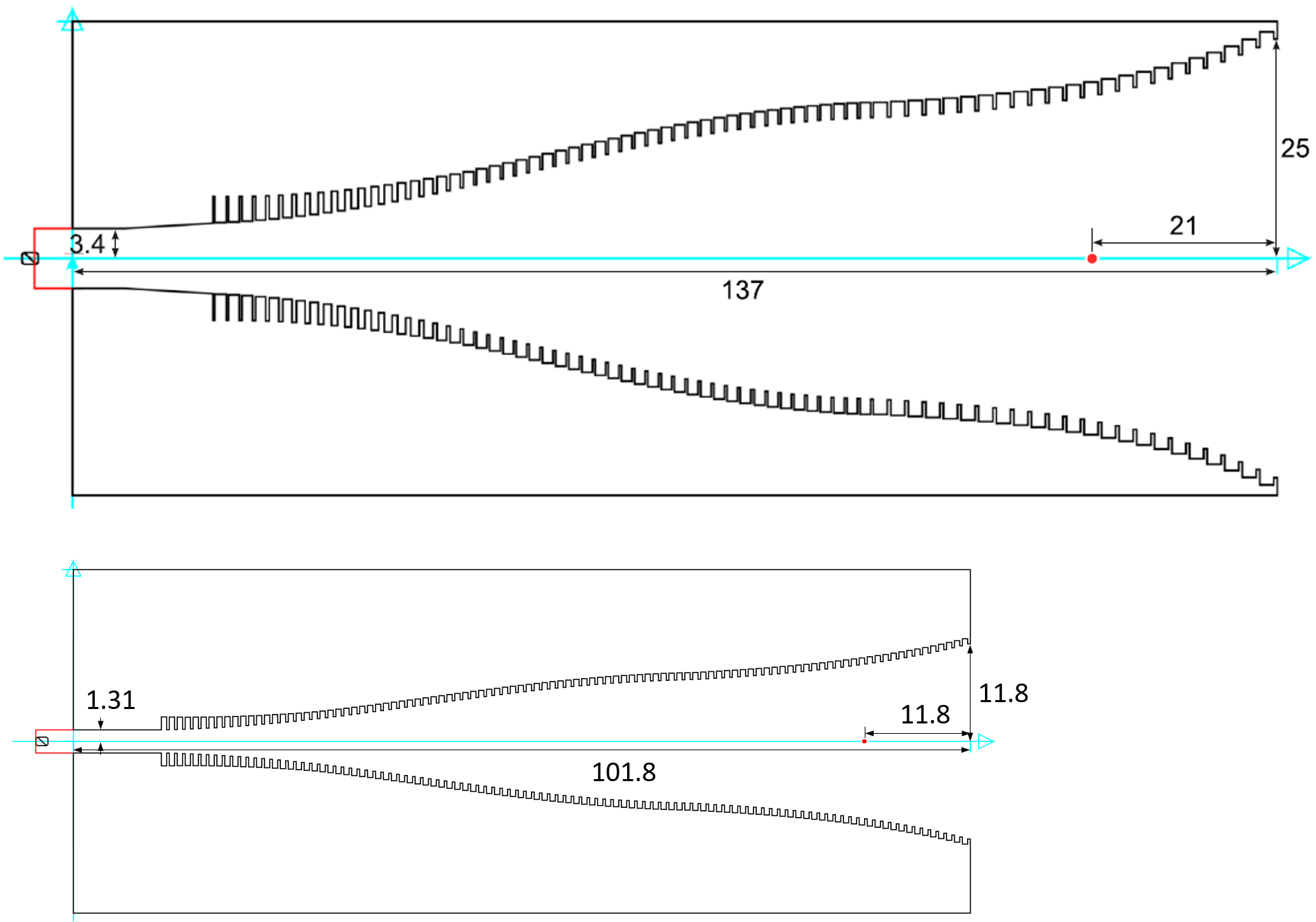}
   \caption{Corrugation profile of the ({\em{top}}) Q-band and ({\em{bottom}}) W-band Strip feed horns. Both profiles, shown on a relative scale, are composed of a $sin^2$ section joined to an exponential one towards the horn aperture. The red dot indicates the phase center position, 21~mm and 11.8~mm from the horn aperture, respectively.}
\label{STRIP_feed_profile}
\end{figure}

In the case of the Q-band design, the first 74~mm from the horn throat has a $sin^2$ profile and determines most of the cross-polarization and matching properties; the exponential section rapidly increases the horn dimension up to the 50~mm aperture, giving the expected full width half maximum (FWHM). The corrugation step is 1.5~mm for the $sin^2$ section and 2~mm for the exponential one. These values result from a trade-off between the electromagnetic performance requirements and the mechanical constraints. Also, the slot width and depth of the corrugations vary along the profile: the slot width is 0.19~mm at the throat and 1.6~mm at the aperture, while the slot depth is 2.97~mm at the throat and 0.81~mm at the aperture. The feed horn's input diameter is 6.8 mm, allowing for a nominal working band 39$-48$~GHz. The Q-band horn profile is shown in the top panel of Figure~\ref{STRIP_feed_profile}.

Concerning the W-band horn profile, it was not possible to achieve the expected aperture diameter to guarantee the same taper as the Q-band. In fact, the resulting horn would have been too long to be housed into the cryostat. For this reason, the horn length has been limited to 101.80~mm, leading to an aperture diameter 22.16~mm. The higher illumination level determined the different resolution of the two frequency channels, $\sim$20\arcmin\ at 43~GHz and $\sim$10\arcmin\ at 95~GHz.
%--- begin rev. #4
As will be clarified in the section~\ref{sec:fh_platelet_W} relating to the manufacturing of the W-band horns, to optimize production times and costs, the metal plates making up the tooth and throat of each corrugation have been processed separately by chemical-etching.
%--- end rev. #4
The corrugation step is set to 0.9~mm: 0.3~mm for the teeth and 0.6~mm for the throats. This fixed step is not optimal in terms of the impedance matching, which required for a variable teeth-to-throat depth ratio on the first $\sim$20 corrugations. It resulted as a compromise between an acceptable impedance matching (better than $-22$~dB over the whole bandwidth) and constructability of the horns. The waveguide radius is 1.31~mm. The W-band horn profile, operative in the 85-105~GHz bandwidth, is shown in the bottom panel of Figure~\ref{STRIP_feed_profile}.

%--- begin rev. #2 #3
%Table~\ref{tab:PerformanceSpecs} summarizes some mechanical specifications, such as aperture diameter and total length of the feed horns, and performance requirements, such as cross-polarization level and impedance matching.
%
%\begin{table}[htbp]
%   \centering
%   \caption[Q-band and W-band feed horn design requirements]{\label{tab:PerformanceSpecs}Q-band and W-band feed horn design specifications.}
%   \begin{tabular}{lcc}
%    \hline
%    %Feature                             & \multicolumn{2}{c}{Value}\\
%                                        & Q-band               & W-band\\
%    \hline
%    Aperture diameter                   & $\leq 50$ mm          & $\leq 22.2$ mm\\
%    Total feed length                   & 137 mm                & 101.8 mm\\
%    Center of phase distance from ap.   & 21 mm                 & 11.8 mm\\
%    Circular waveguide diameter         & 6.8 mm                & 2.62 mm\\
%    Corrugation step at the aperture    & 2 mm                  & 0.9 mm\\
%    Matching over bandwidth             & $<-30$ dB             & $<-30$ dB\\
%    Cross-polarization over bandwidth   & $<-30$ dB             & $<-30$ dB\\
%    Edge taper at center frequency      & $-30$ dB at $20\degr$   & $-20$ dB at $15\degr$\\
%    Frequency range                     & $39-48$ GHz      & $85-105$ GHz\\
%    \hline
%   \end{tabular}
%\end{table}
%--- end rev. #2 #3

The response of the feed horns has been simulated with the software \srsr 4.5 by Orange\footnote{\srsr seems to be no longer maintained, no reference can be provided.}. \srsr provides a rigorous simulation of the electromagnetic performance of any structure with symmetry of revolution consisting of conducting parts and homogeneous dielectric domains. 

%----------------------------------------
\subsubsection{Q-band feed horn simulations}
\label{sec:fh_sim_Q}
\begin{figure}[htbp]
   \centering
   \includegraphics[width=.6\textwidth]{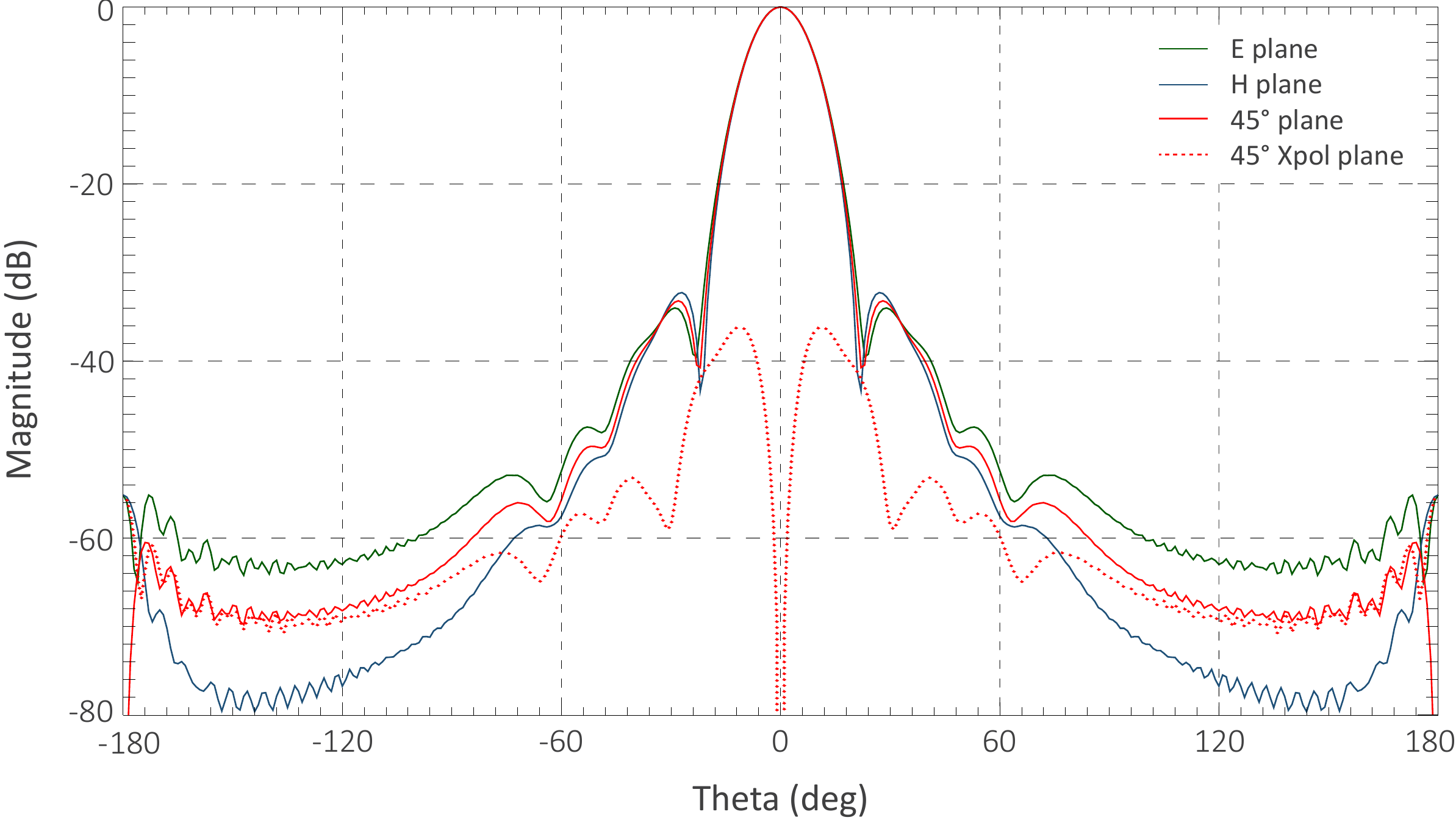}\vspace{0.5 cm}
   \includegraphics[width=.6\textwidth]{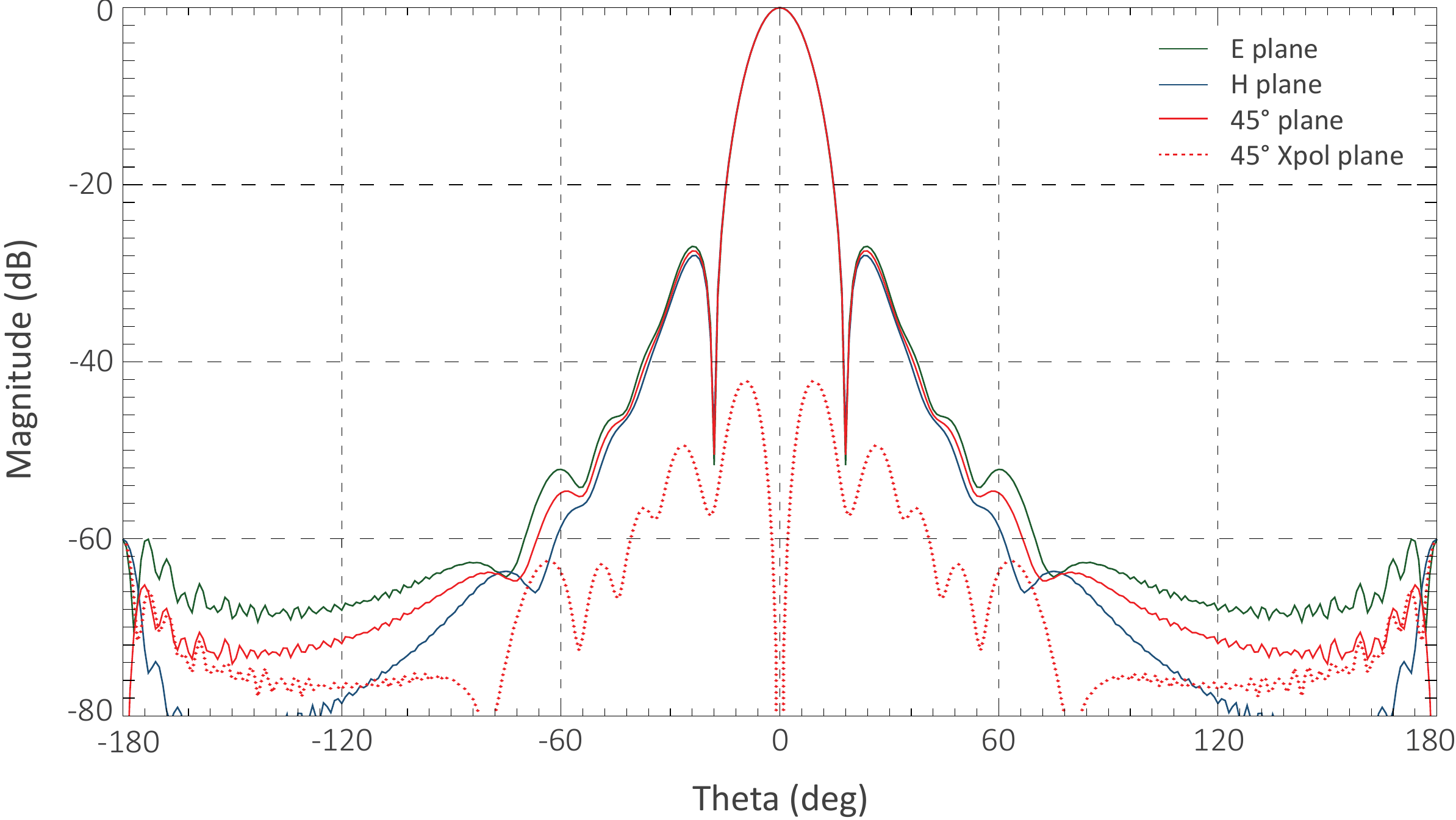}\vspace{0.5 cm}
   \includegraphics[width=.6\textwidth]{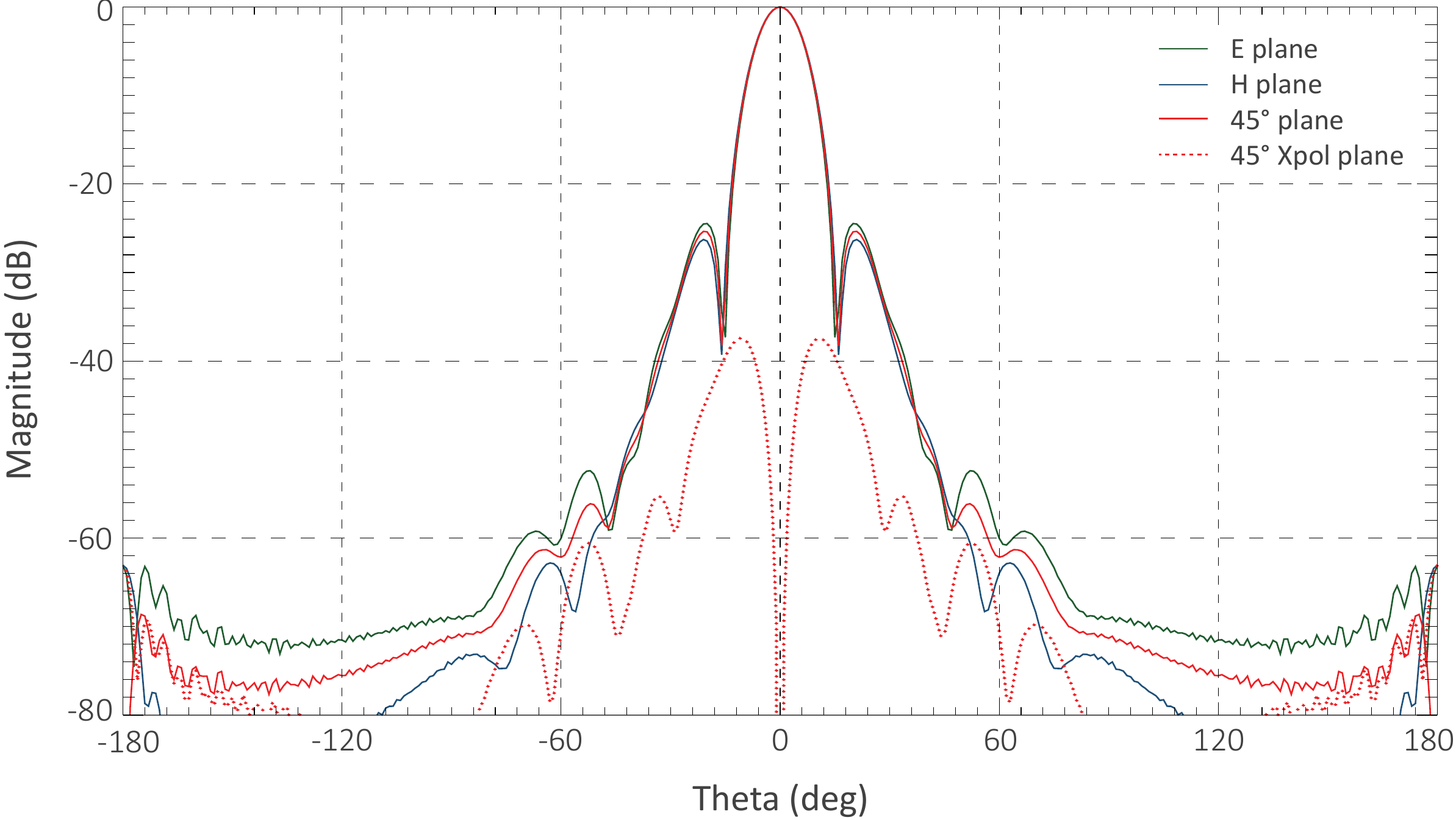}
   \caption{Simulated radiation patterns of the Strip feed horns at 38.7, 43 and 47.3~GHz. The maximum cross-polarization level at the three frequencies is $-36.2$~dB, $-42.2$~dB and $-37.4$~dB, respectively.}
   \label{Q_fh_beams}
\end{figure}

Radiation patterns have been simulated in the required frequency band, with a 0.1~GHz frequency step. As a reference, Figure~\ref{Q_fh_beams} shows the radiation pattern computed at three frequencies in the Q-band, i.e. the center frequency $f_0$ and within $\pm10\%$:
\begin{itemize}
   \item{38.7~GHz ($f_0-10\%$)}
   \item{43~GHz ($f_0$)}
   \item{47.3~GHz ($f_0+10\%$)}
\end{itemize}
Each plot includes the expected radiation pattern on the co-polar principal planes (E-plane, H-plane and $45\degr$-plane) and cross-polar $45\degr$-plane.

\begin{figure}
   \centering
   \includegraphics[width=.7\textwidth]{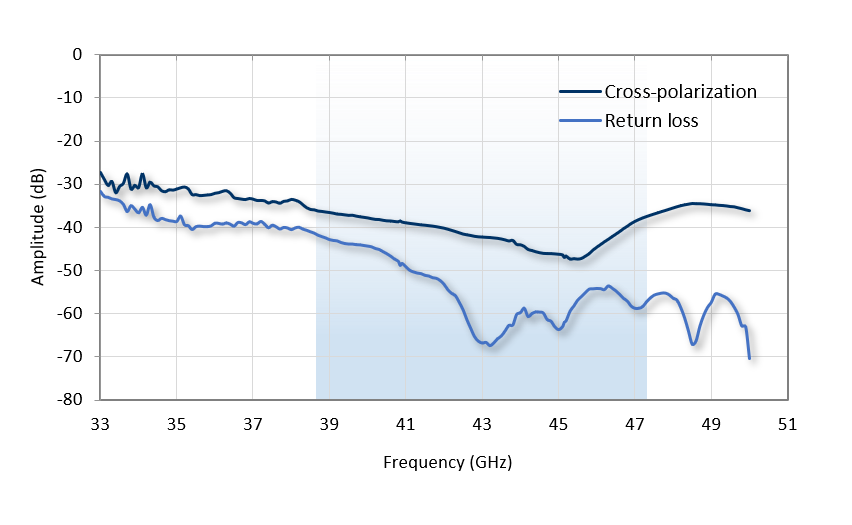}
   \caption{Simulation of the maximum cross-polarization level and return loss as a function of the frequency in the Q-band. At the lower edge of the Q-band the maximum level of cross-polarization is still below $-30$~dB. In the operative bandwidth, between 38.7 and 47.3 GHz (shaded area), cross-polarization is well below $-35$~dB. The feed horn is matched at a level better than $-40$~dB over the whole Q-band, except for the very lower frequencies below 35~GHz.}
   \label{Q_fh_cross_rl}
\end{figure}

Figure~\ref{Q_fh_cross_rl} shows the level of maximum cross-polarization and return loss over the whole Q-band frequencies between 33 and 50 GHz.

%----------------------------------------
\subsubsection{W-band feed horn simulations}
\label{sec:fh_sim_W}
\begin{figure}[htbp]
   \centering
   \includegraphics[width=.6\textwidth]{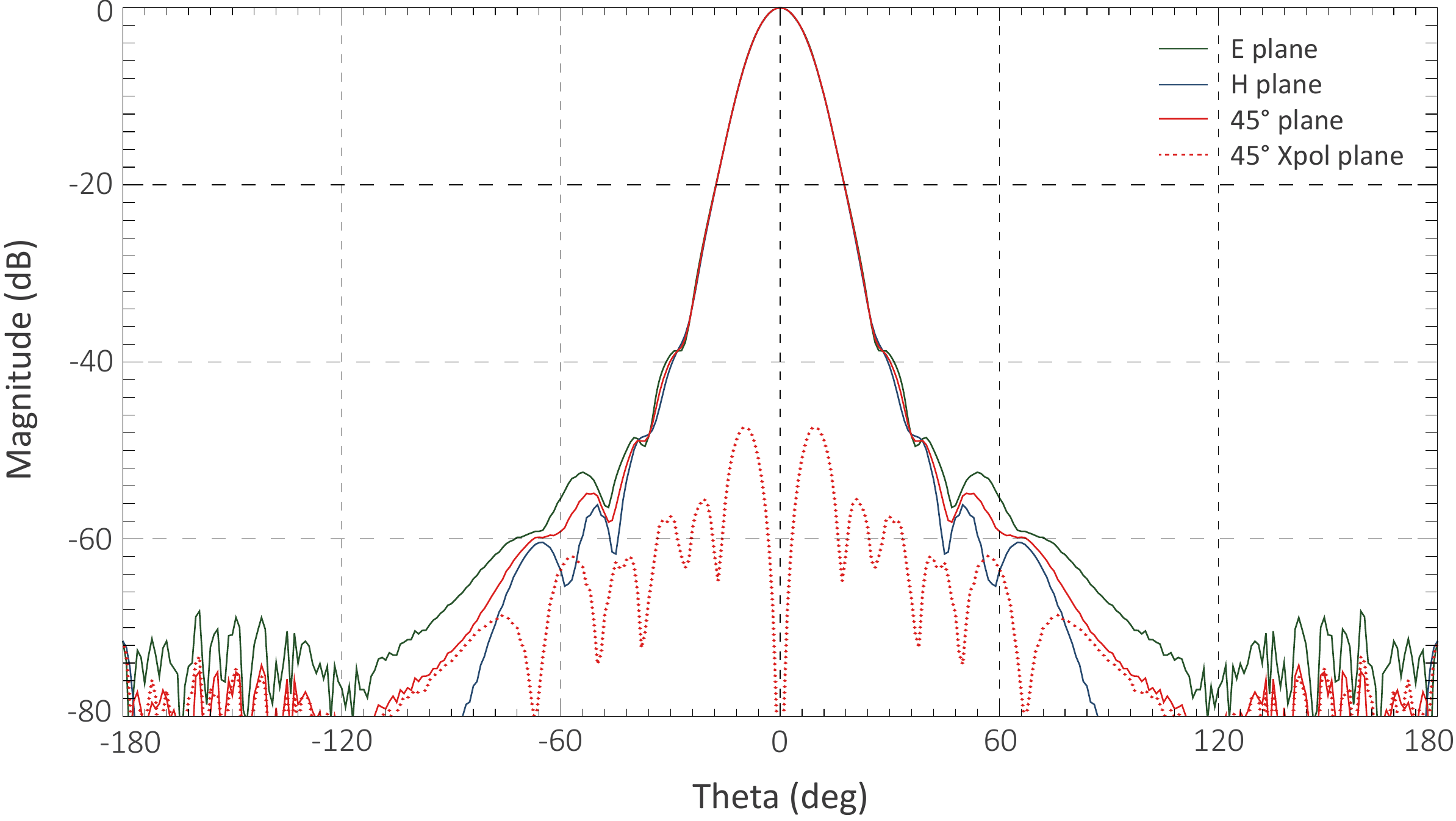}\vspace{0.5 cm}
   \includegraphics[width=.6\textwidth]{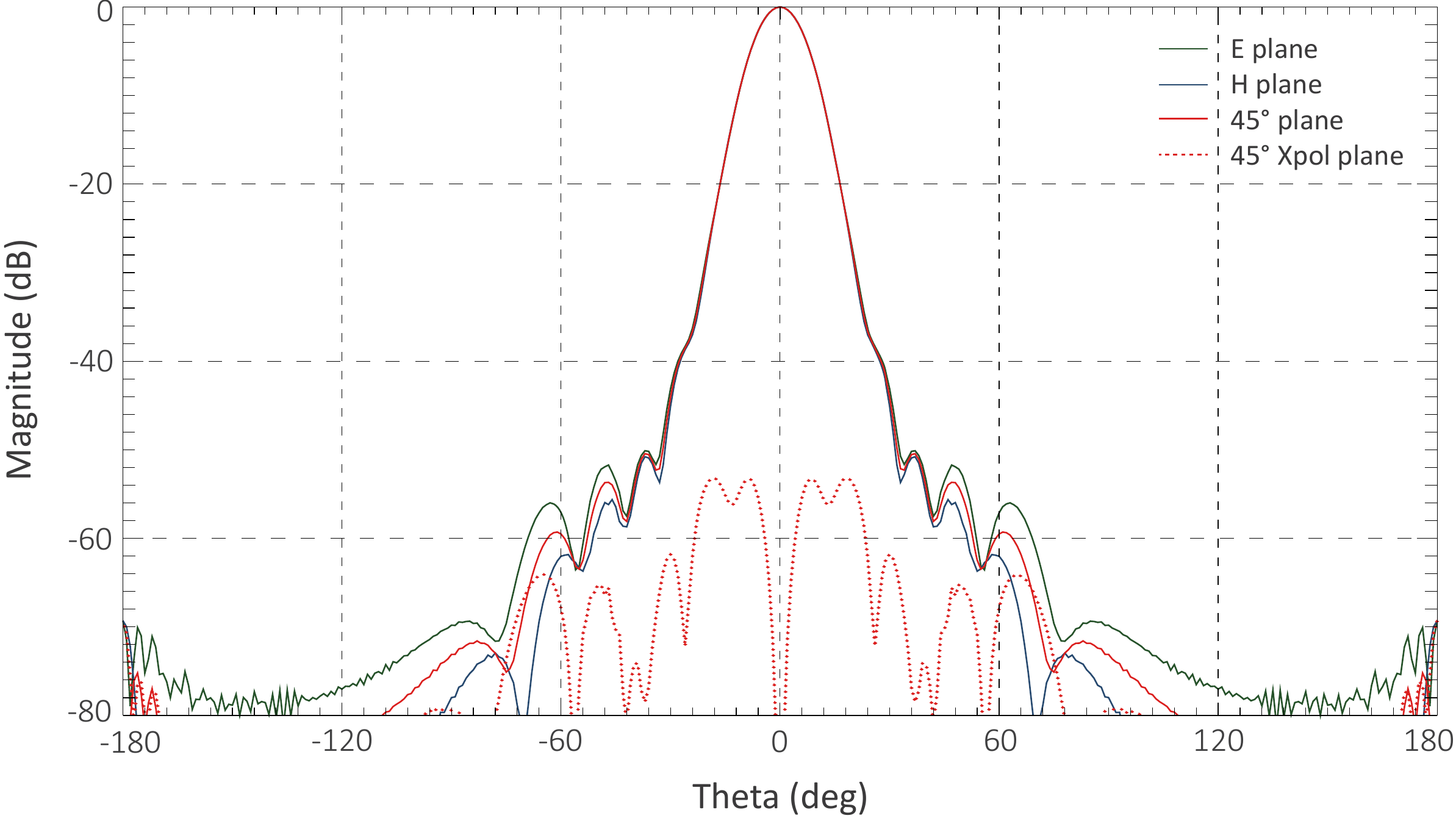}\vspace{0.5 cm}
   \includegraphics[width=.6\textwidth]{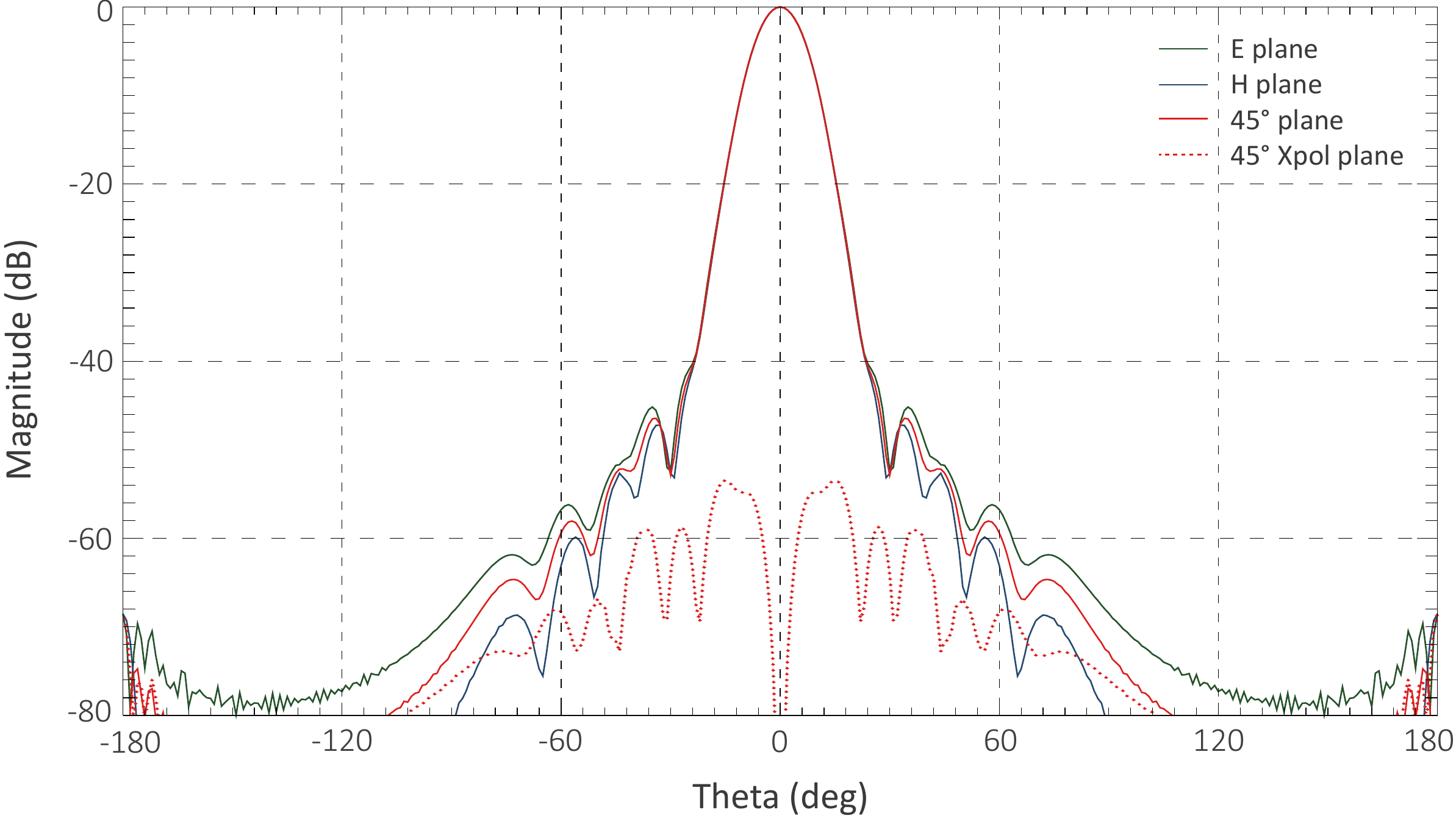}
   \caption{Simulated radiation patterns of the Strip feed horns at 90, 95 and 105~GHz. The maximum cross-polarization level at the three frequencies is $-47.4$~dB, $-53.2$~dB and $-53.5$~dB, respectively.}
   \label{W_fh_beams}
\end{figure}

Radiation patterns have been simulated in the required frequency band, with a 0.1 GHz frequency step. As a reference, Figure~\ref{W_fh_beams} shows the radiation pattern computed at three frequencies in the W-band, i.e. the center frequency $f_0$ and within $\pm10\%$:
\begin{itemize}
   \item{85 GHz ($f_0-10\%$)}
   \item{95 GHz ($f_0$)}
   \item{105 GHz ($f_0+10\%$)}
\end{itemize}
Each plot includes the expected radiation pattern on the co-polar principal planes (E-plane, H-plane and $45\degr$-plane) and cross-polar $45\degr$-plane.

\begin{figure}[htbp]
   \centering
   \includegraphics[width=.7\textwidth]{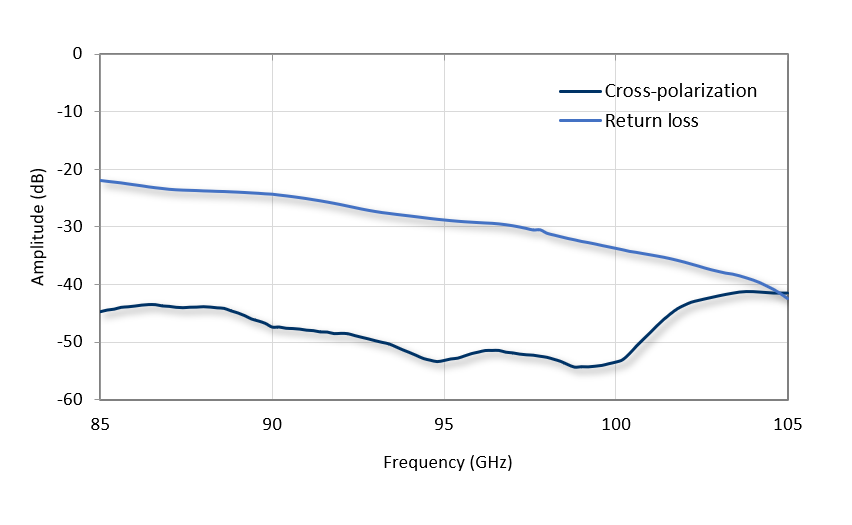}
   \caption{Simulation of the maximum cross-polarization level and return loss as a function of the frequency in the W-band. The cross-polarization level is below $-40$~dB over the whole operative band. The feed horn is matched at a level better than $-22$~dB over the whole 85$-105$~GHz. The return loss is better than $-30$~dB starting from $\sim$97~GHz.}
\label{W_fh_cross_rl}
\end{figure}

Figure \ref{W_fh_cross_rl} shows the level of maximum cross-polarization and return loss over the W-band frequencies between 85 and 105 GHz.

%----------------------------------------
\subsection{Engineering with the platelet technique} %-- CF, FDT
\label{sec:fh_platelet}
All the feed horns have been manufactured by means of the platelet technique~\cite{platelet}, consisting of the overlapping of metal sheets suitably machined to reproduce, when stacked, the feed horn corrugated profile. Q-band modules are engineered so that each hexagonal module includes seven identical feed horns, while the W-band are constructed as single elements by stacking ring plates.

The main advantages of this manufacturing technique, compared to traditional methods such as electroforming, is the drastic reduction in time and cost (up to 90\% for large arrays), while still guaranteeing good electromagnetic performance. Furthermore, platelet horns can easily accommodate any kind of corrugation profile (sometimes virtually impossible with other techniques) with no added cost or fabrication complexity. As a drawback, a platelet array will have a planar aperture plane, thus requiring a telescope with a highly flat focal surface for good matching.
%--- begin rev. #5
This limitation clearly relates to horns belonging to the same module, necessarily sharing the same plane where horns apertures lie. We suitably positioned the hexagonal modules of the Q-band channel on the focal surface of the telescope and oriented them by means of a mechanical support structure.
%--- end rev. #5
Furthermore, a critical requirement of platelet technique is strict control of the platelet assembly and alignment system. For space applications, a further concern is the relatively high mass of the array, which can be mitigated by machining additional cavities in the staked metal package to lighten the whole structure.

%----------------------------------------
\subsubsection{Manufacturing of the Q-band modules}
\label{sec:fh_platelet_Q}
The design of the LSPE/Strip Q-band 7-element modules is based on the W-band array \cite{ICEAA2013-6632377} we developed within the framework of a technological development funded by the Italian Space Agency (ASI). An extensive test campaign including thermal and structural analyses was performed on the W-band prototype, resulting in a potentially space-qualified unit. This led to a similar mechanical engineering of the Strip Q-band modules.

\begin{figure}[htbp]
   \centering
   \includegraphics[width=1.0\textwidth]{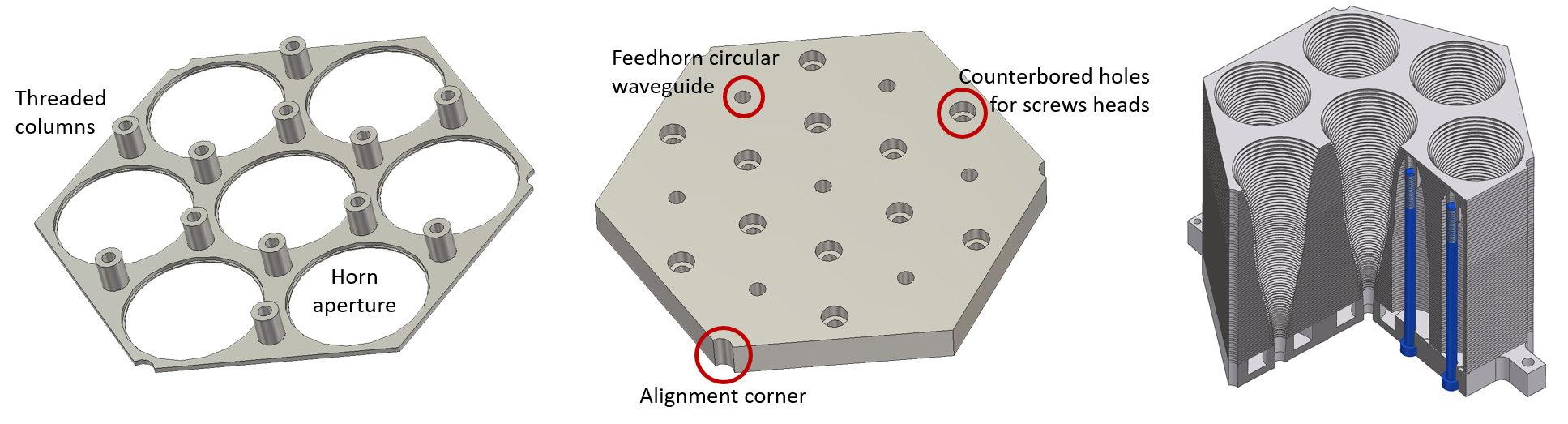}
   \caption{Drawing of a Strip Q-band feed horn module. \textit{Left}: the 16~mm thick base plate. The back of the plate includes the mechanical interface to the polarizer flanges, with pin holes for the alignment (not represented in this sketch). \textit{Center}: the module head plate includes the \quotes{bolts} (i.e. internally threaded cylinders) for the screws on its back side, providing a clean ground plane for the seven feed horn apertures. \textit{Right}: a section of an assembled hexagonal module, showing the M6 screws passing through the lightening cavities of the plates.}
   \label{Strip_Q_module}
\end{figure}

The Strip hexagonal modules are built with aluminum alloy (Anticorodal, Al6082) plates as follows:
\begin{itemize}
	\item{a 16~mm thick base plate to accommodate the screws heads and provide the interface flange to seven polarizers (see left panel of Figure~\ref{Strip_Q_module});}
	\item{72 plates with seven dual diameter holes, i.e. a corrugation for each horn, and twelve additional cavities for a ${\sim}50\%$ mass reduction. Plates 1-50 ($sin^2$ profile) are 1.5~mm thick; plates 51-72 ($exp$ profile) are 2~mm thick (see right panel of Figure~\ref{Strip_Q_module}).
	%--- begin rev. #7
    All hexagonal plates are manufactured using rectified calibrated Al6082 sheets not subjected to a stress relieving process, suitably machined in order to avoid deformations;}
    %--- end rev. #7
	\item{a 2~mm thick head plate to accommodate the threads for clamping the whole assembly (see center panel of Figure~\ref{Strip_Q_module}).}
\end{itemize}
The plates, fabricated with a ${\sim}0.02$~mm tolerance, are tightened together by twelve M6 screws (Ergal, Al7075) passing through the lightening cavities and directly screwed to the head plate. Before assembly, each plate was measured with a CNC metrology machine Werth$\copyright$ Scope Check 200 (1.8~$\mu$m accuracy in xyz axes) to verify manufacturing tolerances.

We aligned the plates externally, using alignment pins at the plate corners (see center panel of Figure~\ref{Strip_Q_module}). The twelve screws are tightened with a torque wrench and the alignment tool removed. The aluminum alloys chosen for the plates (Al6082) and screws (Al7075) are suitable for a cryogenic environment at 20~K, ensuring homogeneous thermal contractions. In fact, they have an almost identical thermal expansion coefficient, 2.34~$10^{-5}$~K$^{-1}$ and 2.35~$10^{-5}$~K$^{-1}$ for Al6082 and Al7075, respectively. Furthermore, the main alloying agent of Al7075 is zinc, which favors the self-hardening of the alloy, providing a high tensile strength, comparable to that of some steels, at a density about three times lower.

\begin{figure}[htbp]
   \centering
   \includegraphics[width=6.6 cm]{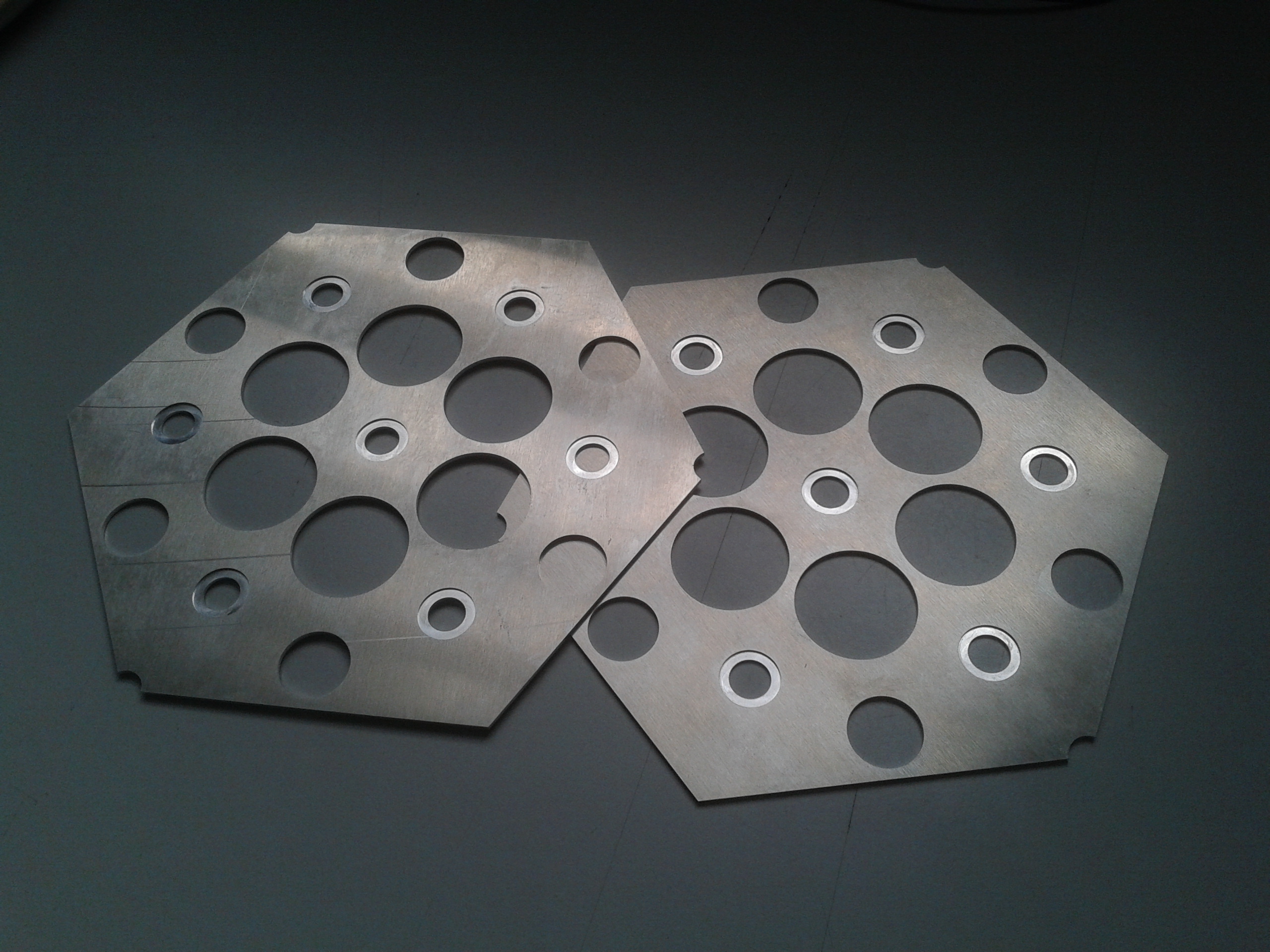}\hspace{5 pt}
   \includegraphics[width=6.6 cm]{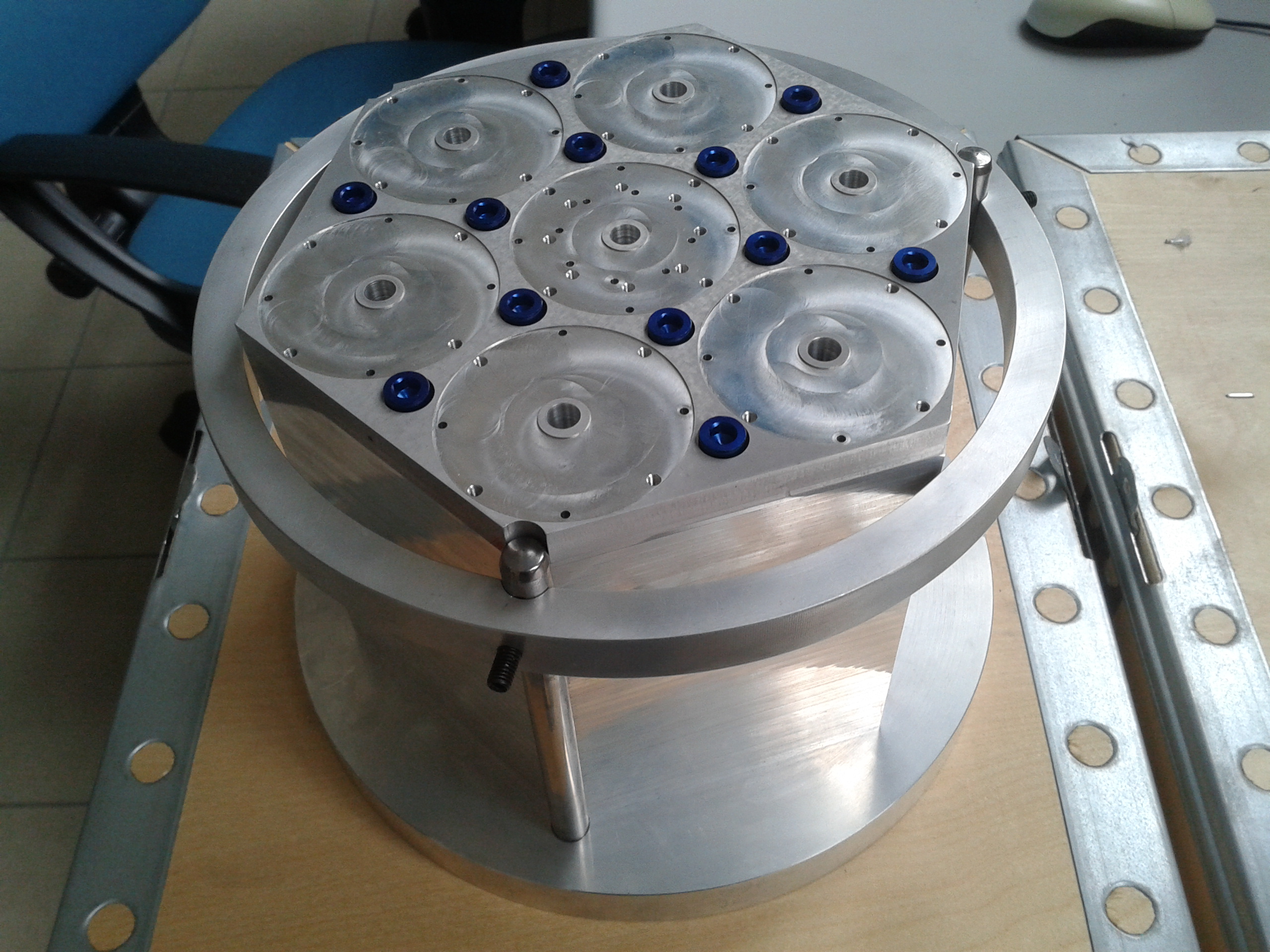}\vspace{5 pt}
   \includegraphics[width=6.6 cm]{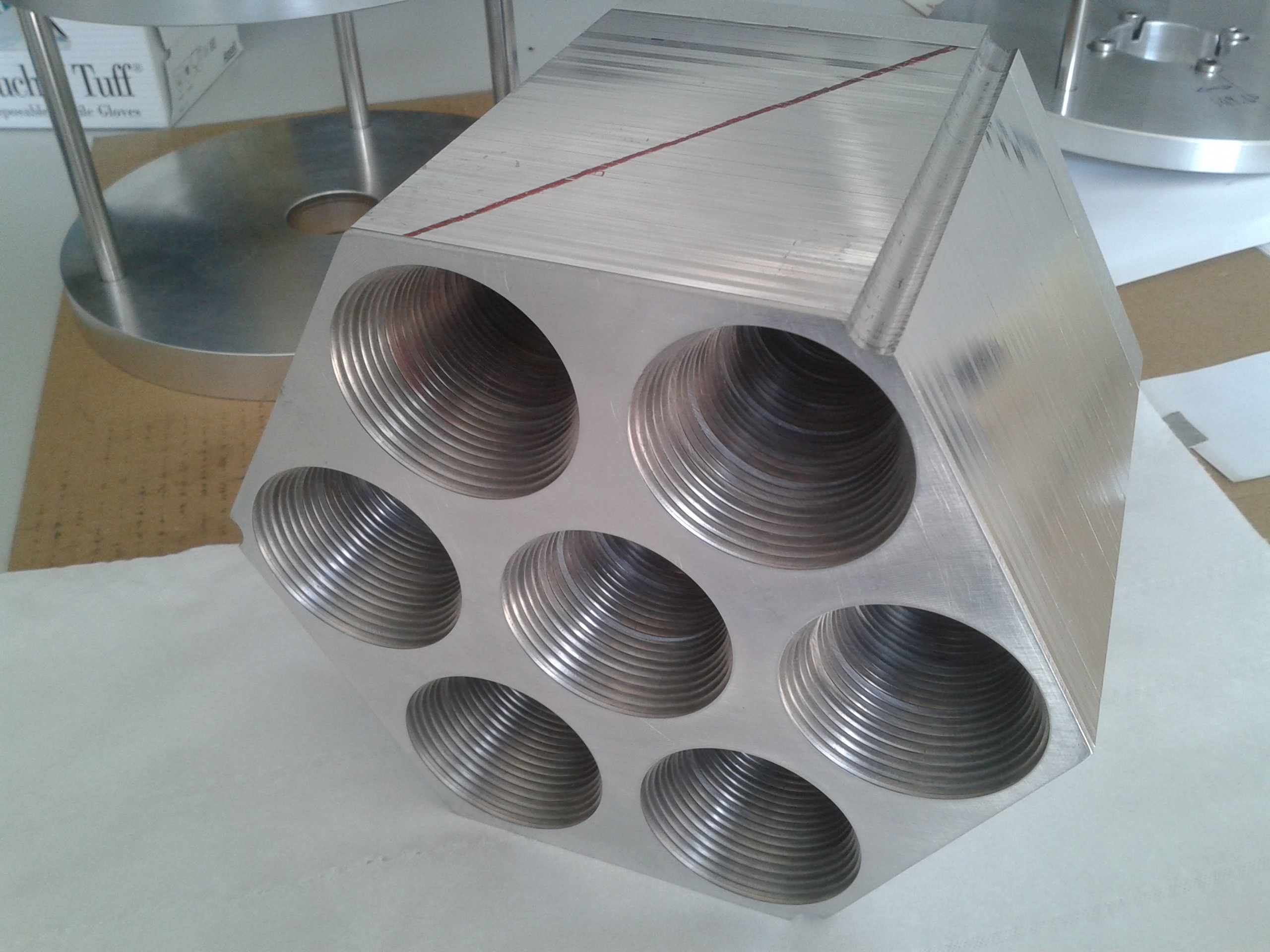}\hspace{5 pt}
   \includegraphics[width=6.6 cm]{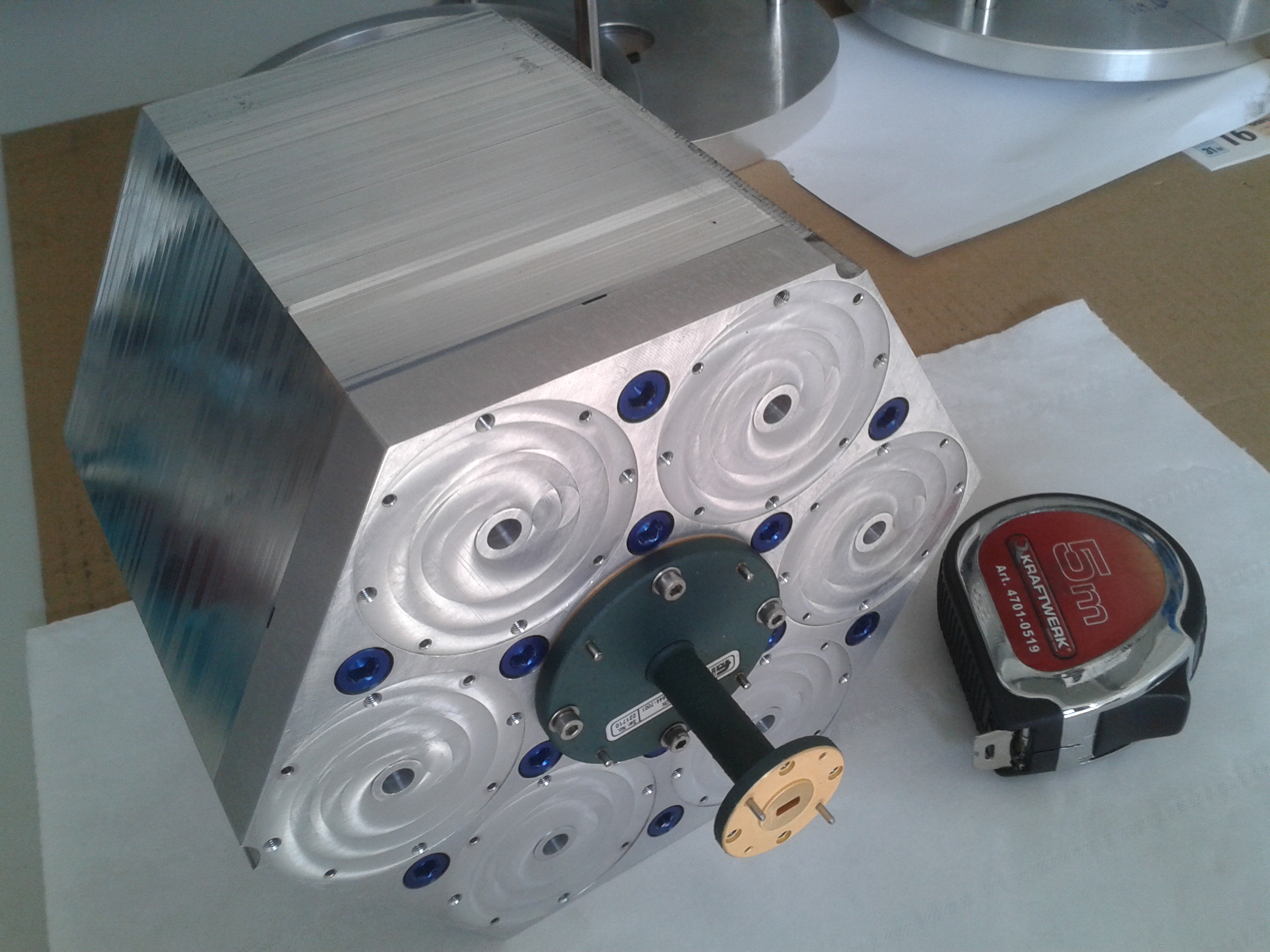}\vspace{5 pt}
   \includegraphics[width=6.6 cm]{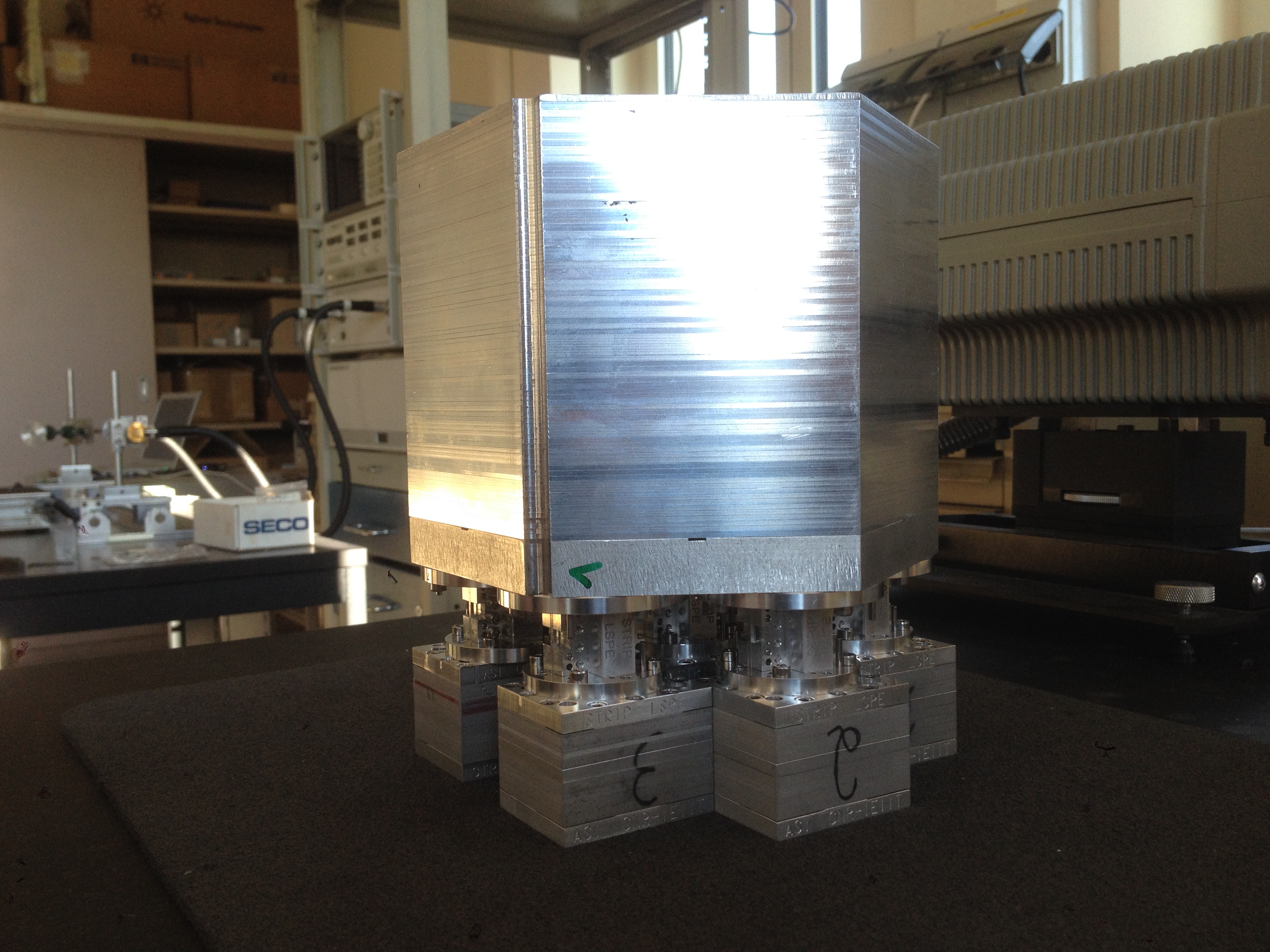}\hspace{5 pt}
   \includegraphics[width=6.6 cm]{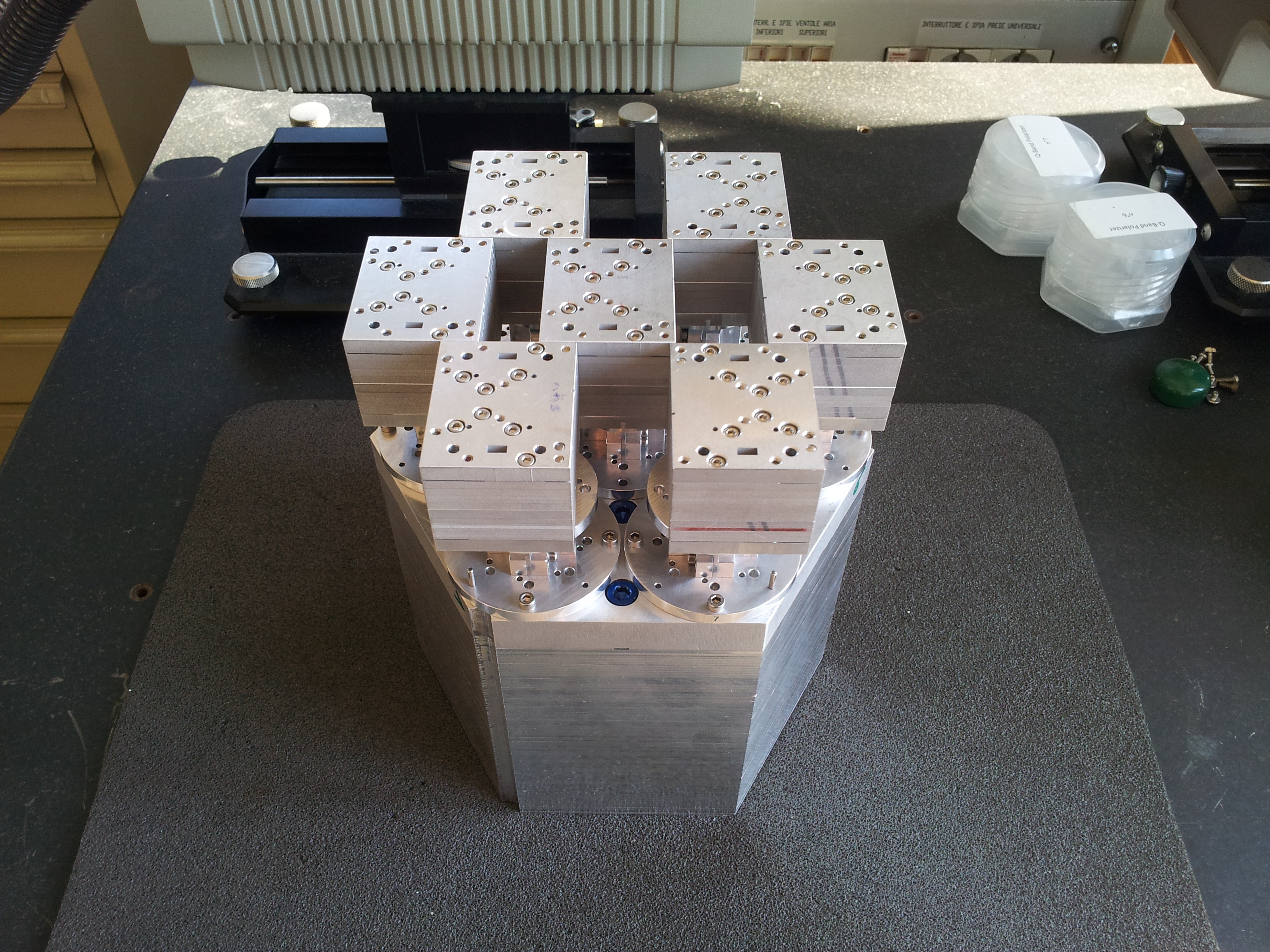}
   \caption{\textit{Top-left}: two plates of an hexagonal module. The seven dual-diameter holes of the corrugation are distinguishable from the twelve bigger lightening cavities. \textit{Top-right}: the assembly and alignment system with the three thick pins; the circular tool on top of the assembly prevents pins from moving and keeps them perfectly aligned and parallel. \textit{Middle-left}: a front view of the first realized module showing the seven feed horns apertures. \textit{Middle-right}: a rear view of the same module, with a circular-to-rectangular transition assembled. The transition flange is similar to the polarizer's one. \textit{Bottom}: the module is assembled with polarizers and orthomode transducers (OMTs) for return loss measurements.}
   \label{STRIP_HexPhoto}
\end{figure}

In Figure~\ref{STRIP_HexPhoto} we show several views of one of our Q-band modules, including single platelets, a fully integrated module, and the mating with the dual circular polarization waveguide components.

%----------------------------------------
\subsubsection{Manufacturing of the W-band modules}
\label{sec:fh_platelet_W}
The W-band feed horn assemblies were engineered with the platelet technique, by overlapping 197 chemical etched aluminum rings, each of them representing a 0.3~mm thick tooth or a 0.6~mm thick slot of a corrugation of the horn. A 3~mm thick head plate, including three whole corrugations and a 0.3~mm tooth, was machined and provided the mechanical interface for the assembly (six threaded holes) and alignment (pin holes) of the horn plates. The 10~mm thick base plate provided the counterbore holes for the screws and the holes for the alignment pins, as well as the 2.62~mm diameter waveguides of the antennas.

\begin{figure}[htbp]
   \centering
   \includegraphics[width=0.7\textwidth]{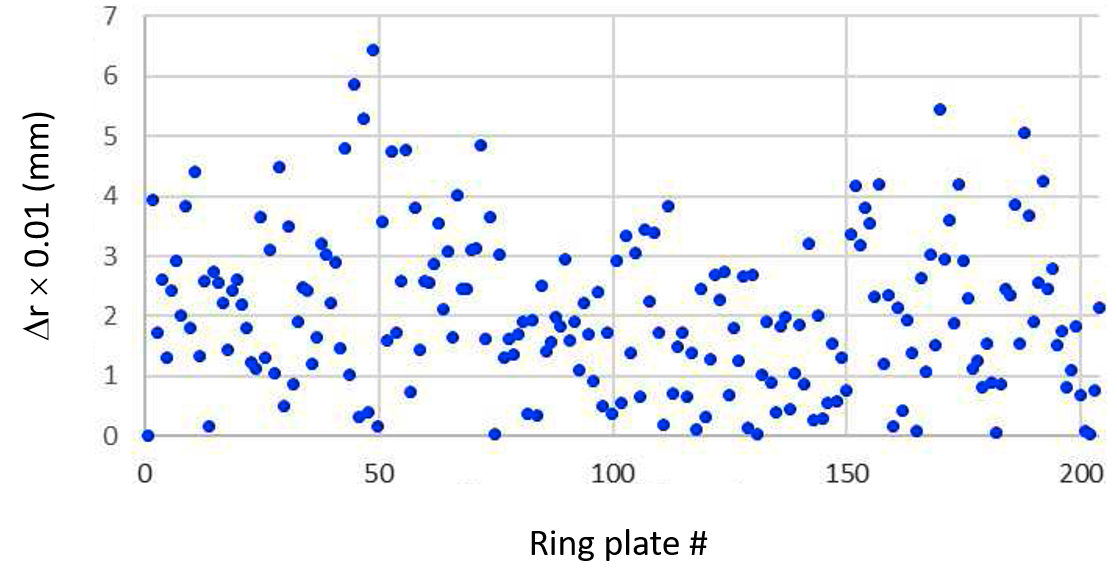}
   \caption{Absolute difference between nominal and measured values of holes radius of the ring-plates of the first W-band feed horn unit.}
   \label{fig:radius}
\end{figure}

The hole radii of the aluminum rings were measured with the multi-sensor coordinate measuring machine Scope Check 200. Figure~\ref{fig:radius} shows the discrepancy between measured and nominal hole radii for the ring-plate of our first W-band unit, with values within $\sim$0.06~mm. While acceptable, this tolerance is significantly worse than what we find for platelets obtained via traditional machining
%--- begin rev. #7
%($\sim$0.03~mm)
($\sim$0.02~mm),
%--- end rev. #7
indicating a somewhat lower quality in the chemical etching process.
Moreover, ring-plates thickness measurements showed a maximum discrepancy of about 0.006~mm (2\% and 1\% of the nominal 0.3 mm and 0.6 mm plates thickness, respectively). As discussed in section \ref{sec:W-meas-pattern}, this resulted in a decrease of the horns total length by about 2.25 mm, given the large number of plates making up each antenna.
%--- begin rev. #7
We did not observe deformations of the plates due to the chemical etching process.
%--- end rev. #7

\begin{figure}[htbp]
   \centering
   \includegraphics[width=0.95\textwidth]{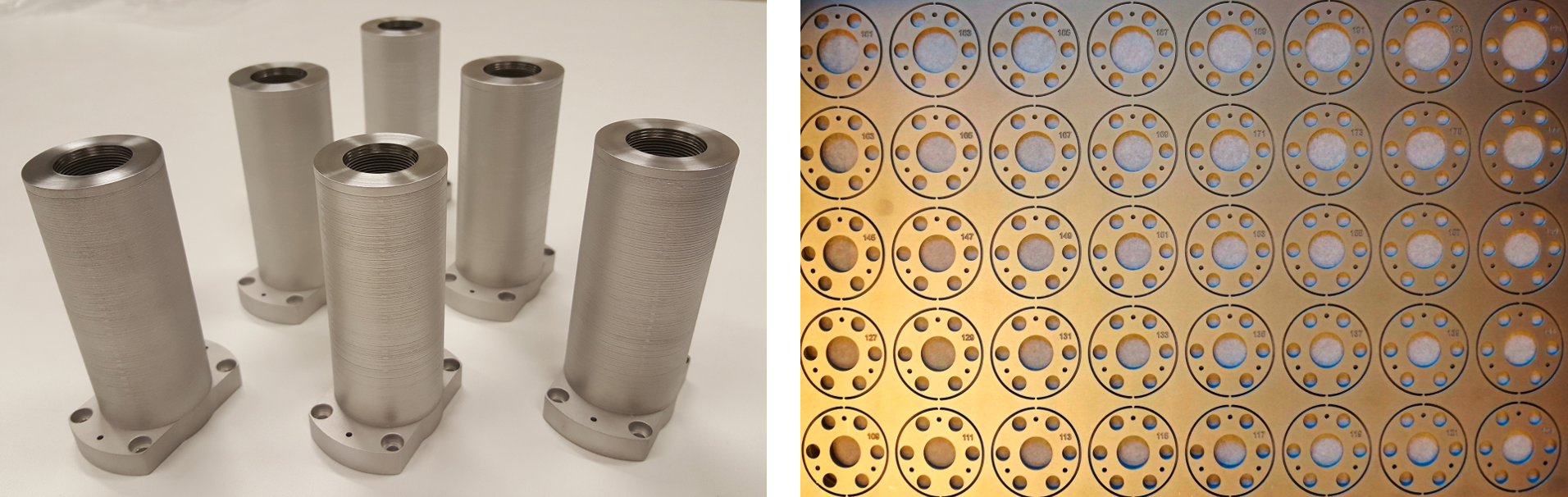}
   \caption{\textit{Left}: picture of the six W-band feed horn assemblies. \textit{Right}: detail of the chemical etched aluminum rings. The number of each ring is also engraved on the plate surface.}
   \label{fig:feed_W}
\end{figure}

The left panel of Figure~\ref{fig:feed_W} shows a picture of the six W-band feed horns. The right panel of the same figure shows a detail of the chemical etched aluminum rings, 0.6~mm thick, representing the throats of the horn corrugations. The six through holes for the screws and the three holes for the alignment pins are clearly visible in each ring plate.

%----------------------------------------
\section{Feed horn characterization}
\label{sec:measurements}
The forty-nine Q-band and the six W-band feed horns have been fully characterized in terms of their radiation pattern and return loss in their operating frequency band. All measurements have been performed in the anechoic chamber of the Physics Department at the Università degli Studi di Milano, with the exception of the return loss measurements in the Q-band, which were performed at the CNR--IEIIT laboratories in Torino.

Hereafter we report details and results of the Q-band and W-band test campaigns.

%----------------------------------------
\subsection{Q-band measurements}
\label{sec:Q-meas}

%----------------------------------------
\subsubsection{Radiation patterns}
\label{sec:Q-meas-pattern}
Radiation pattern measurements were performed in the far-field regime. The aperture of the Q-band feed horns is 50~mm, leading to a far-field distance $d_{\rm ff}=716.6$~mm at 43~GHz. During the measurements, the distance between the source antenna and the antenna under test (AUT) was set to $\sim$1860~mm, i.e. ${\sim}2.6 \times d_{\rm ff}$.

\begin{figure}[htbp]
   \centering
   \includegraphics[width=0.7\textwidth]{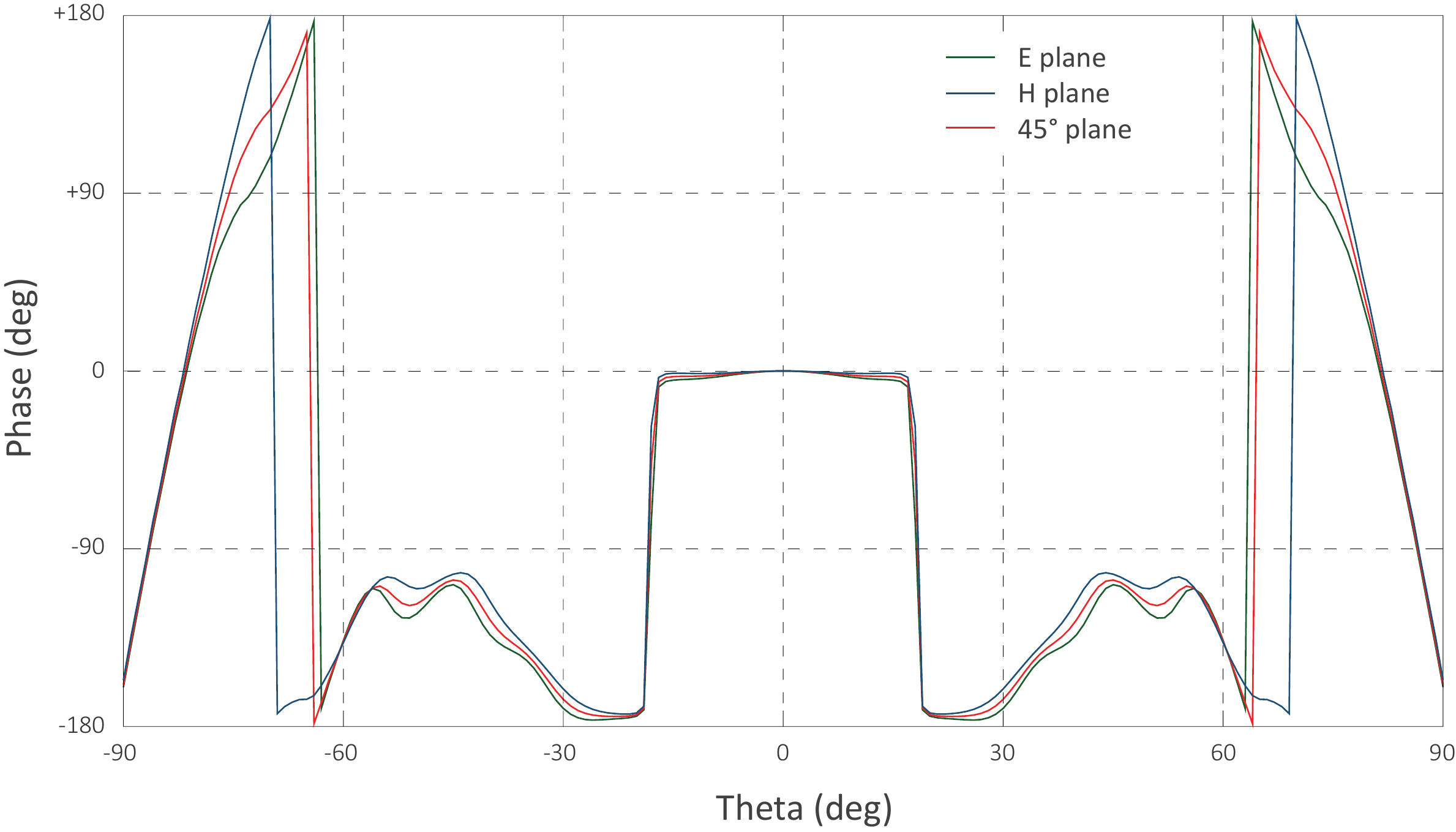}
   \caption{Simulated phase diagram at 43~GHz, when the phase center is 21 mm inside the horn aperture plane. Phase is stable in the main beam angular range (about $\pm15\degr$) in all co-polar planes.}
   \label{STRIP_FH_phase}
\end{figure}

The AUT was positioned in the facility with its phase center aligned to the azimuth axis. In principle, since the phase center changes with frequency, its position is not uniquely defined within the horn band, producing a slightly elongated \quotes{phase center region} rather than a single well-defined point. Nevertheless we used \srsr simulations to optimize the position of the phase center at the center frequency $f_0$, by minimizing the phase diagrams (relevant to the three planes E, H and $45\degr$) variations in the angular region of the main beam, resulting in a position 21~mm inside the horn aperture.
Figure~\ref{STRIP_FH_phase} shows the phase diagram in the angular range $\pm90\degr$: phase is stable in the main beam angular range (about $\pm15\degr$) in all co-polar planes.\par

\begin{figure}[htbp]
   \centering
   \includegraphics[width=0.5\textwidth]{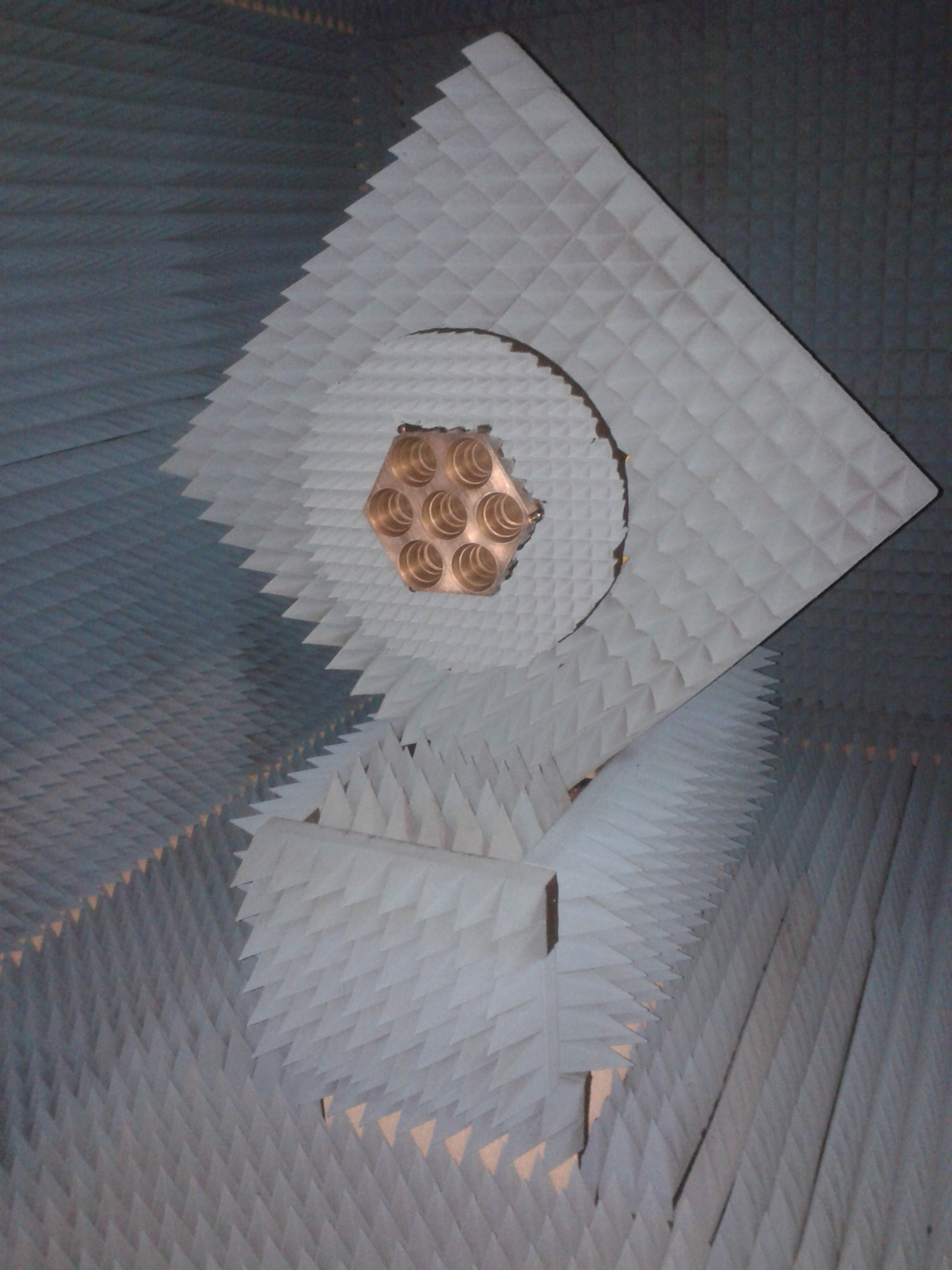}
   \caption{A Q-band hexagonal module is mounted on the positioner of the anechoic chamber at the Physics Dept. of the Università degli Studi di Milano. The AUT is covered with the Eccosorb$^{\circledR}$ to minimize spurious reflections inside the chamber without altering the horn response.}
   \label{STRIP_Q_setup}
\end{figure}

Figure~\ref{STRIP_Q_setup} shows the experimental setup with an hexagonal module mounted on the azimuth/polarization positioner of the anechoic chamber. The source antenna signal is fed by a Microwave Signal Generator MG3690C by Anritsu$\copyright$, amplified by a ERZ-HPA-3300-4700-27E microwave amplifier by Erzia$\copyright$, and detected by a 85025D power detector by Agilent$\copyright$ at the AUT. The readout is performed by means of the scalar network analyzer HP8757D by Agilent$\copyright$.

Antenna radiation patterns were measured at six frequencies over the operative bandwidth:
\begin{itemize}
	\item{38.7~GHz ($f_0-10\%$)}
	\item{40.85~GHz ($f_0-5\%$)}
	\item{42~GHz ($f_0-2.3\%$)}
	\item{43~GHz ($f_0$)}
	\item{45.15~GHz ($f_0+5\%$)}
	\item{47.3~GHz ($f_0+10\%$)}
\end{itemize}
on the four co-polar principal planes (E, H and $\pm$45\degr) and the two cross-polar $\pm$45\degr\ planes.

\begin{figure}[htbp]
   \centering
   \includegraphics[width=6.8 cm]{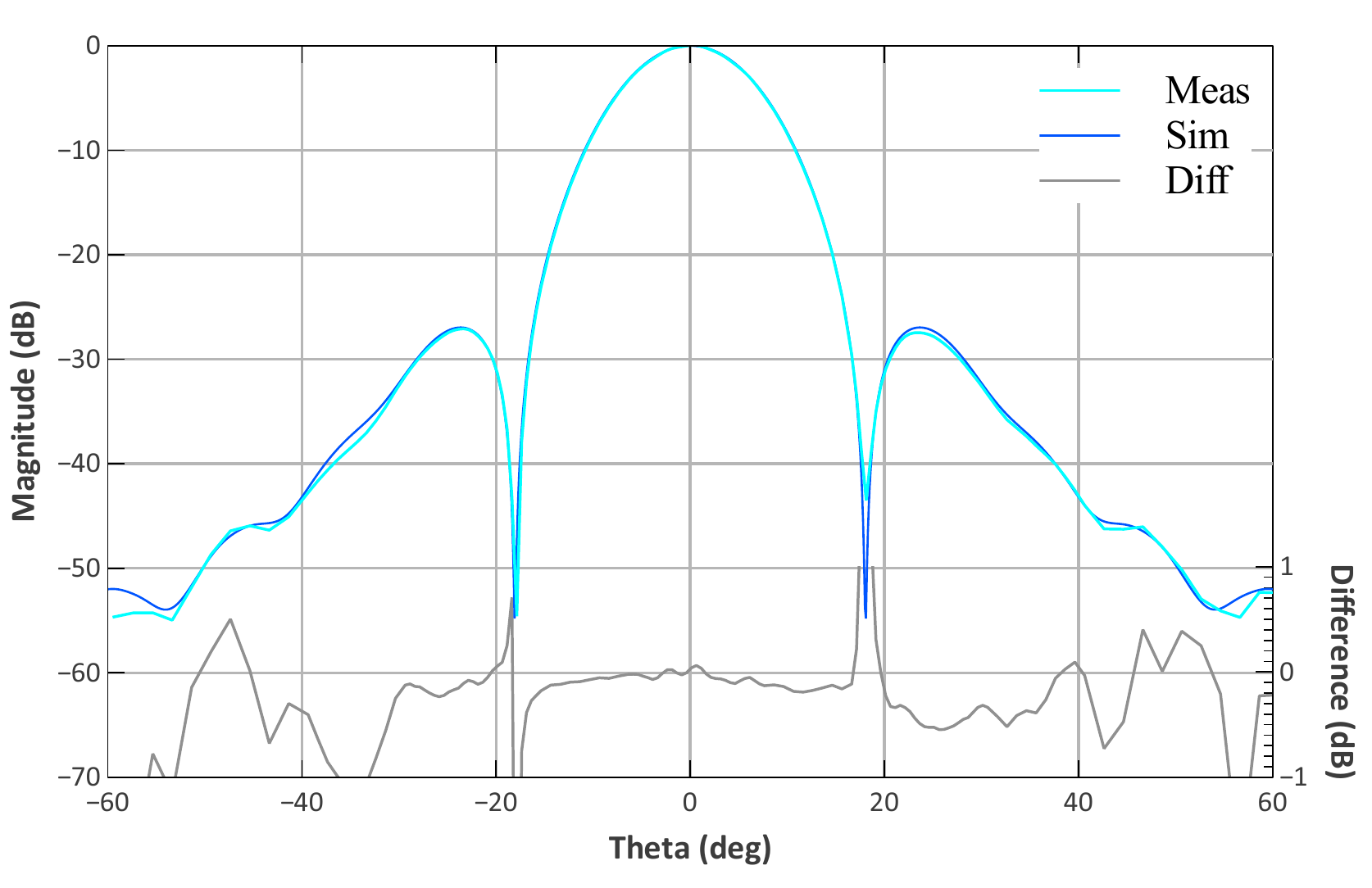}%\hspace{1 pt}
   \includegraphics[width=6.8 cm]{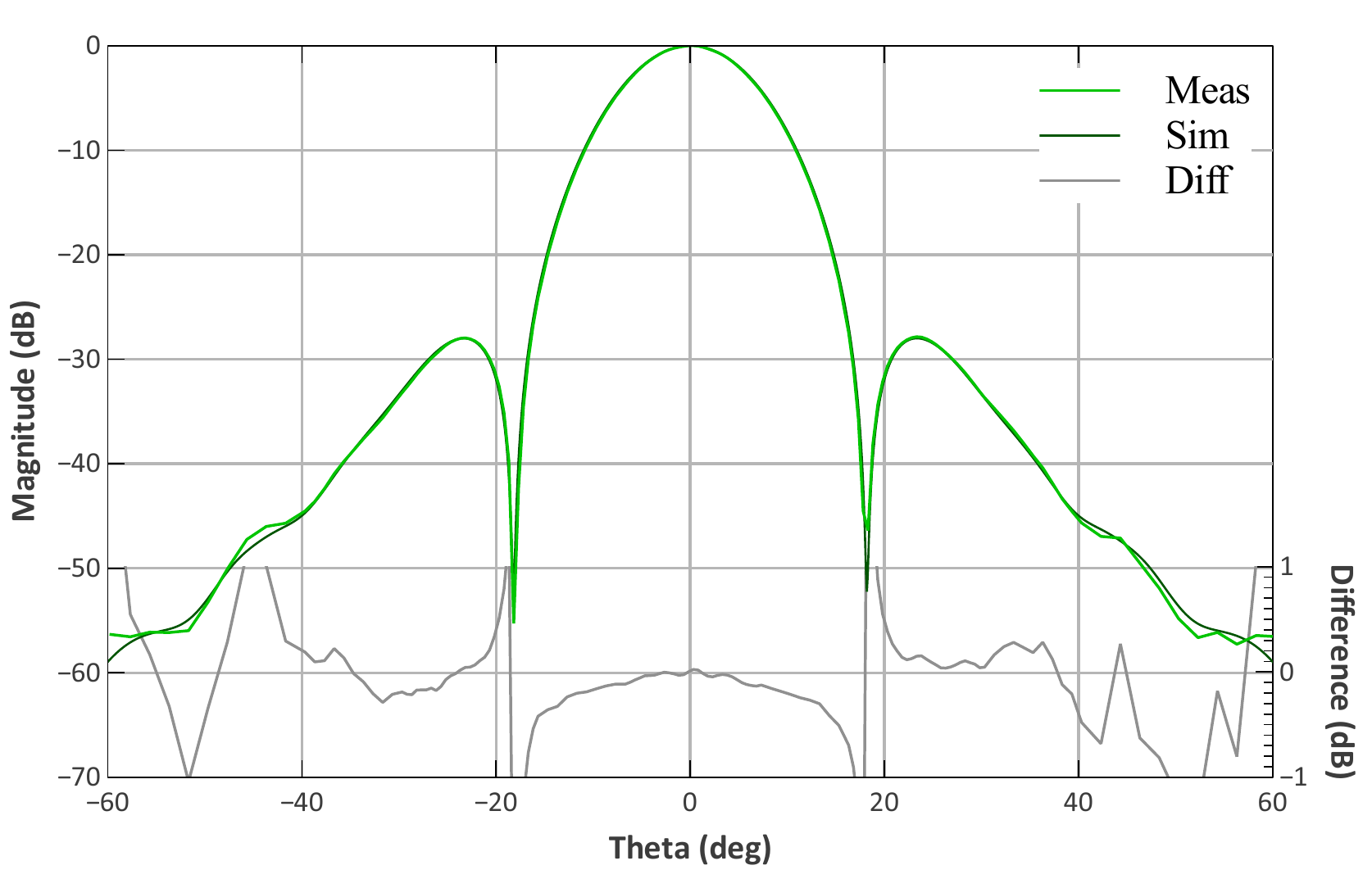}\vspace{20 pt}
   \includegraphics[width=6.8 cm]{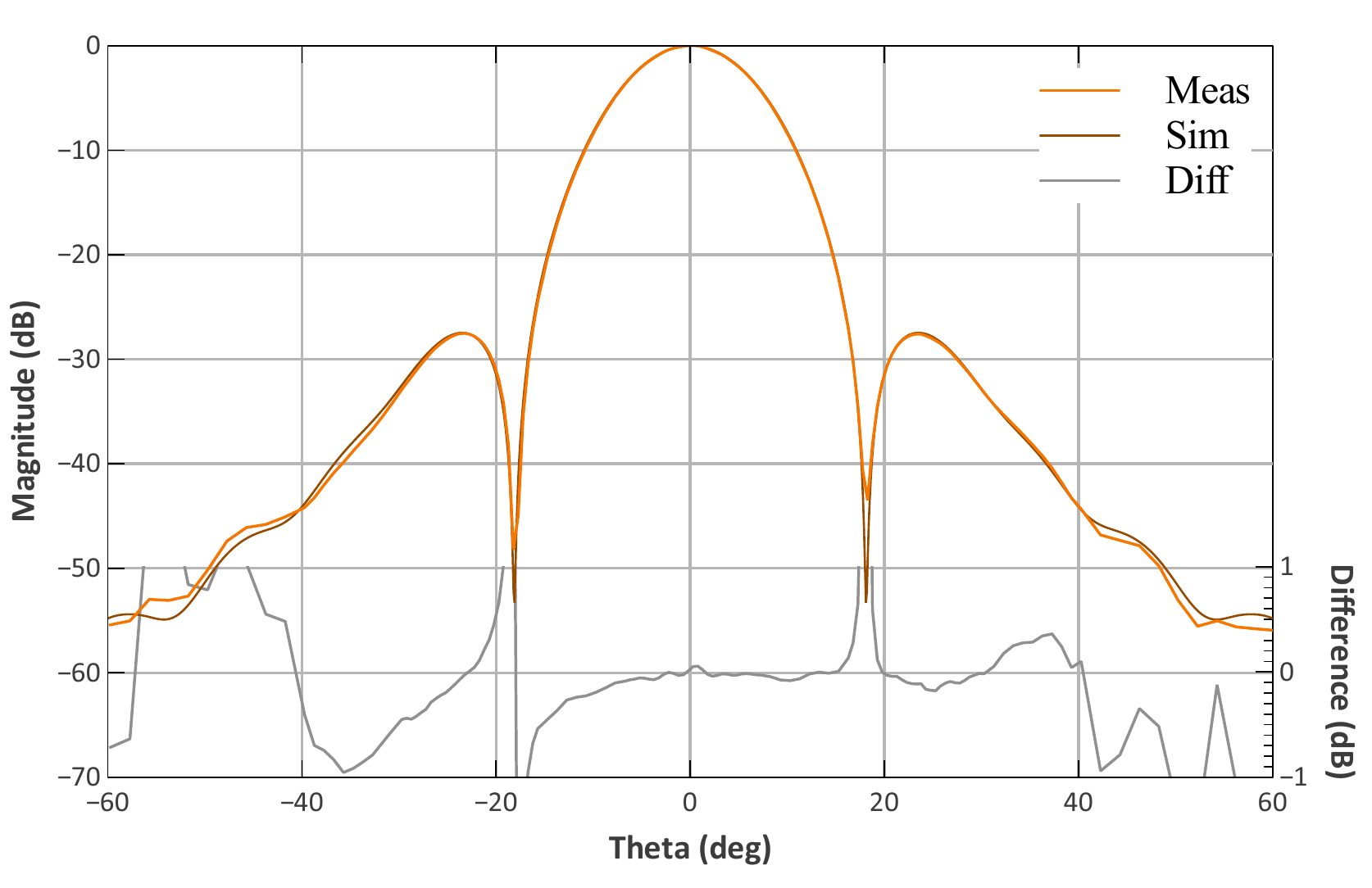}%\hspace{1 pt}
   \includegraphics[width=6.8 cm]{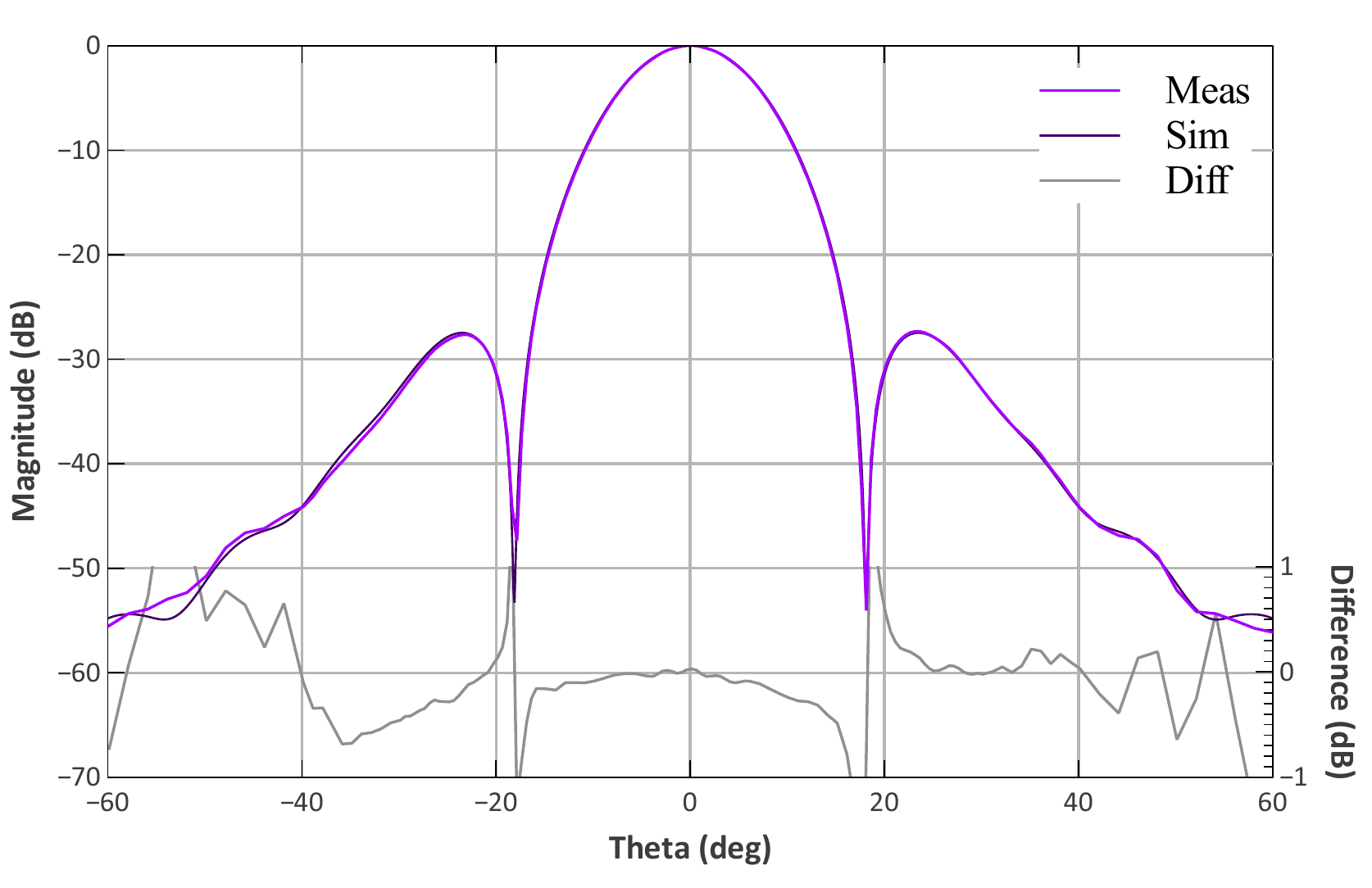}\vspace{20 pt}
   \includegraphics[width=6.5 cm]{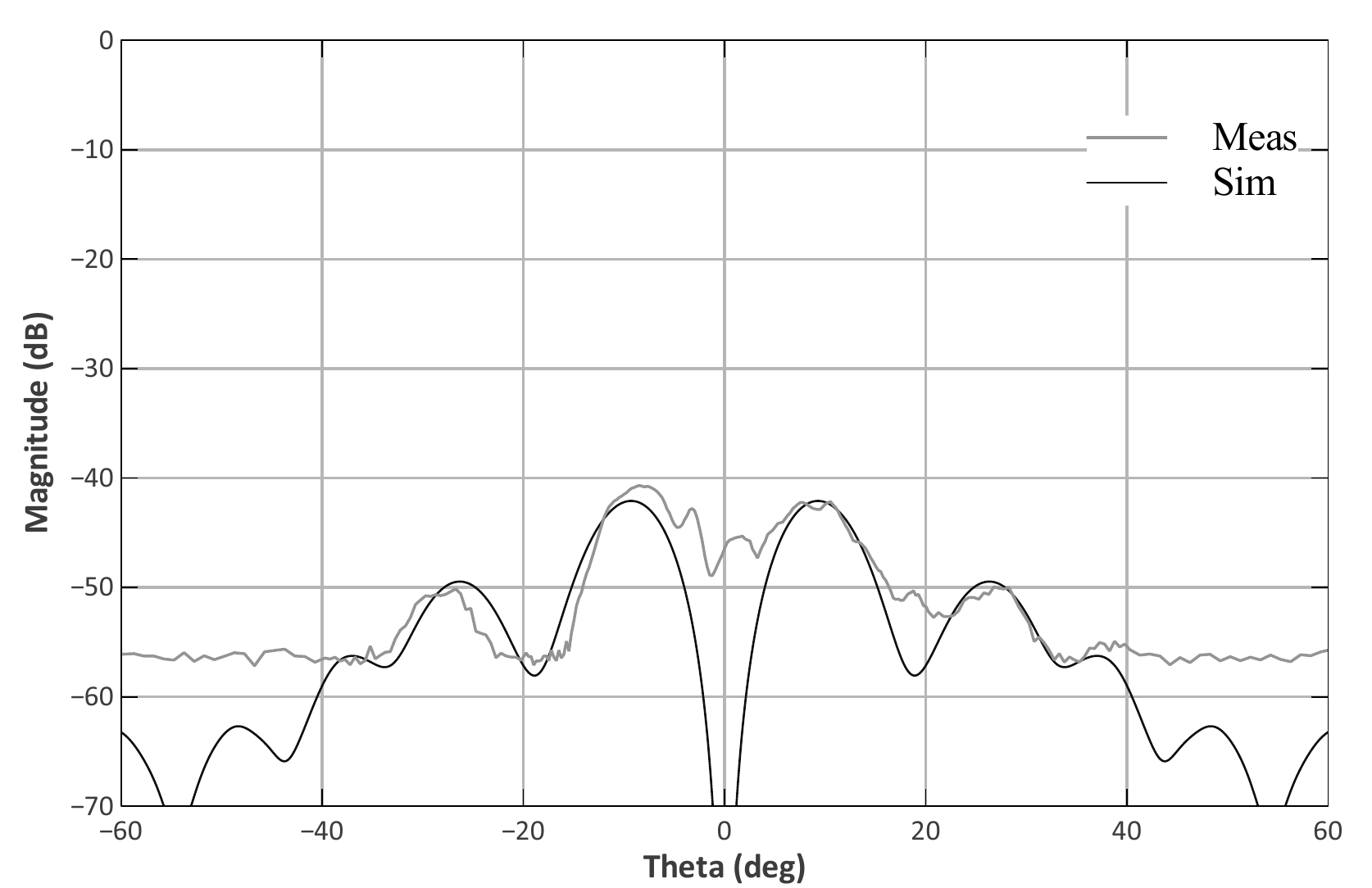}\hspace{10 pt}
   \includegraphics[width=6.5 cm]{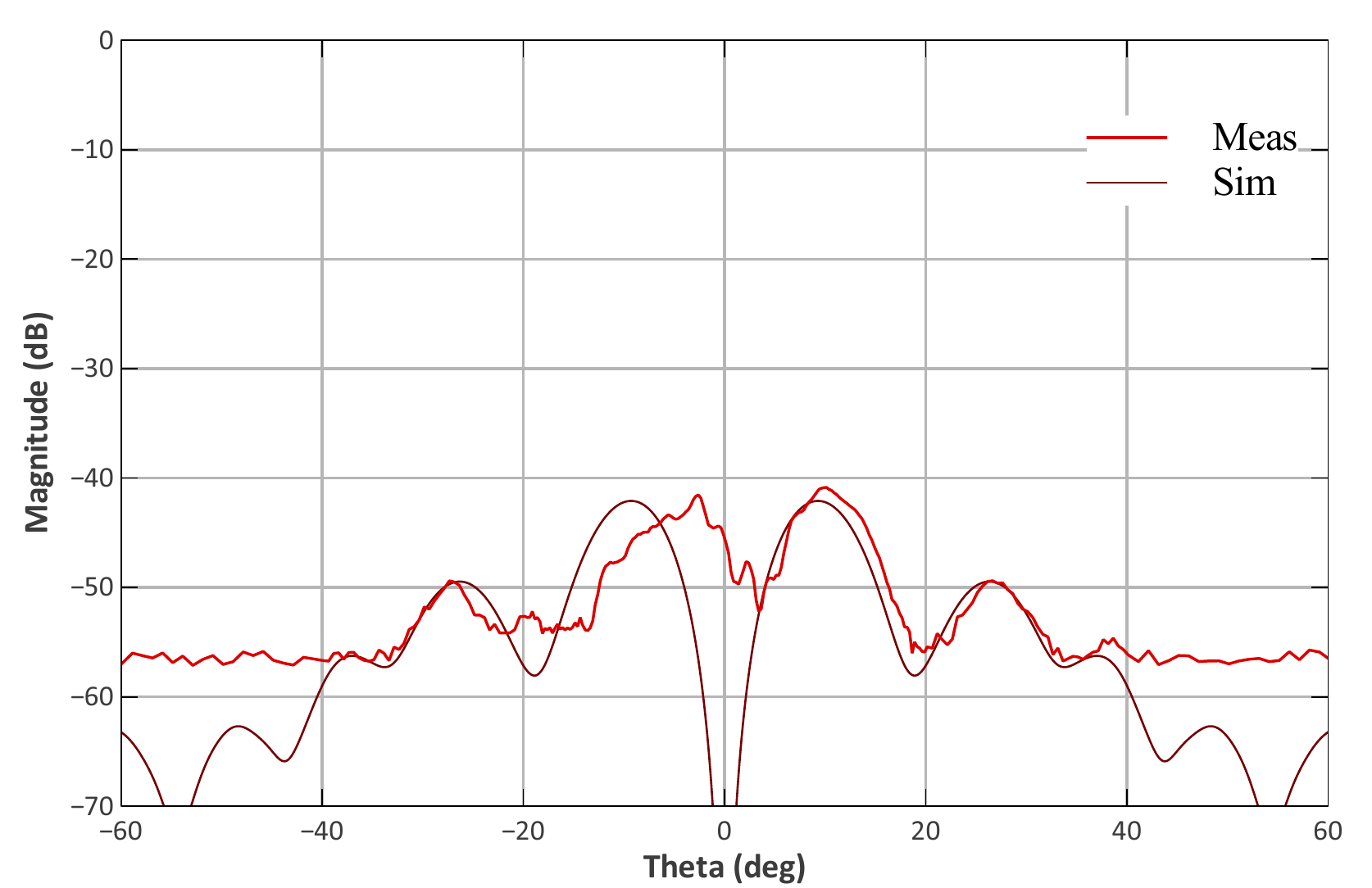}%\hspace{10 pt}
   \caption{Strip V$_0$ (see text) measured radiation patterns at 43~GHz ($f_0$). The difference in magnitude (dB) between measured and simulated patterns is reported at the bottom of each co-polar plot. From left to right, \textit{Top}: co-polar E-plane and co-polar H-plane. \textit{Middle}: co-polar $45\degr$ plane and co-polar $-45\degr$ plane. \textit{Bottom}: cross-polar $45\degr$ plane and cross-polar $-45\degr$ plane.}
   \label{STRIP_FH_V0}
\end{figure}

\begin{figure}[htbp]
   \centering
   \includegraphics[width=6.7 cm]{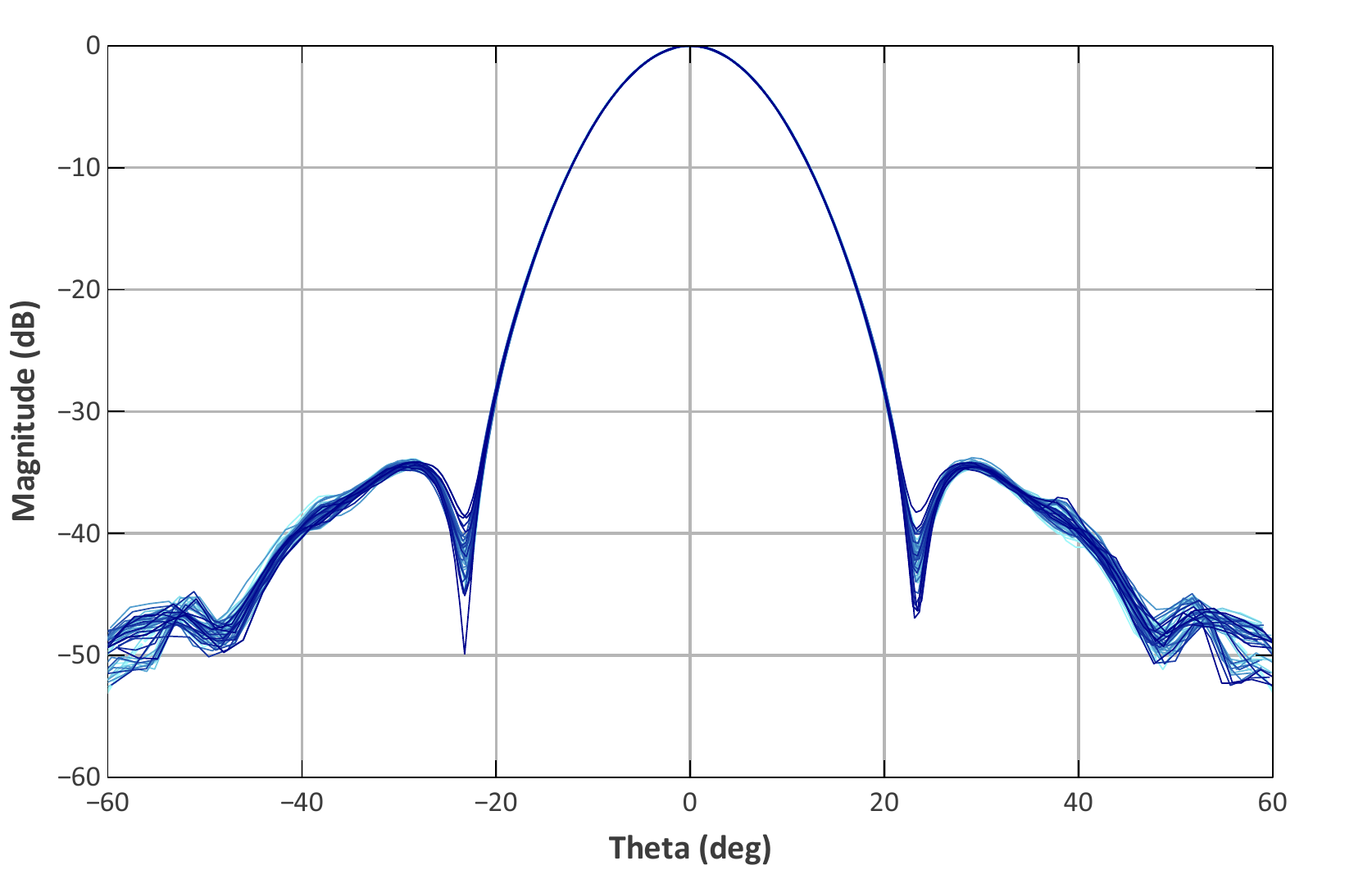}
   \includegraphics[width=6.7 cm]{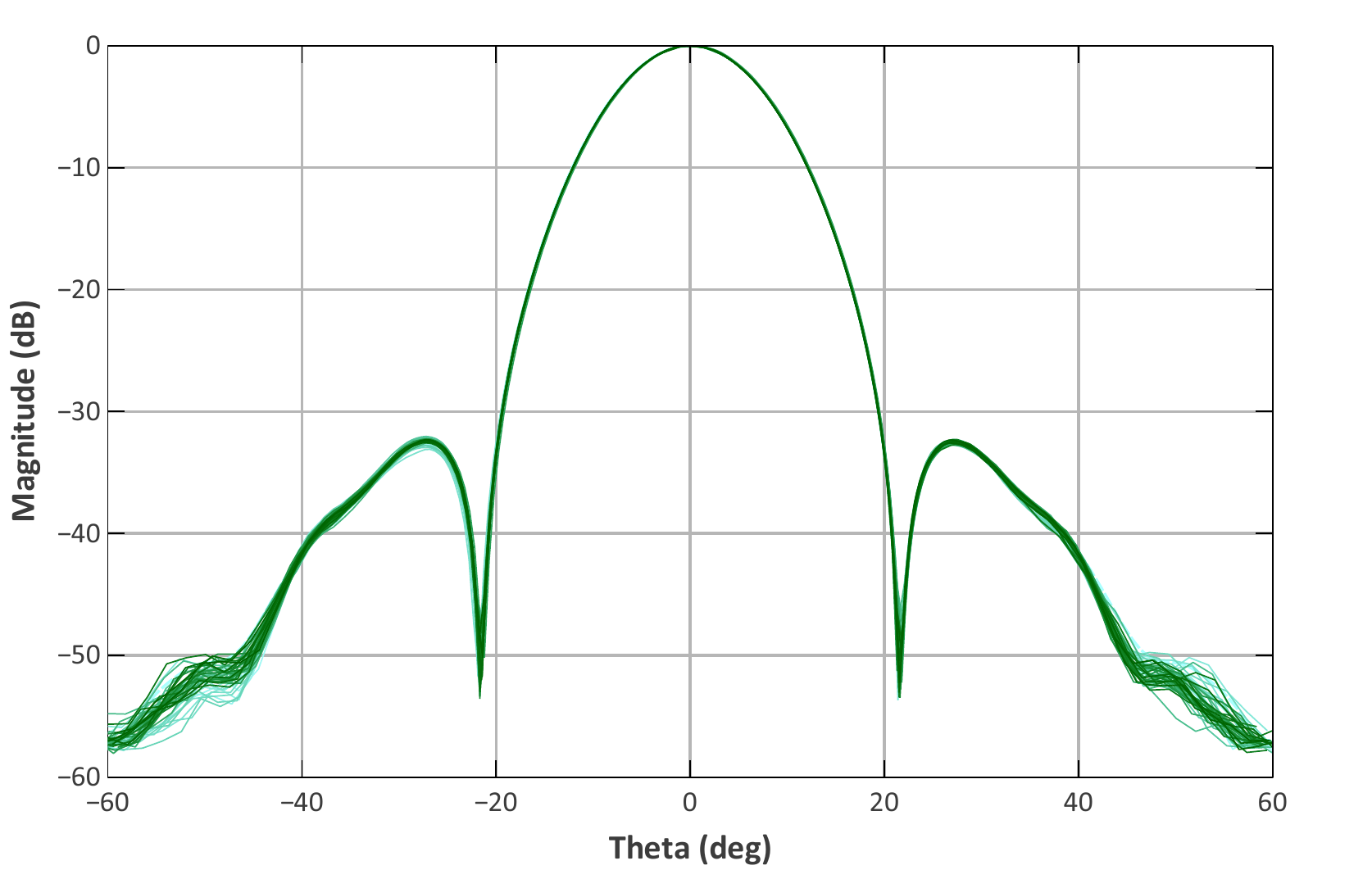}\vspace{20 pt}
   \includegraphics[width=6.7 cm]{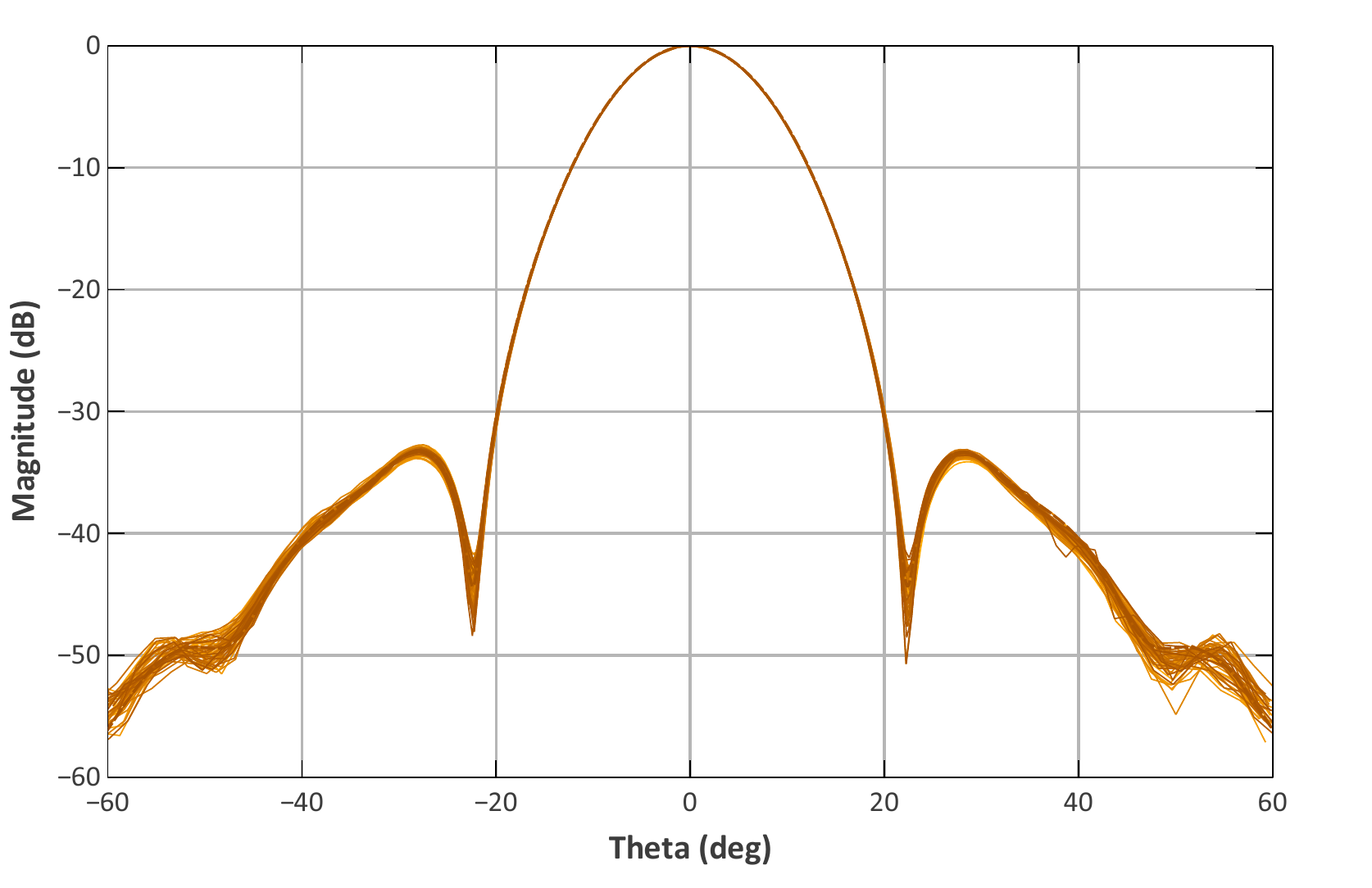}
   \includegraphics[width=6.7 cm]{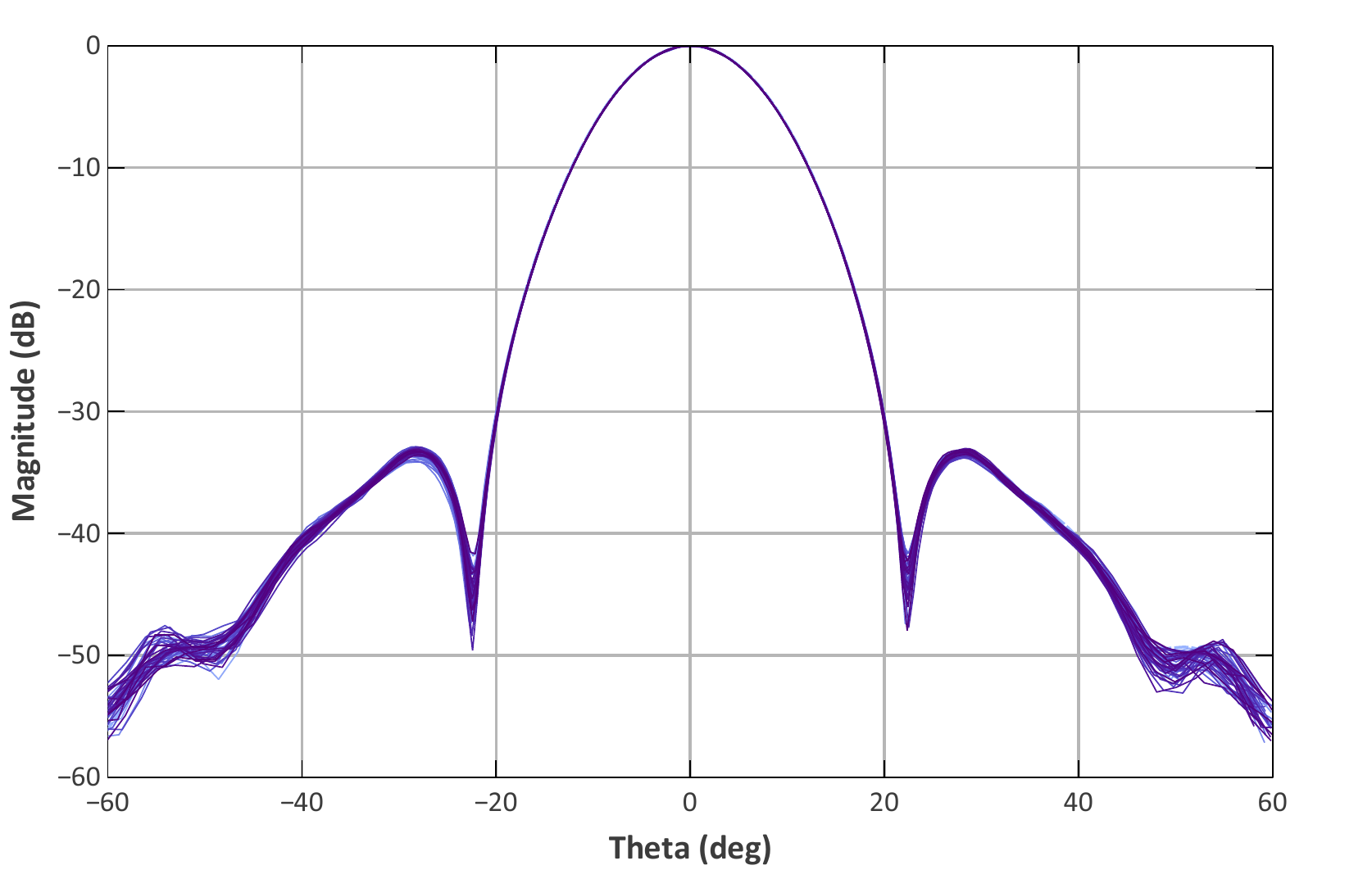}\vspace{20 pt}
   \includegraphics[width=6.7 cm]{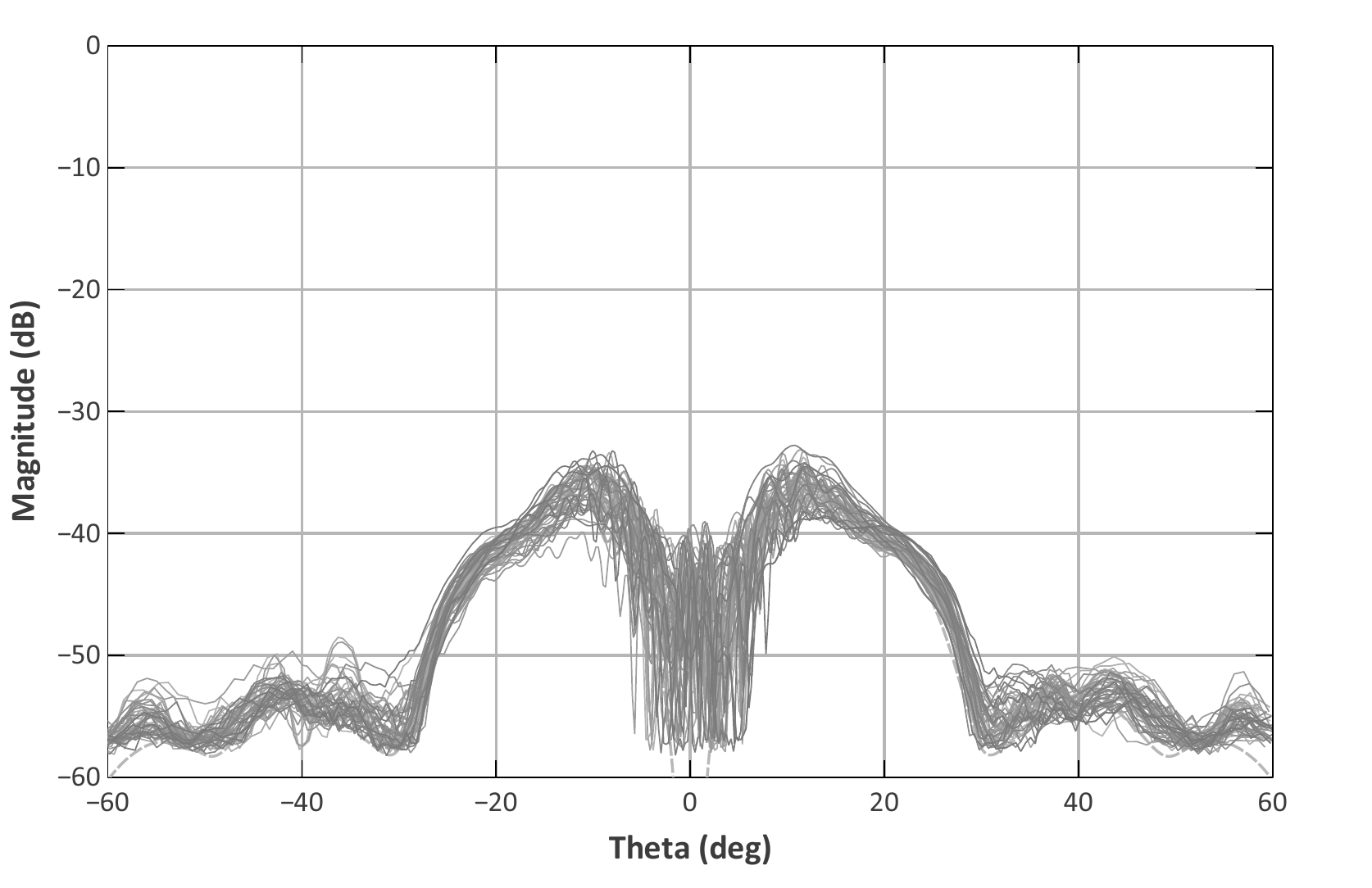}
   \includegraphics[width=6.7 cm]{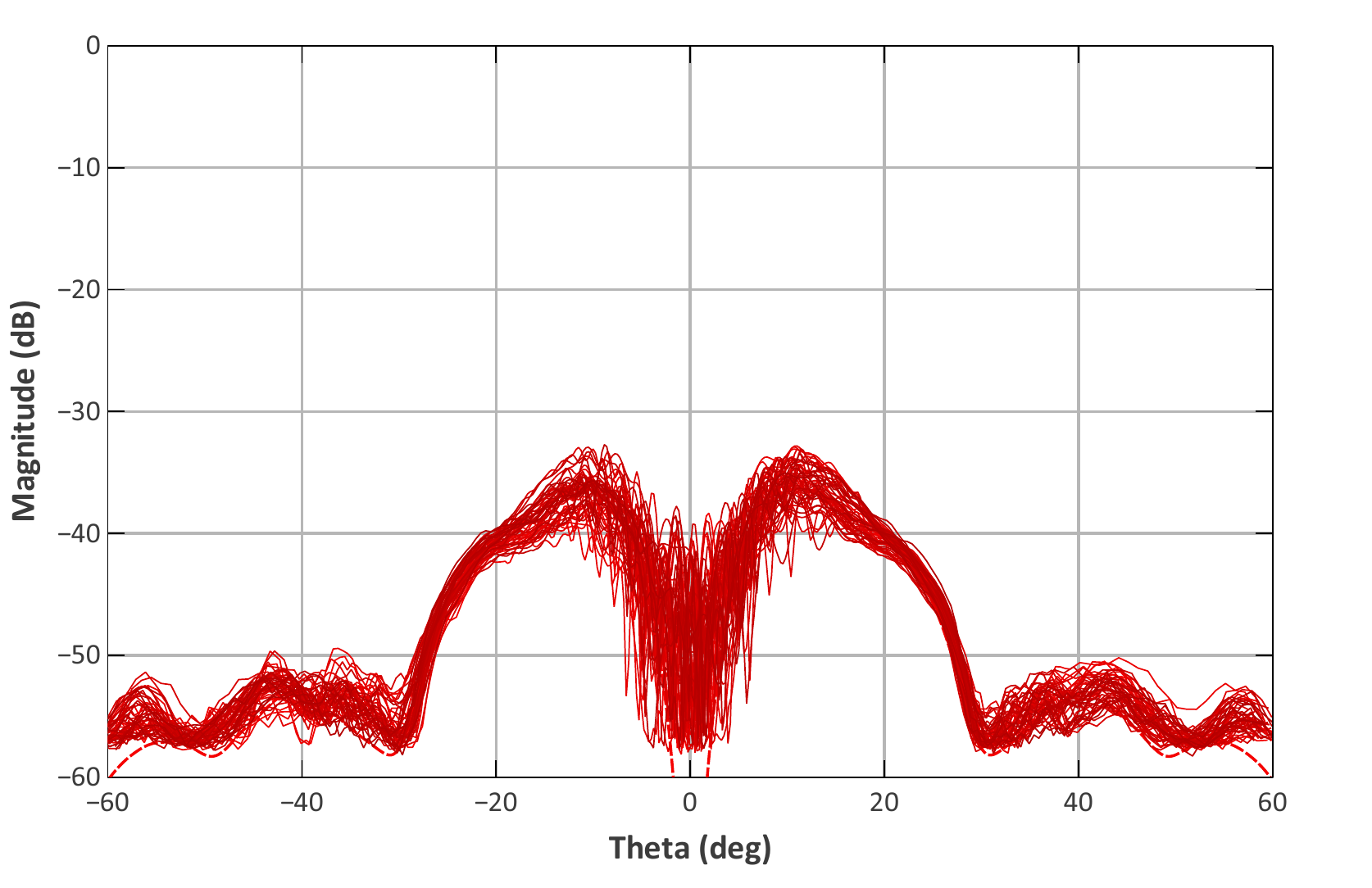}
   \caption{Measured radiation patterns of all forty-nine feedhorns at 38.7~GHz ($f_{0}-10\%$). From left to right, \textit{Top}: co-polar E-plane and co-polar H-plane. \textit{Middle}: co-polar $45\degr$ plane and co-polar $-45\degr$ plane. \textit{Bottom}: cross-polar $45\degr$ plane and cross-polar $-45\degr$ plane.}
   \label{STRIP_FH_f00}
\end{figure}

\begin{figure}[htbp]
   \centering
   \includegraphics[width=6.7 cm]{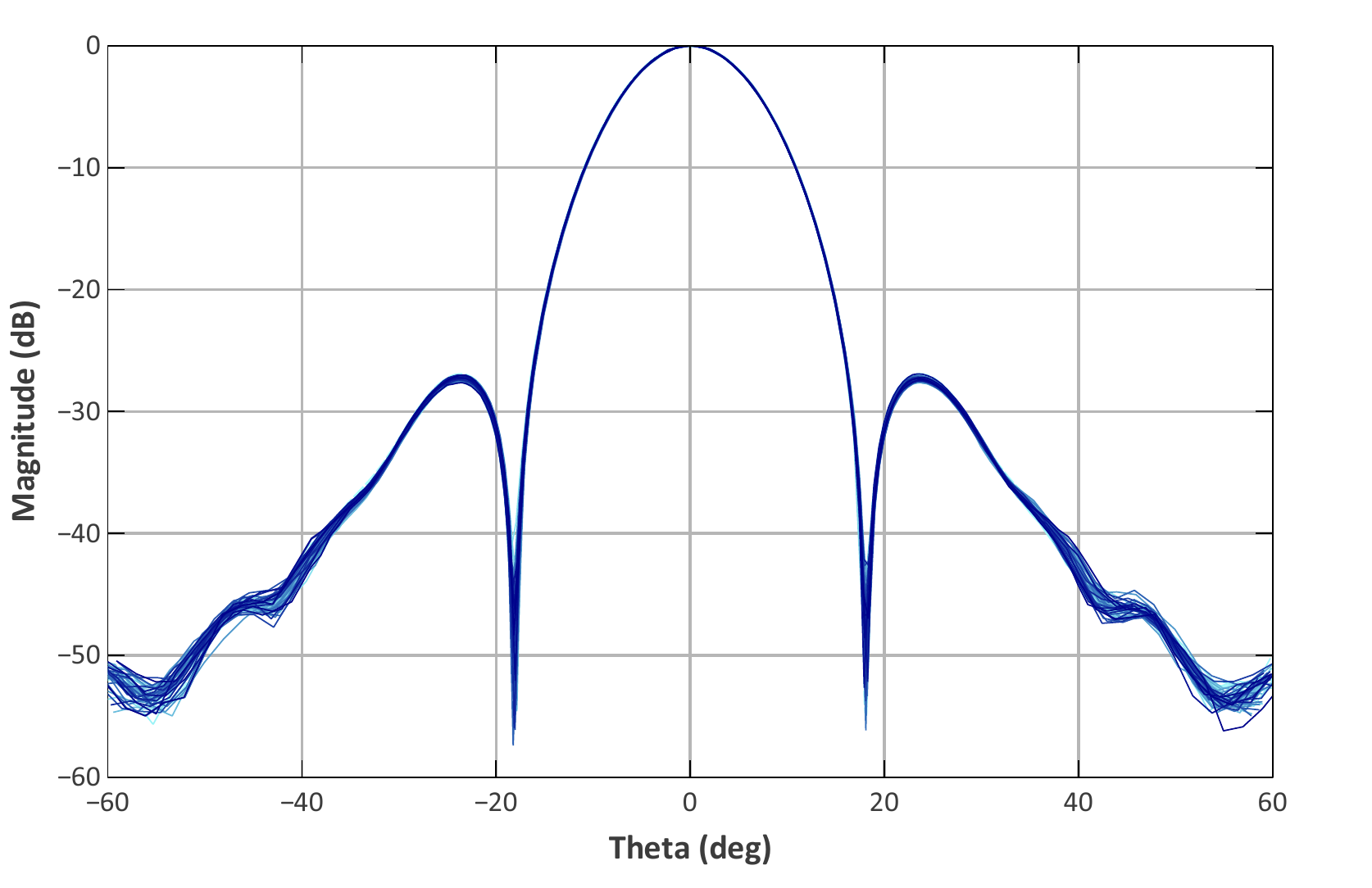}
   \includegraphics[width=6.7 cm]{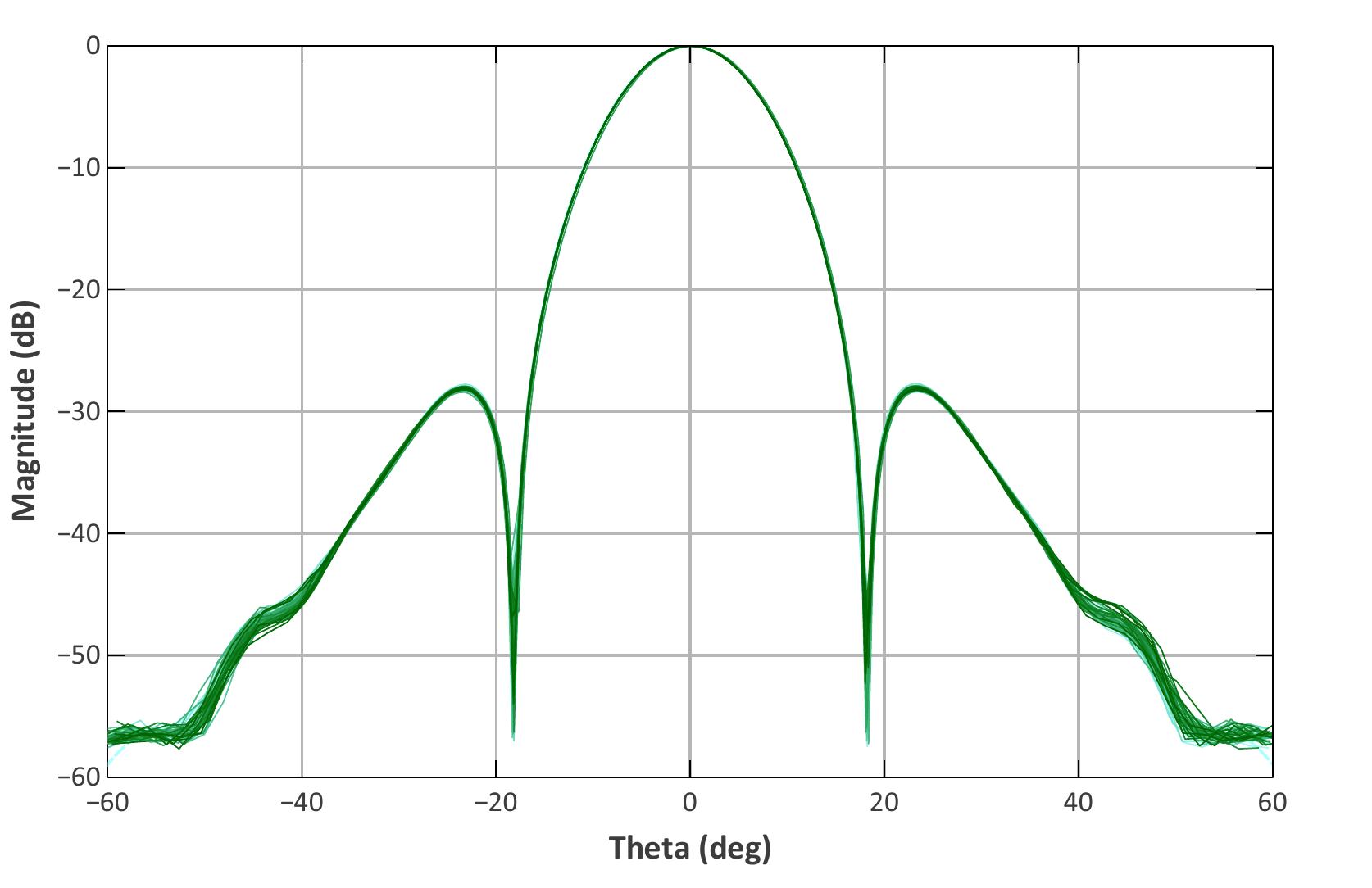}\vspace{20 pt}
   \includegraphics[width=6.7 cm]{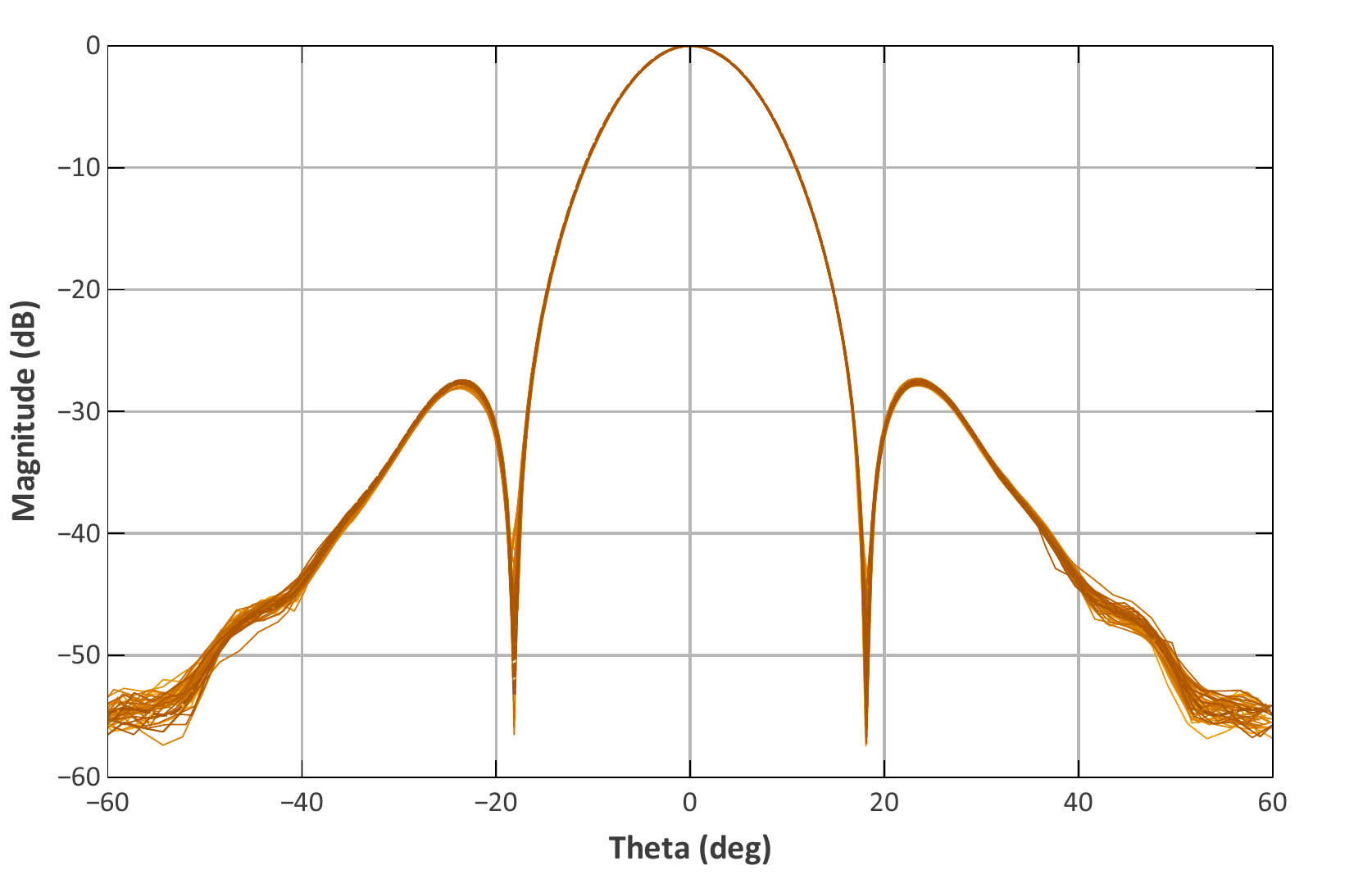}
   \includegraphics[width=6.7 cm]{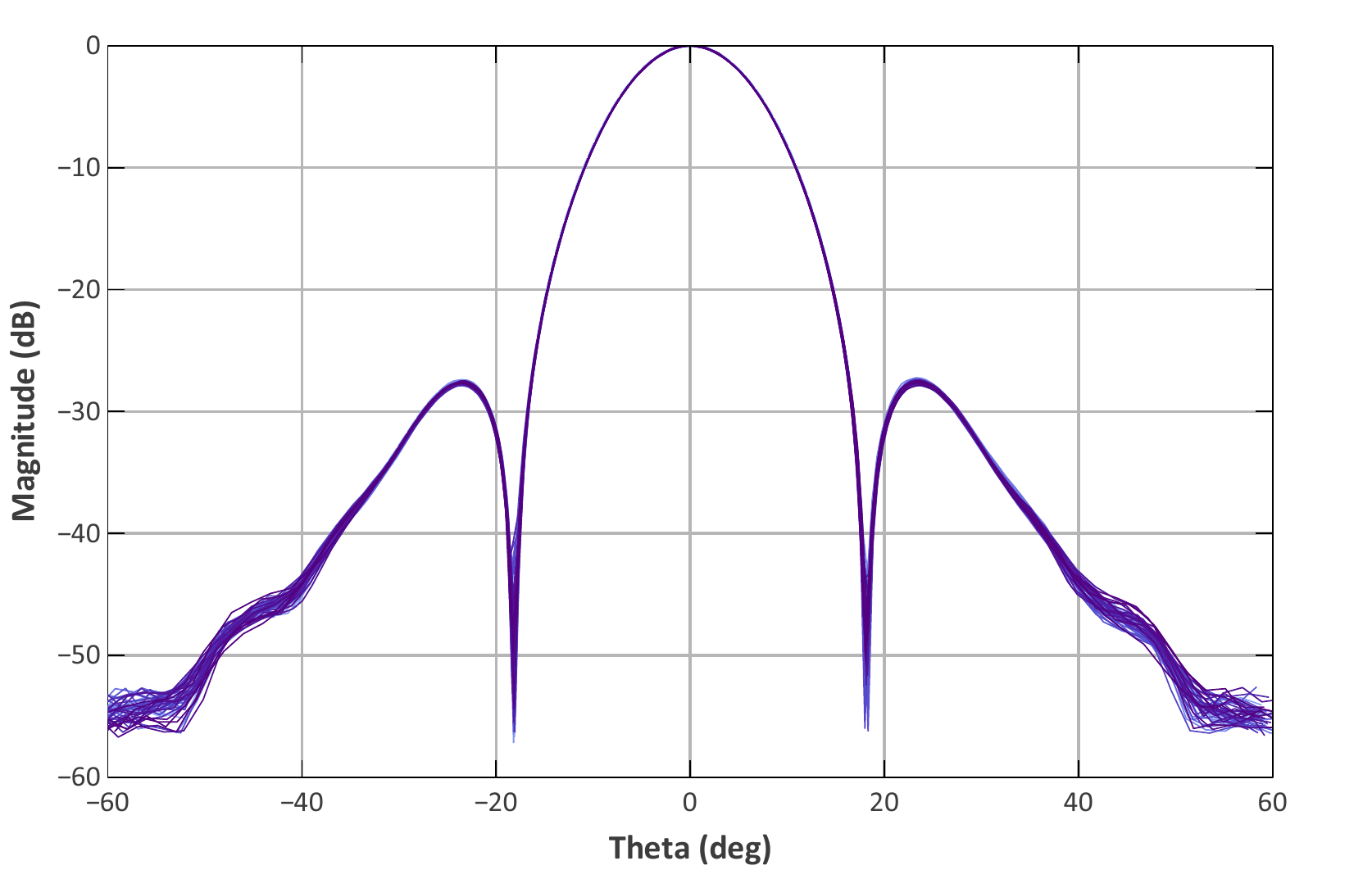}\vspace{20 pt}
   \includegraphics[width=6.7 cm]{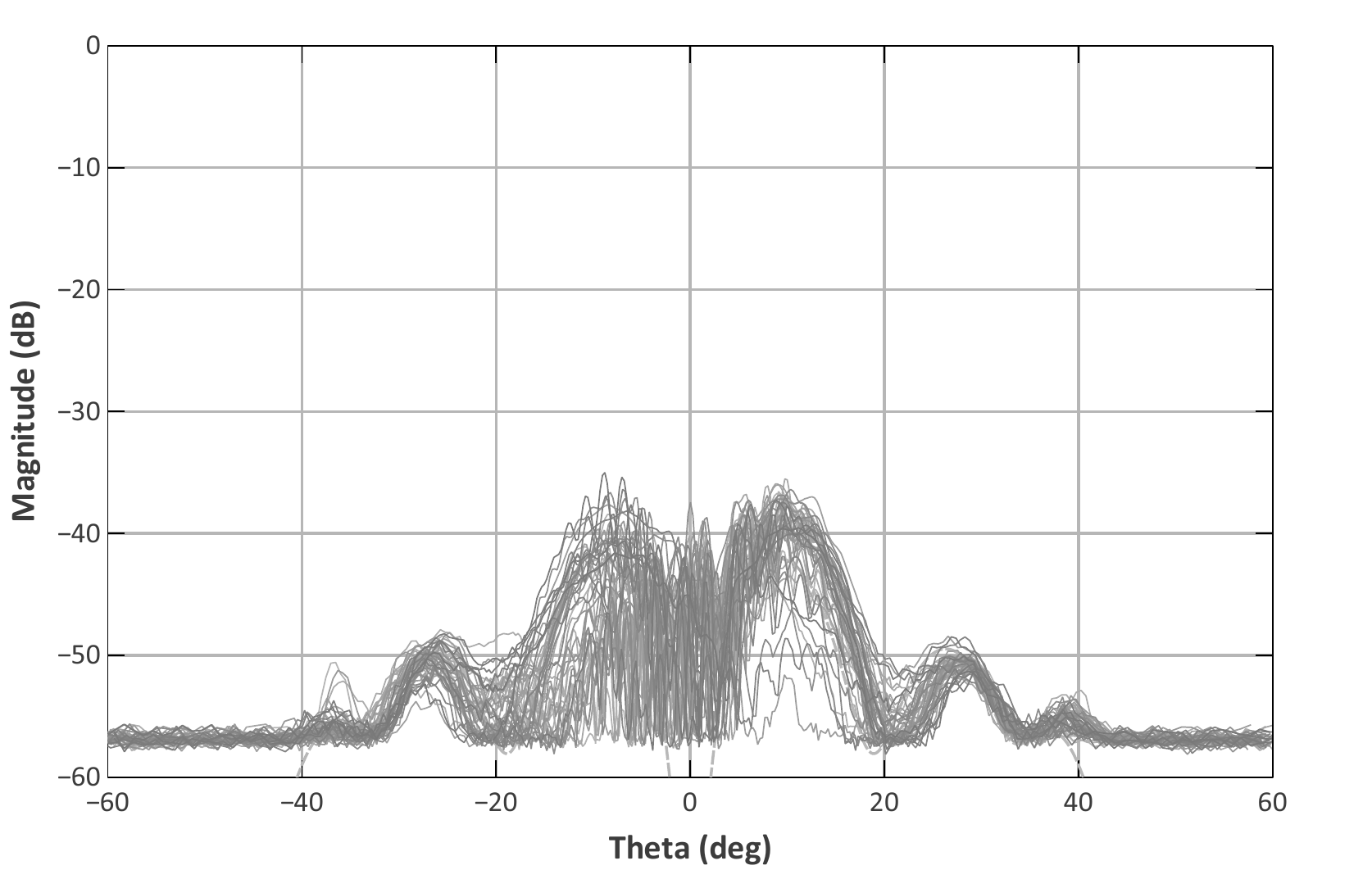}
   \includegraphics[width=6.7 cm]{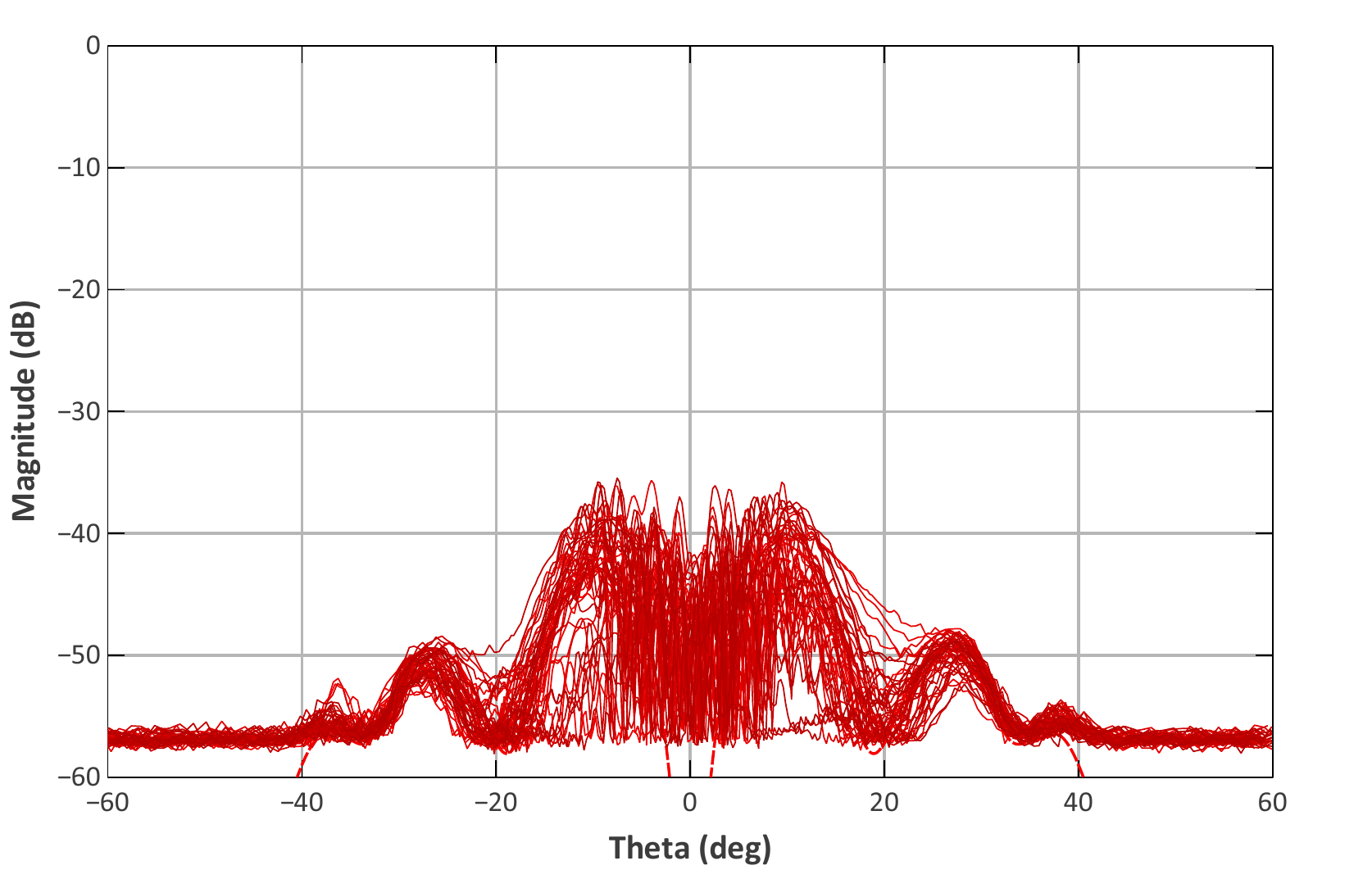}
   \caption{Measured radiation patterns of all forty-nine feedhorns at 43~GHz ($f_{0}$). From left to right, \textit{Top}: co-polar E-plane and co-polar H-plane. \textit{Middle}: co-polar $45\degr$ plane and co-polar $-45\degr$ plane. \textit{Bottom}: cross-polar $45\degr$ plane and cross-polar $-45\degr$ plane.}
   \label{STRIP_FH_f03}
\end{figure}

\begin{figure}[htbp]
   \centering
   \includegraphics[width=6.7 cm]{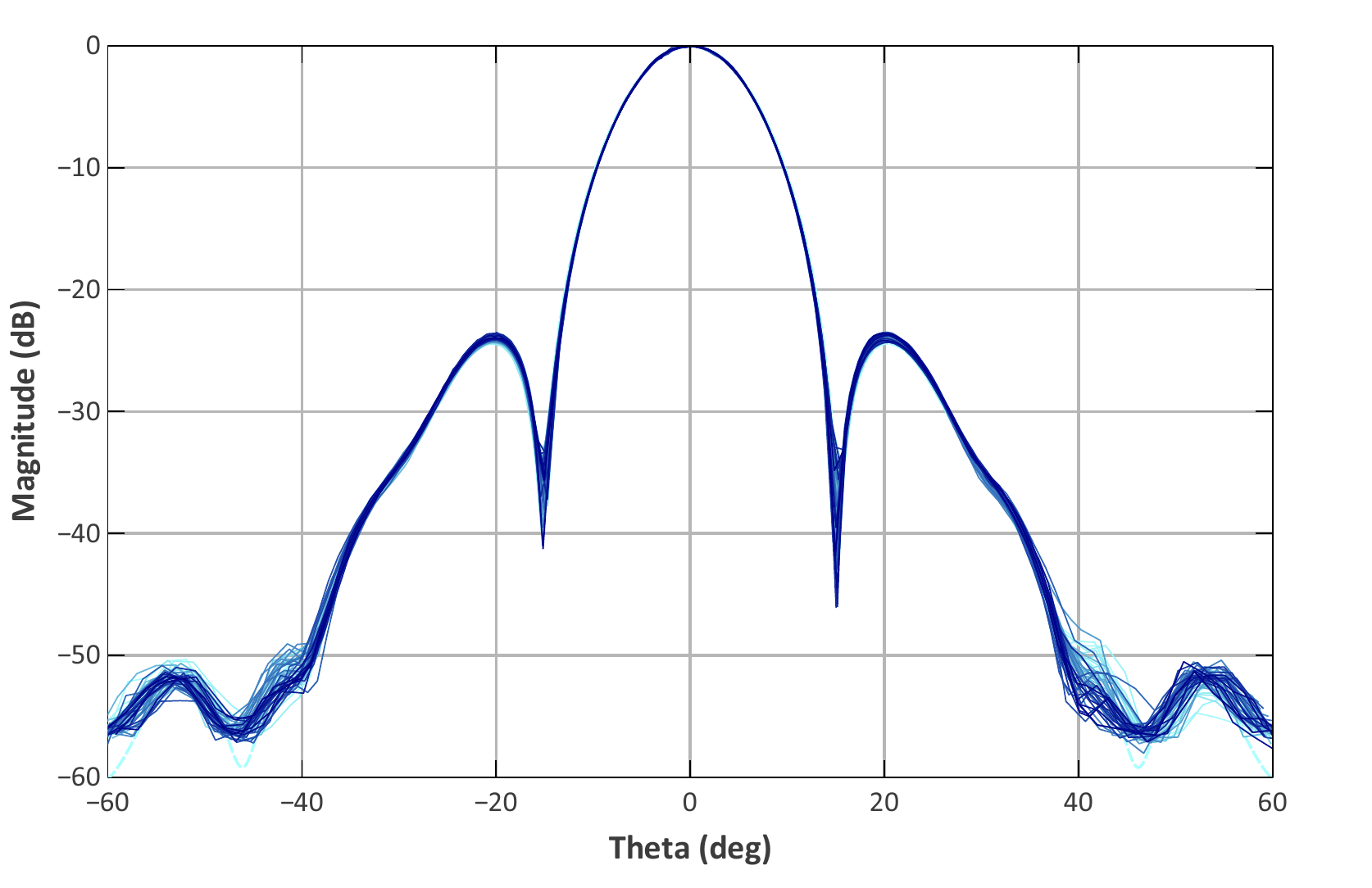}
   \includegraphics[width=6.7 cm]{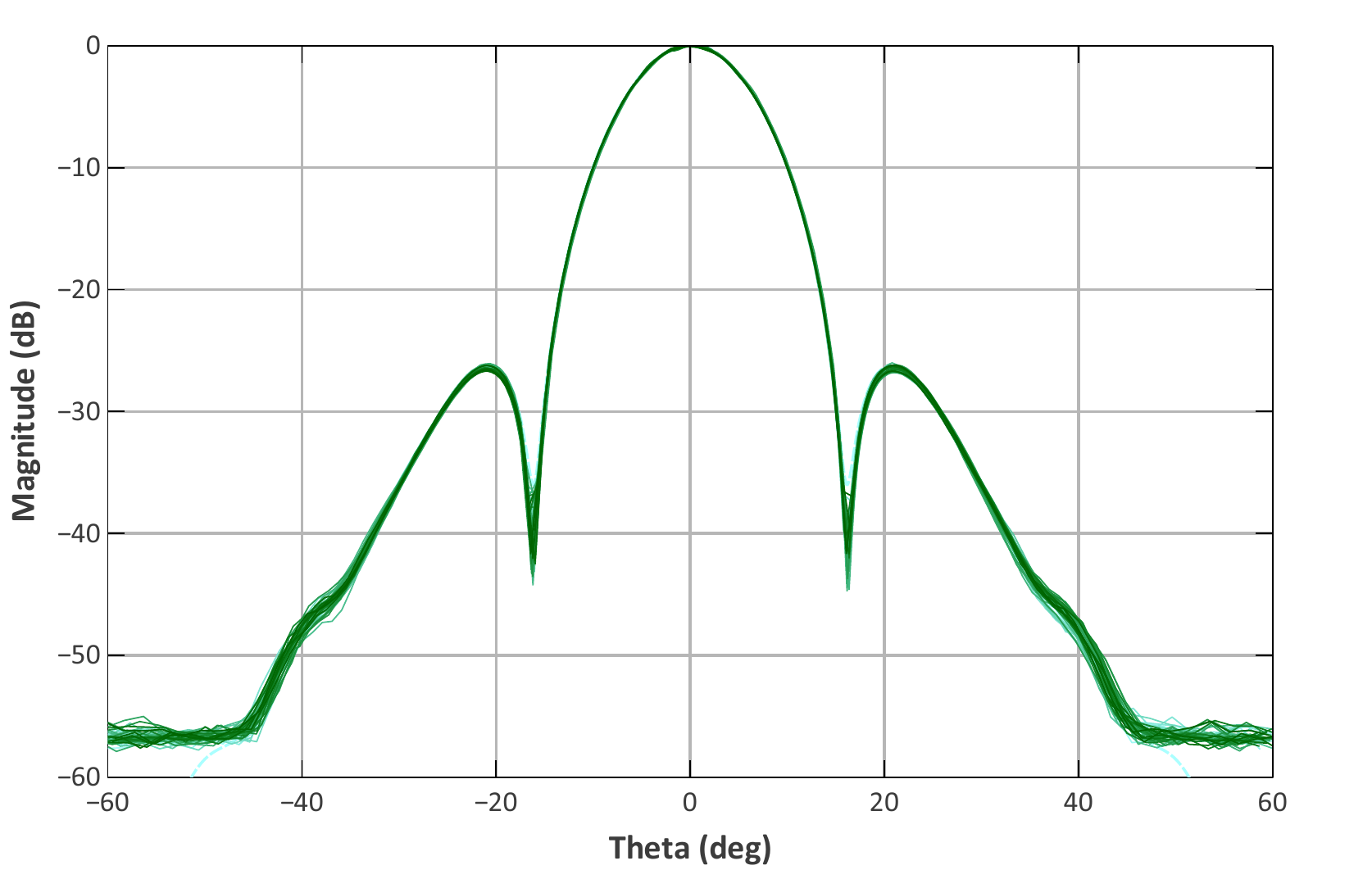}\vspace{20 pt}
   \includegraphics[width=6.7 cm]{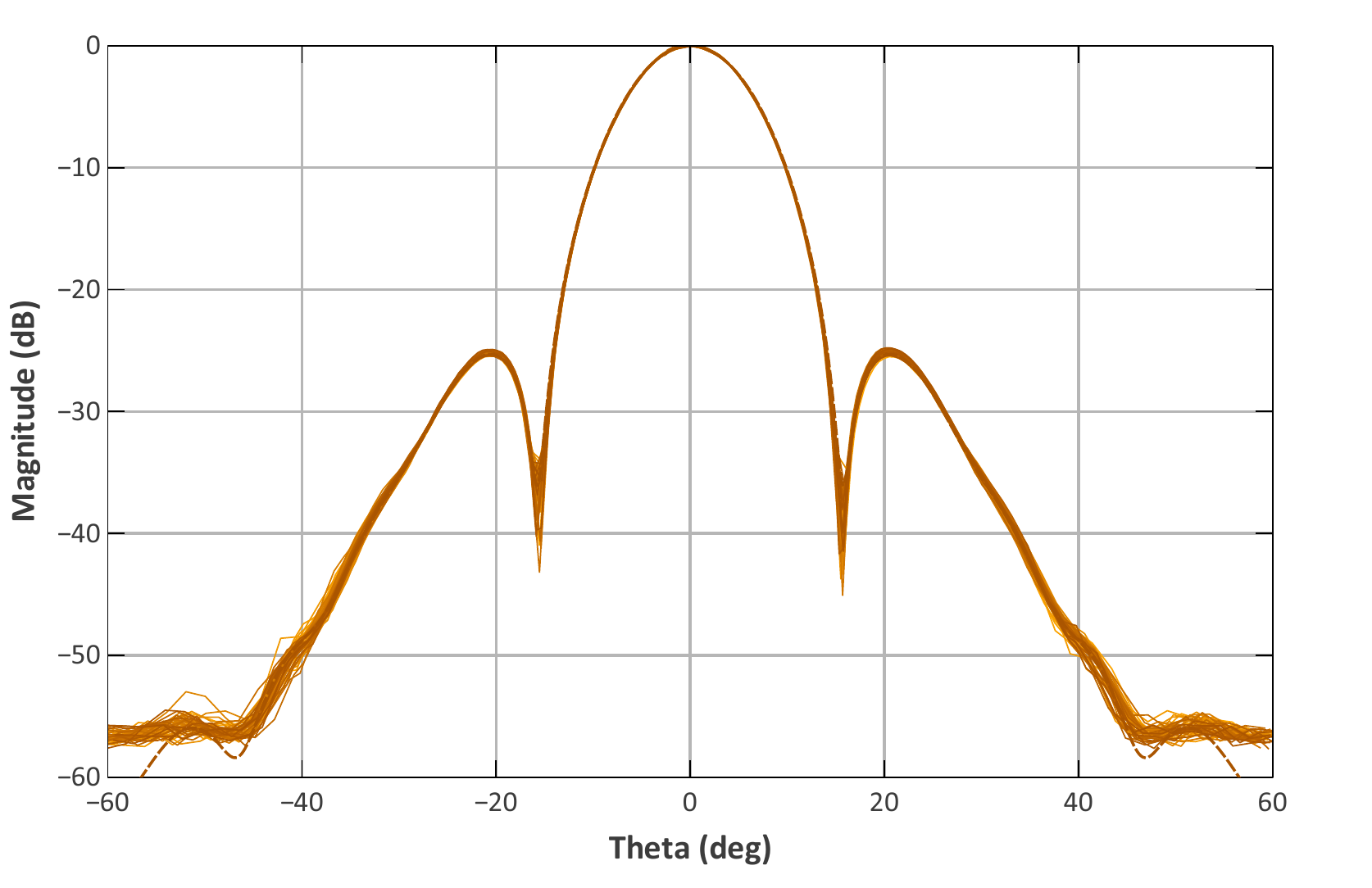}
   \includegraphics[width=6.7 cm]{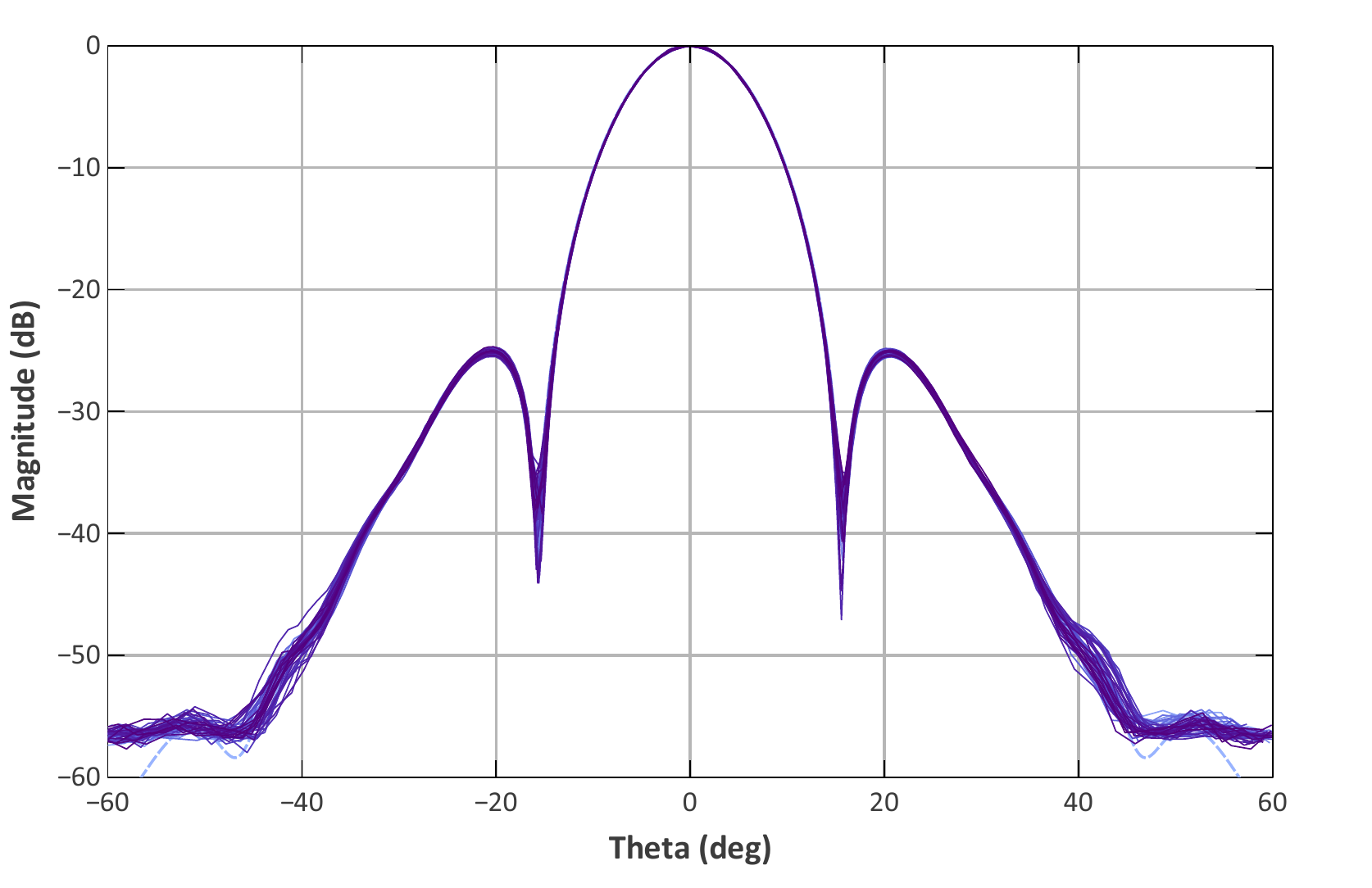}\vspace{20 pt}
   \includegraphics[width=6.7 cm]{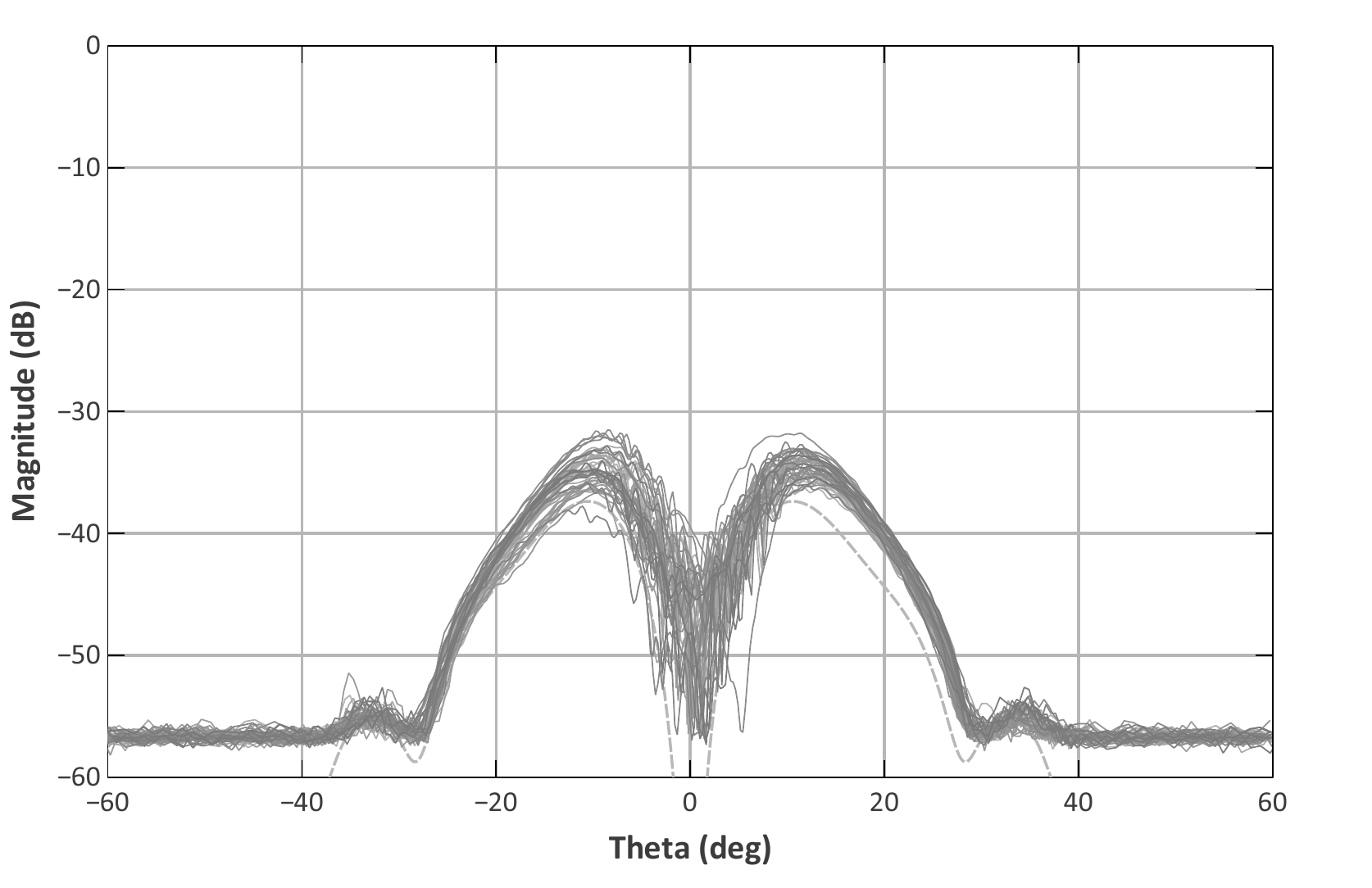}
   \includegraphics[width=6.7 cm]{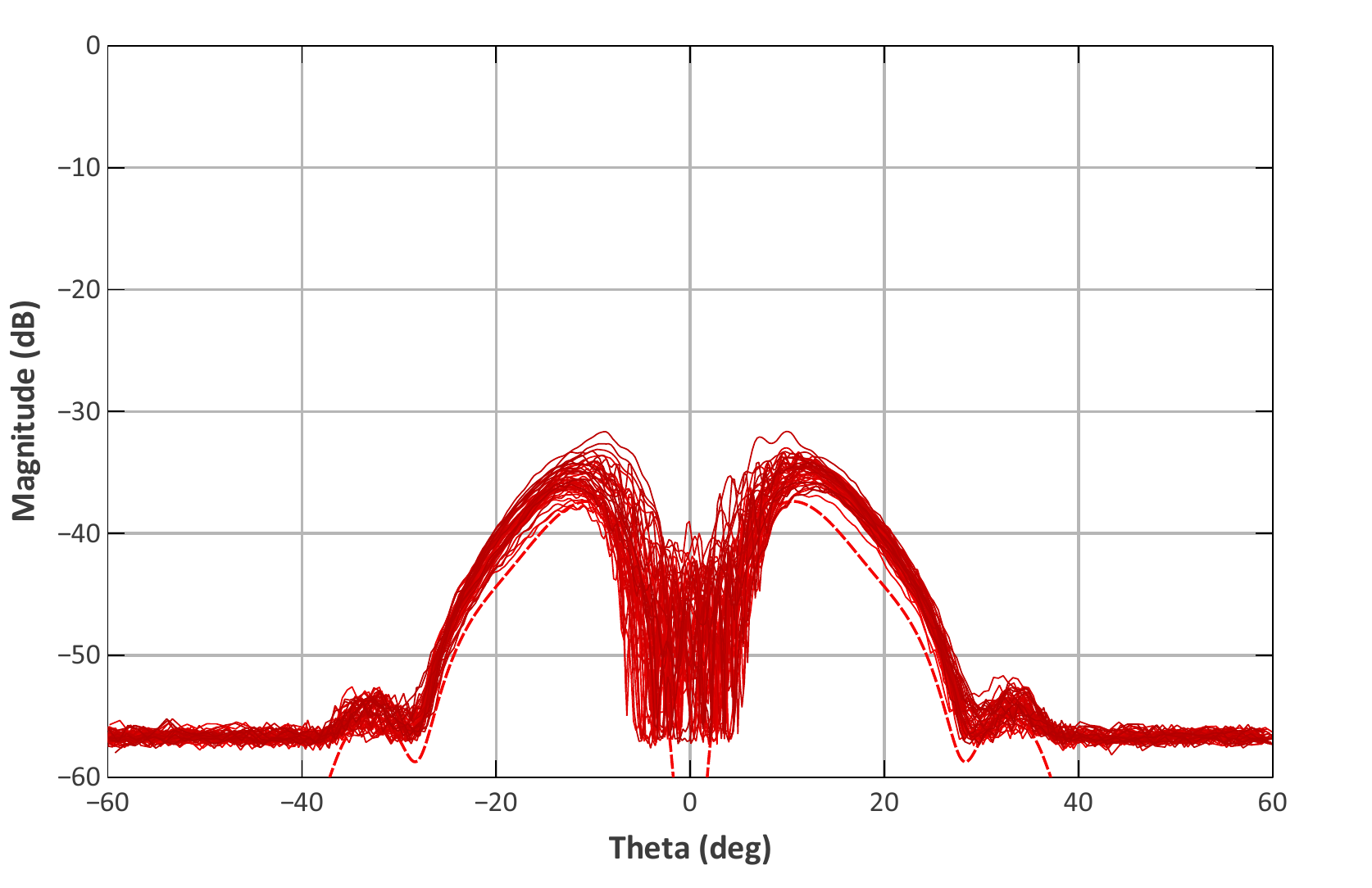}
   \caption{Measured radiation patterns of all forty-nine feedhorns at 47.3~GHz ($f_{0}+10\%$). From left to right, \textit{Top}: co-polar E-plane and co-polar H-plane. \textit{Middle}: co-polar $45\degr$ plane and co-polar $-45\degr$ plane. \textit{Bottom}: cross-polar $45\degr$ plane and cross-polar $-45\degr$ plane.}
   \label{STRIP_FH_f05}
\end{figure}

A total of 1764 radiation patterns were collected. As an example, Figure~\ref{STRIP_FH_V0} shows a comparison between measured and simulated radiation patterns at $f_{0} = 43$~GHz for the central feed horn of a lateral module, placed in the right-bottom region of the Strip focal plane (named V$_0$). The level of consistency shown here is typical of all our Q-band measurements. Moreover, a comparison between the angular response of all forty-nine feed horns is evaluated at each measured frequency, to assess the repeatability and reliability in the manufacturing of the feed horns. As an example, Figure~\ref{STRIP_FH_f00} to \ref{STRIP_FH_f05} show the co-polar and cross-polar radiation patterns on the principal planes of all forty-nine Q-band horns at 38.7, 43 and 47.3~GHz, respectively. The simulated patterns are superimposed on all plots.

Measurements and simulations of the co-polar planes exhibit an agreement within a fraction of a dB in the angular region up to the first sidelobe. Limiting to the main beam region, in the azimuth range $\pm15\degr$ that mostly contributes to telescope illumination, the measurement-simulation agreement is better than 0.5~dB. This further improves by neglecting the radiation pattern null-regions, where the signal magnitude rapidly drops by 30~dB or more: in these regions (at about $\pm18\degr$ for the center frequency $f_0$), simulations predict about $-50$~dB level, so that measurements are limited by instrumental dynamic range.

Similar considerations apply to cross-polar measurements: in addition to the low signal-to-noise ratio, these measurements are very sensitive to the experimental setup, in particular to the non-ideal alignment of the source and AUT antennas polarization planes, and to the polarization purity of the circular-to-rectangular waveguide transition downstream of the antenna circular interface. Furthermore, the anechoic chamber well approximates an open space environment, but it is not perfectly anechoic so that spurious reflection at the chamber walls contributes to the signal received by the feed horn under test. This introduces some residual contamination in both the cross-polar and co-polar measurements, particularly at large azimuth positions and near the radiation pattern nulls. However, the envelopes reported in the bottom panel of Figures~\ref{STRIP_FH_f00} to \ref{STRIP_FH_f05} show that the requirement on the maximum cross-polarization level ($<-30$~dB) is satisfied over the whole operative bandwidth, being better than $-40$~dB at the center frequency.

%--- begin rev. #6 and #10
To verify the impact of possible misalignments during platelets stacking on the feed horns performance, we measured and compared the radiation pattern of a few horns, after several assembly-disassembly cycles. The repeatability of the measurement was excellent and no appreciable differences were observed down to the level of the far sidelobes. In principle, an assessment of the effects of misalignments should be performed also with optical simulations, but our reference simulator, SRSR-D\copyright, is restricted to axisymmetric structures, so that misalignments between platelets and similar imperfections cannot be implemented in the model.
%--- end rev. #6 and #10

%----------------------------------------
\subsubsection{Return loss}
\label{sec:Q-meas-rl}
\begin{figure}[htbp]
   \centering
   \includegraphics[width=6.6 cm]{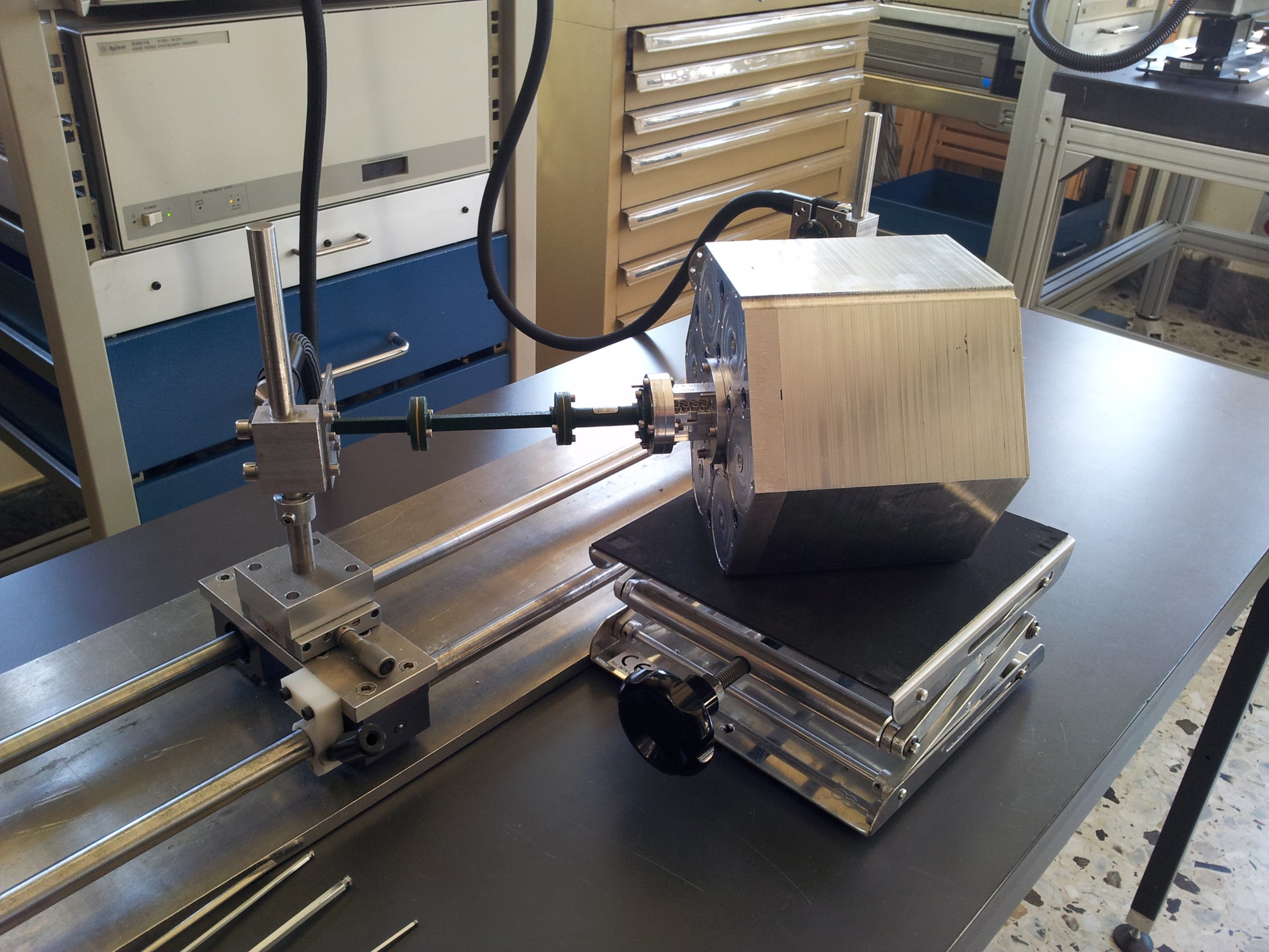}\hspace{5 pt}
   \includegraphics[width=6.6 cm]{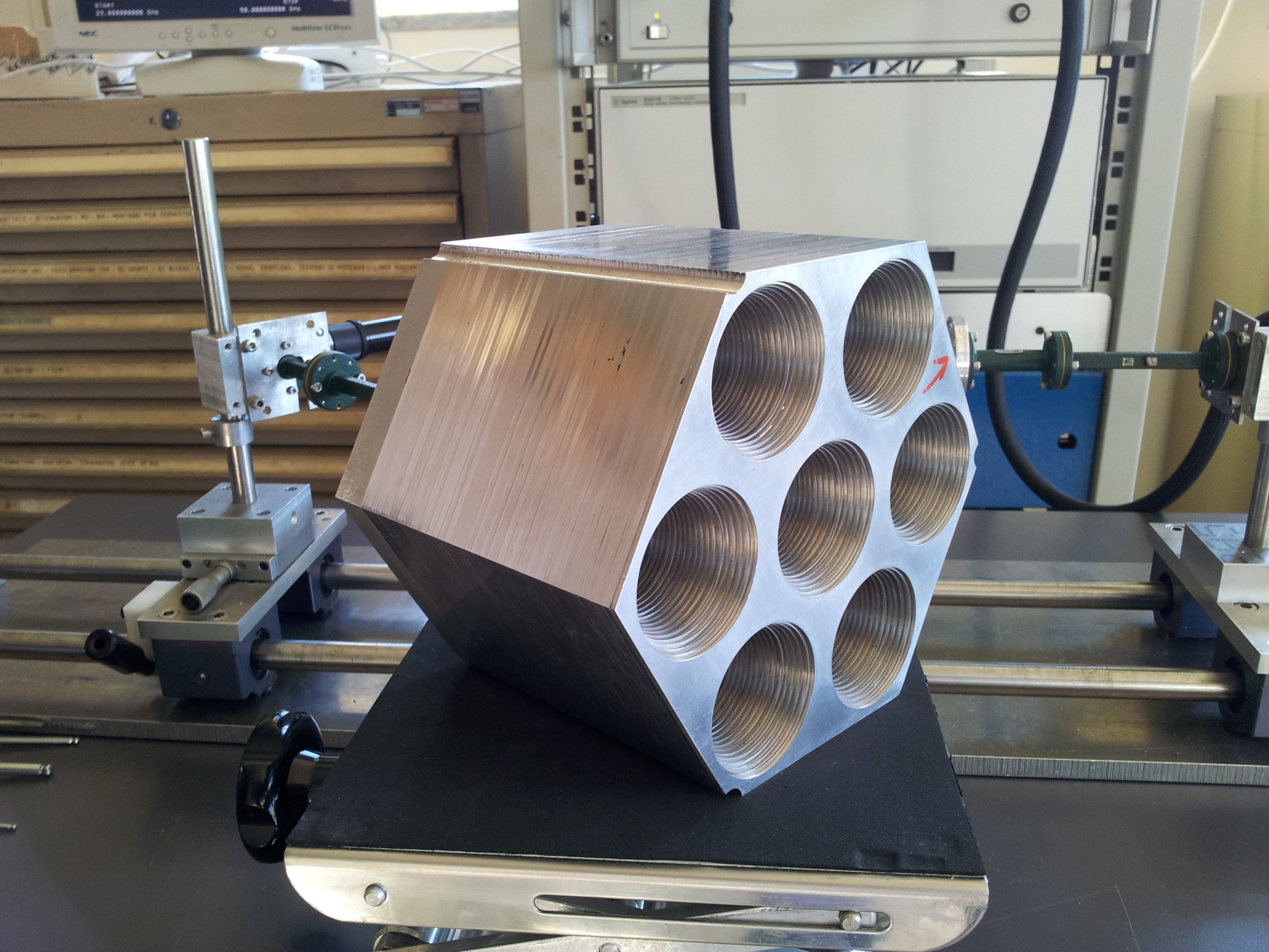}\vspace{5 pt}
   \includegraphics[width=6.6 cm]{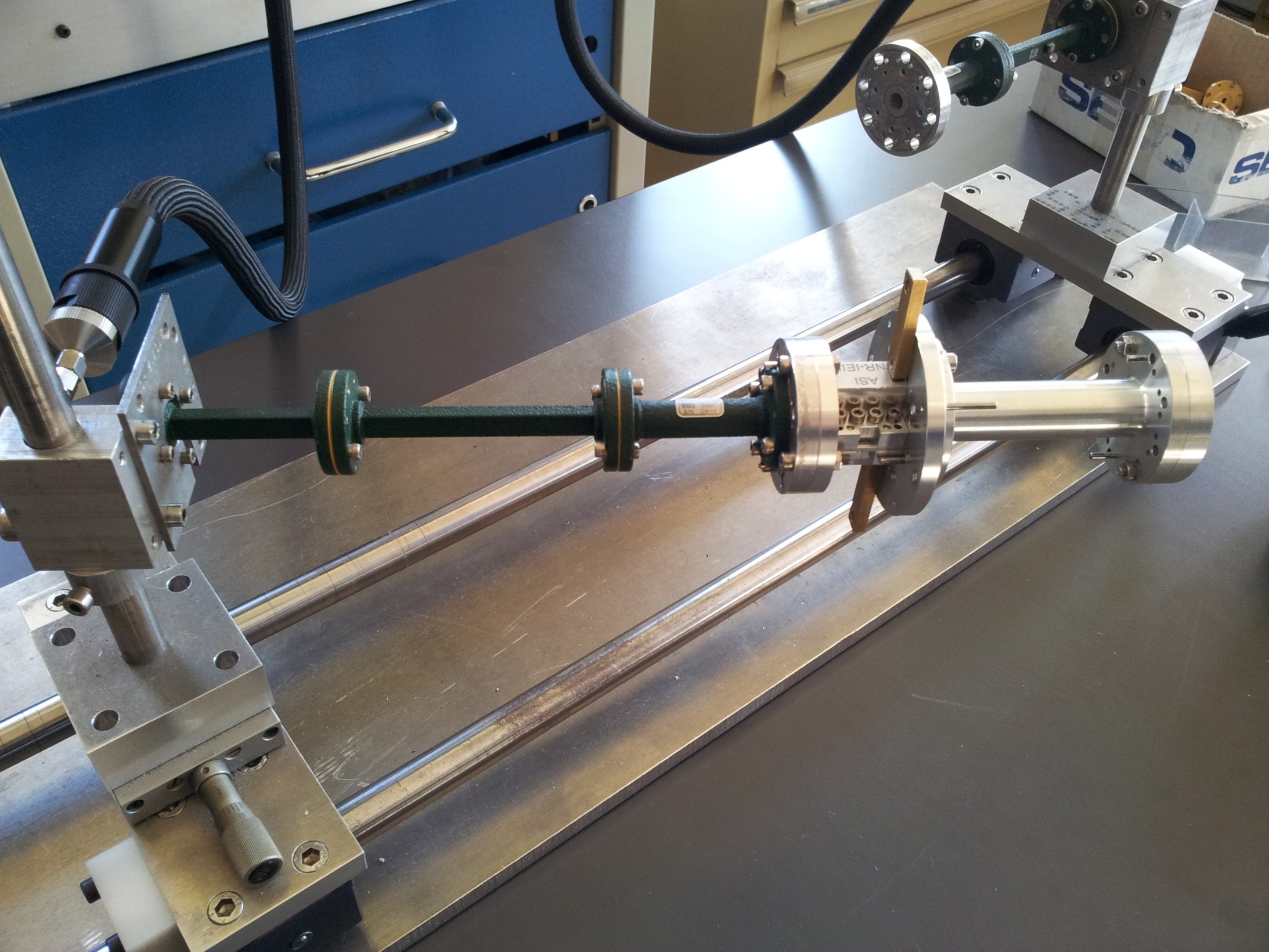}\hspace{5 pt}
   \includegraphics[width=6.6 cm]{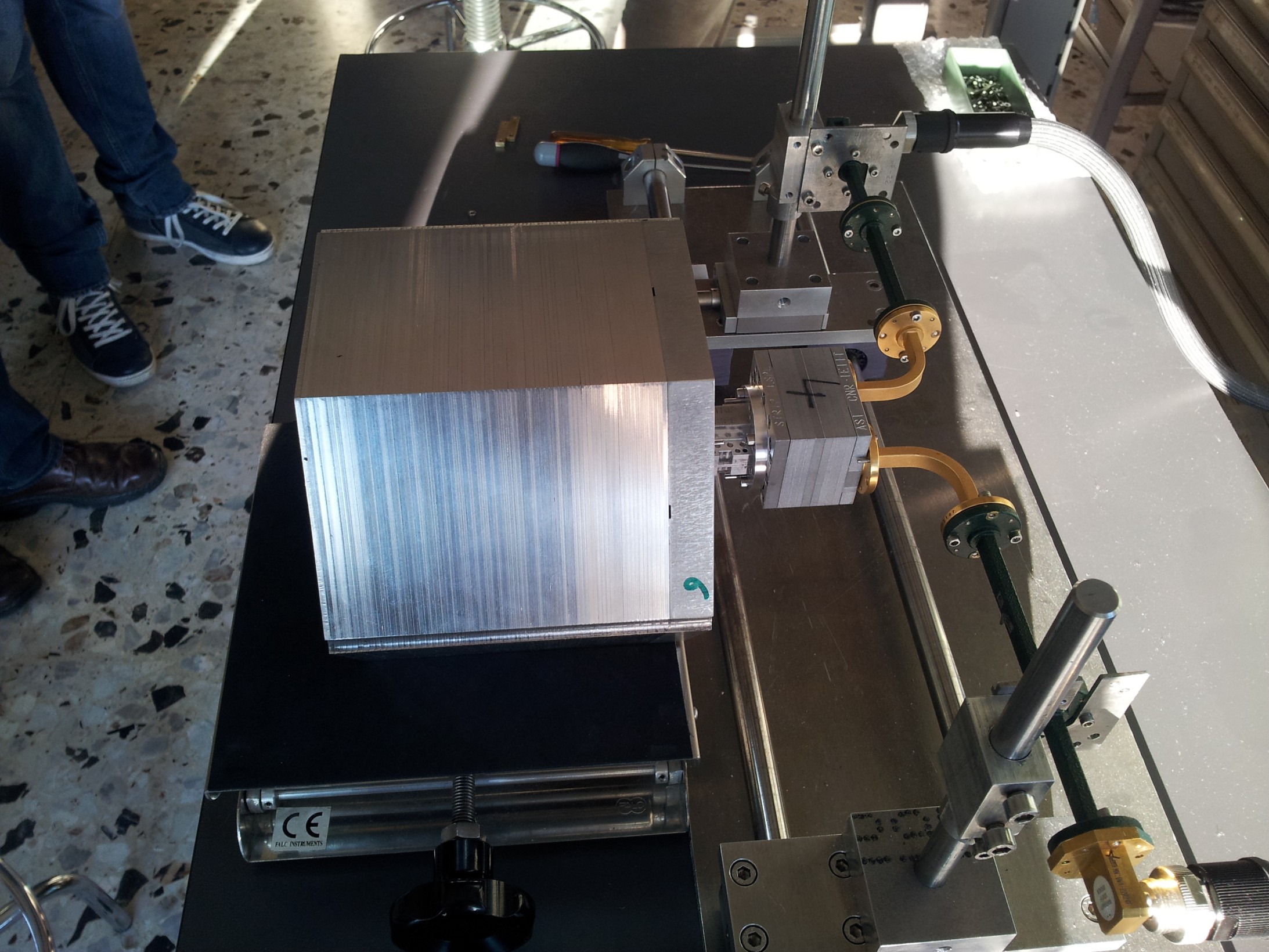}%\vspace{5 pt}
   \caption{\textit{Top}: two photographs of the experimental setup for the return loss measurement, taken at CNR-IEIIT laboratories in Torino. On the left the passive chain is shown: the central feed horn of the hexagonal module is connected to a dummy polarizer, providing the correct mechanical interface; a circular waveguide to WR22 transition adapts the path to the standard rectangular waveguide. \textit{Bottom-left}: a detail of the calibration, with the polarizer connected to the waveguide load at the feed horn interface. \textit{Bottom-right}: feed horns have been used as matched load to characterize the isolation between the two circular polarizations when the Strip passive components (feed horn, polarizer and OMT) are assembled.}
   \label{STRIP_FH_RLsetup}
\end{figure}

In addition to the feed horn under test, the experimental setup for the measurement of the return loss included a dummy polarizer providing the required mechanical interface (a non-standard flange interfaces each feed horn to the polarizer), connected to a circular-to-rectangular waveguide (WR22) transition. By means of a waveguide-to-cable transition, the passive chain was connected to an Agilent$\copyright$ HP8510 vector network analyzer. Figure~\ref{STRIP_FH_RLsetup} shows some pictures of the experimental setup.

\begin{figure}[htbp]
   \center
   \includegraphics[width=.7\textwidth]{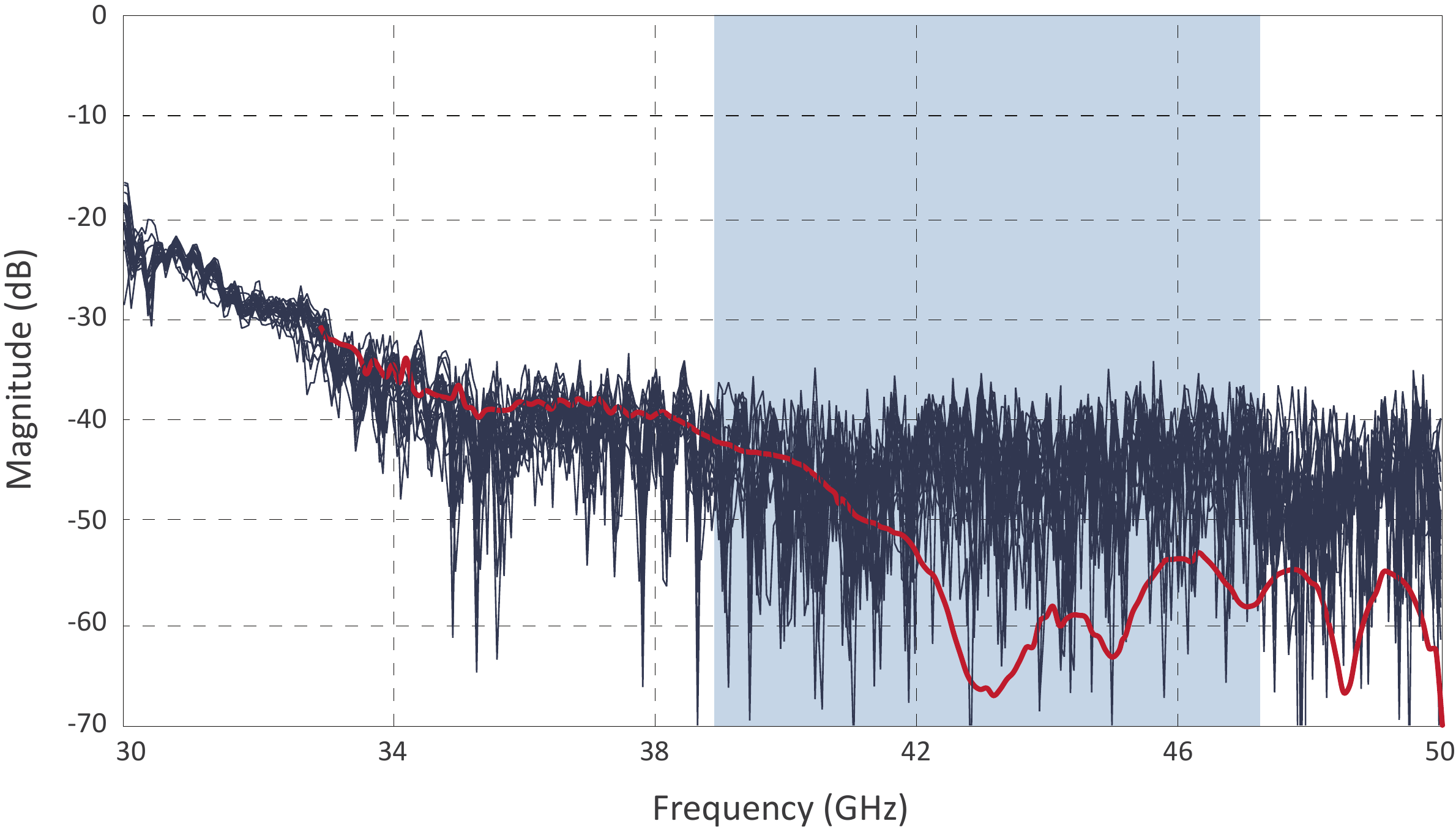}
   \caption{Measured levels of reflection coefficient of all forty-nine feed horns. Return loss levels are better than $-40$~dB on the whole 20\% bandwidth (shaded area).}
   \label{STRIP_FH_RLmeas}
\end{figure}

The measured return loss over the whole Q-band for all forty-nine feed horns has been compared to the \srsr simulations. Figure~\ref{STRIP_FH_RLmeas} shows that the measured level, although the envelope is within the requirements ($<-40$~dB over the 20\% operative bandwidth), is about 10~dB higher than the simulated level for frequencies greater than 41~GHz. Waveguide calibration could be an issue in these measurements, but also the laboratory environment not being properly shielded could have introduced spurious reflections.
%--- begin rev. #8
In our case, the measurements were not carried out in an anechoic chamber. We placed an absorber panel in front of the feed horn apertures: in addition to the effect of the reflection coefficient of the absorber, there could be a weak spillover from the far sidelobe region of the horns, due to reflections at the walls of the room. Furthermore, the rapid changes observed in the measurements suggest that the reflections are occurring very far from the input port, thus supporting the hypothesis of an environmental contribution. For these measurements, we calibrated the VNA with the TRL (Thru-Reflect-Line) technique in circular waveguide, therefore without the aid of matched loads that could have determined the observed plateau.
%--- end rev. #8
Given the measured levels, the characterization has been considered successful and no further efforts to improve the experimental setup have been performed.

In order to check reliability for cryogenic operation, we repeated the full characterization on a spare 7-element module after a cool down-warm up cycle in liquid nitrogen, finding no measurable variation in performance.

%----------------------------------------
\subsection{W-band measurements}
\label{sec:W-meas}

%----------------------------------------
\subsubsection{Radiation patterns}
\label{sec:W-meas-pattern}
W-band antenna radiation patterns were measured in the far-field regime, at about 1850~mm from the source antenna, at five frequencies over their operative band:
\begin{itemize}
\item 85~GHz ($f_0$-10\%)
\item 90~GHz ($f_0$-5\%)
\item 95~GHz ($f_0$)
\item 100~GHz ($f_0$+5\%)
\item 105~GHz ($f_0$+10\%)
\end{itemize}
Each feed horn was characterized on three co-polar planes (E, H and +45\degr) and one cross-polar plane (+45\degr). These also include repeated measurements on the same E and H planes when the horn is flipped by 180\degr\ around its axis to break any degeneracy between possible asymmetries in the antenna and in the anechoic chamber. The six W-band feed horns (plus a seventh prototype as a spare unit) were measured in the bigger anechoic chamber at the University of Milano, which is equipped with a vector network analyzer MS4647B by Anritsu$\copyright$ with a 3739C broadband test-set and millimeter-wave extension modules in the W-band.

\begin{figure}[htbp]
   \centering
   \includegraphics[width=.7\textwidth]{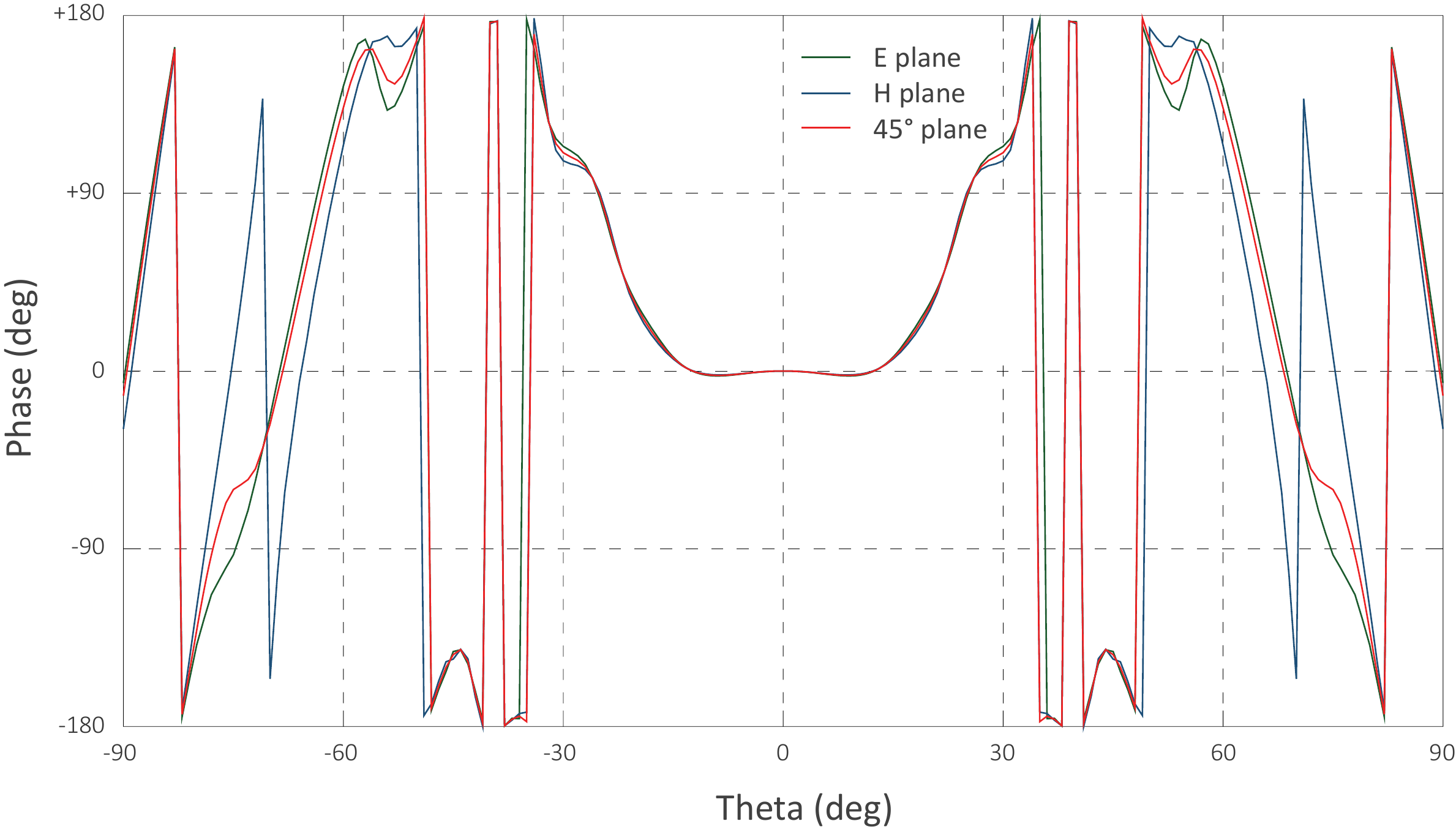}
   \caption{Simulated phase diagram at 95~GHz, when the phase center is 11.8~mm inside the horn aperture plane. Phase is stable in the main beam angular range (about $\pm18\degr$) in all co-polar planes.}
   \label{fig:faseW}
\end{figure}

\srsr simulations provided the position of the phase center at $f_0=95$~GHz, i.e. 11.8~mm inside the horn aperture plane, as shown in Figure~\ref{fig:faseW}.

\begin{figure}[htbp]
   \centering
   \includegraphics[width=6.8 cm]{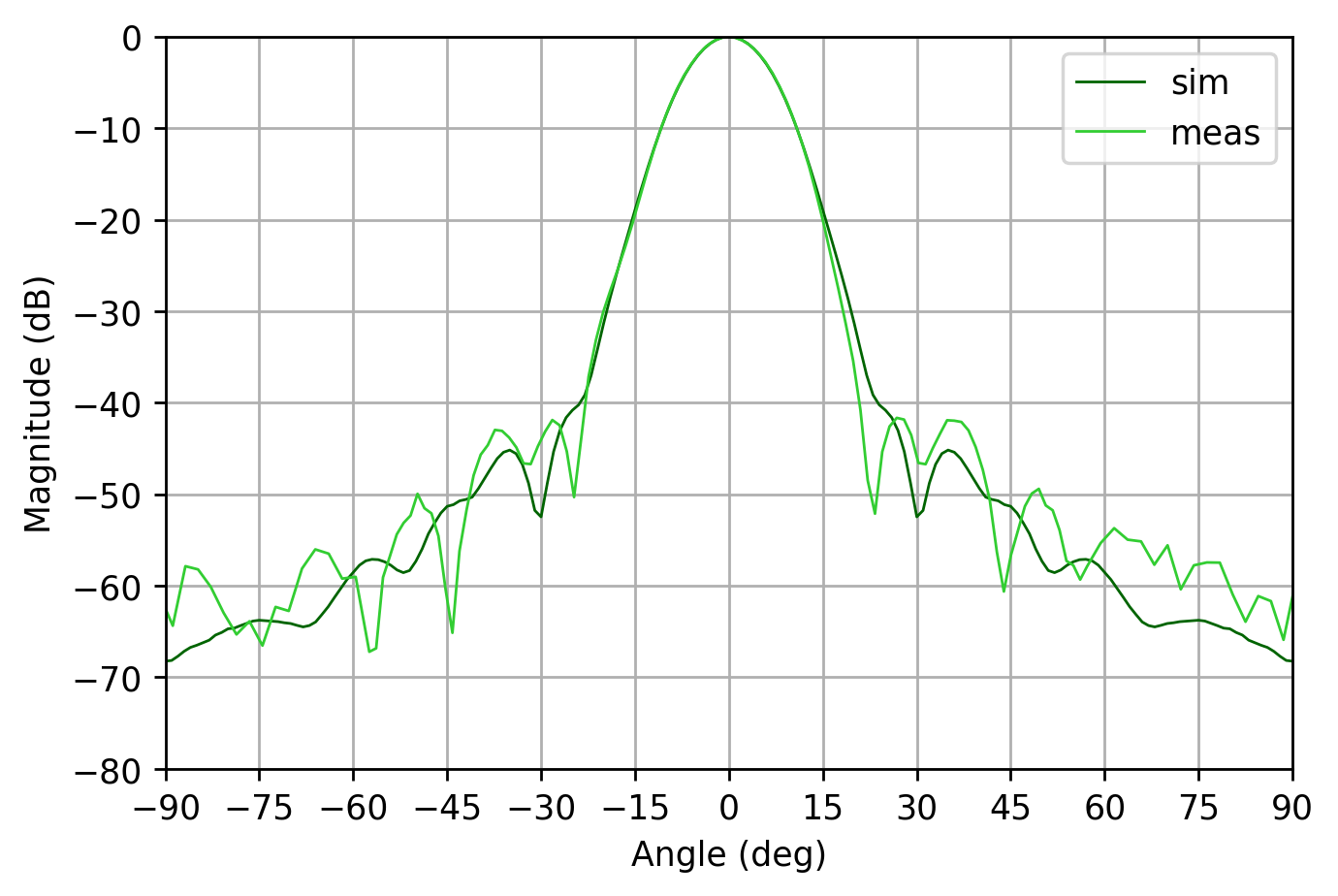}%\hspace{1 pt}
   \includegraphics[width=6.8 cm]{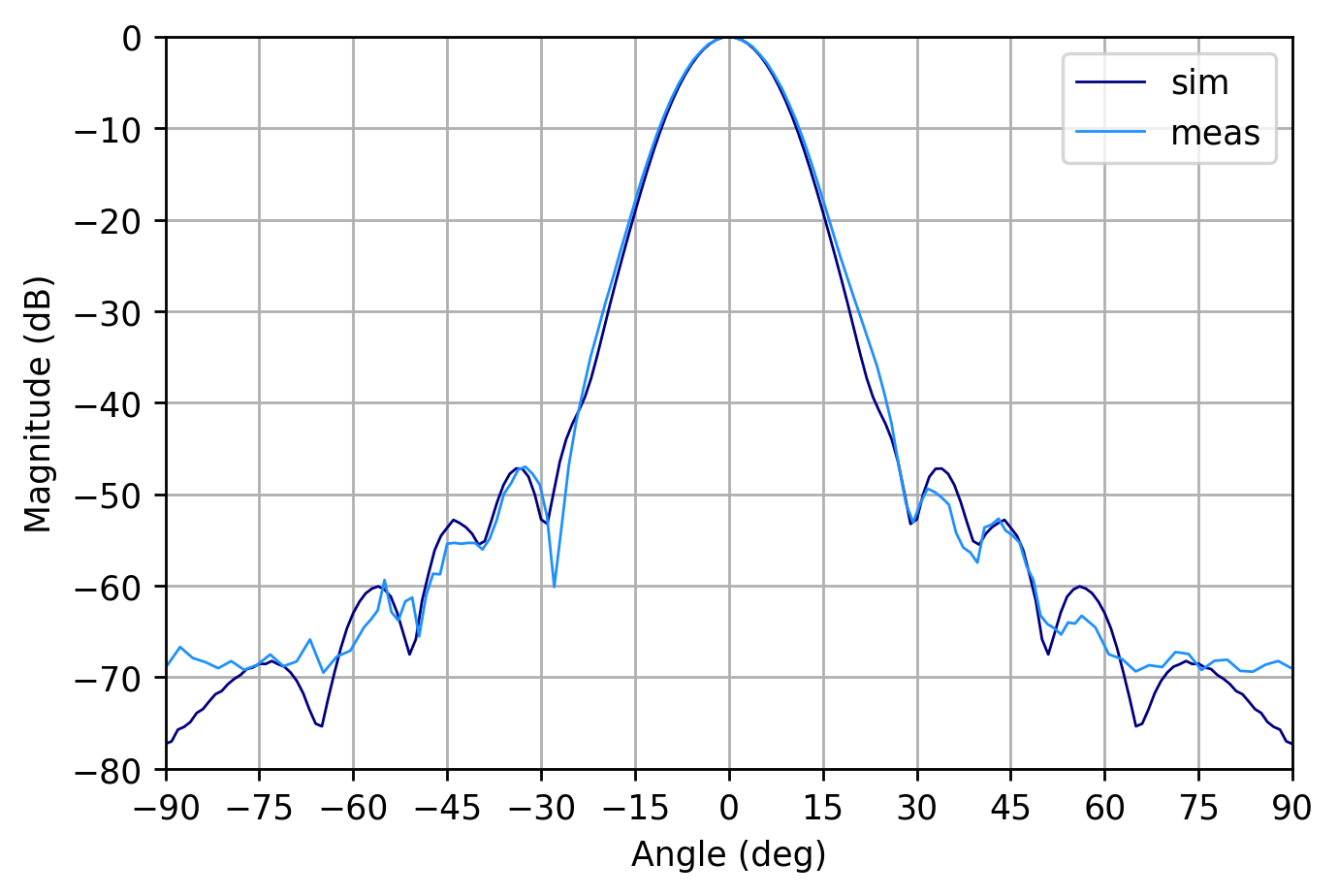}\vspace{10 pt}
   \includegraphics[width=6.8 cm]{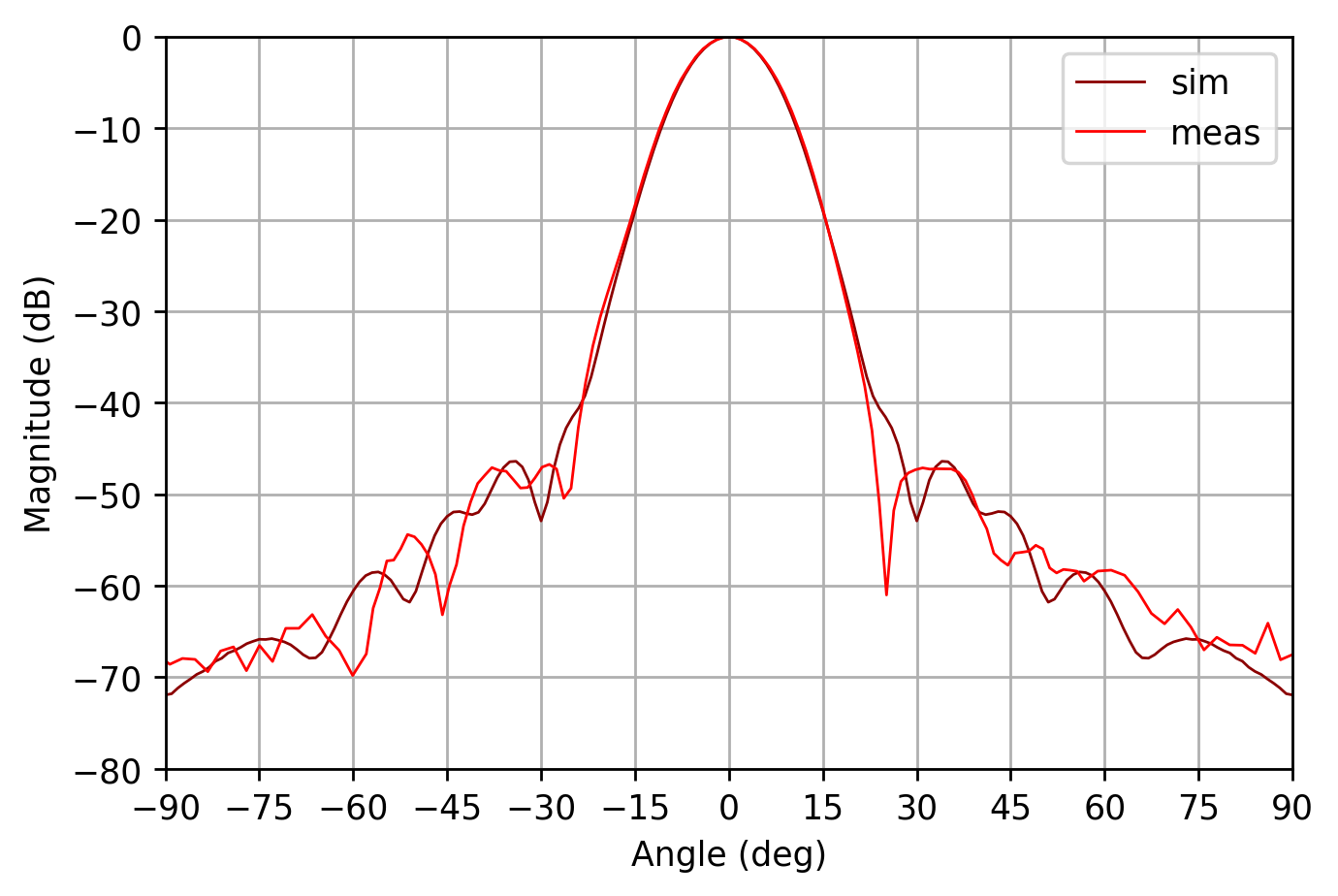}%\hspace{1 pt}
   \includegraphics[width=6.8 cm]{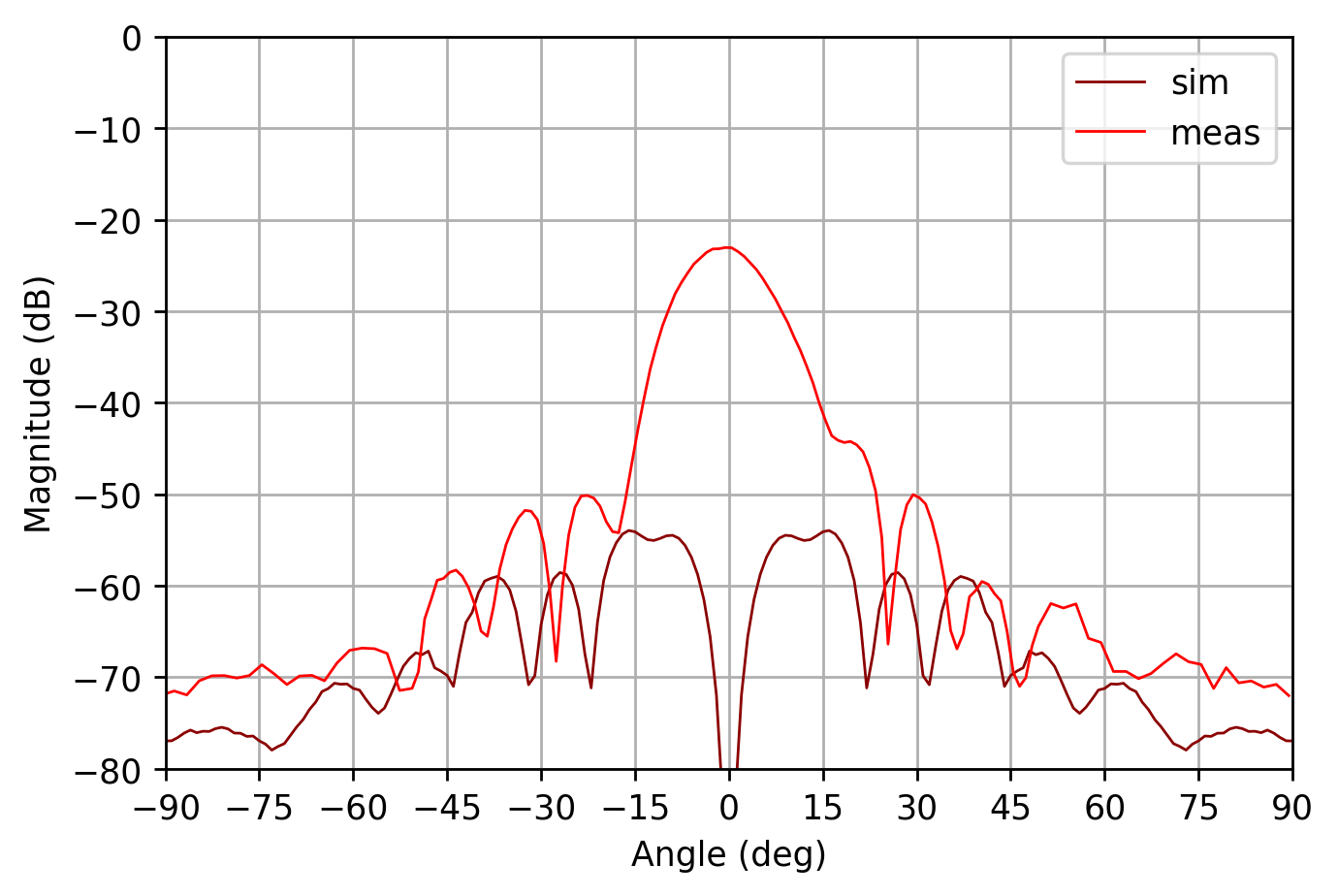}\vspace{10 pt}
   \caption{Measured radiation patterns at 100~GHz of a W-band feed horn assembly, as compared to simulations. From left to right, \textit{Top}: co-polar E-plane and co-polar H-plane. \textit{Bottom}: co-polar $45\degr$ plane and cross-polar $-45\degr$ plane. A good agreement, at the level of fractions of a dB (down to about $-20$~dB) are observed only in the main beam region of co-polar patterns.}
   \label{STRIP_WC}
\end{figure}

\begin{figure}[htbp]
   \centering
   \includegraphics[width=.5\textwidth]{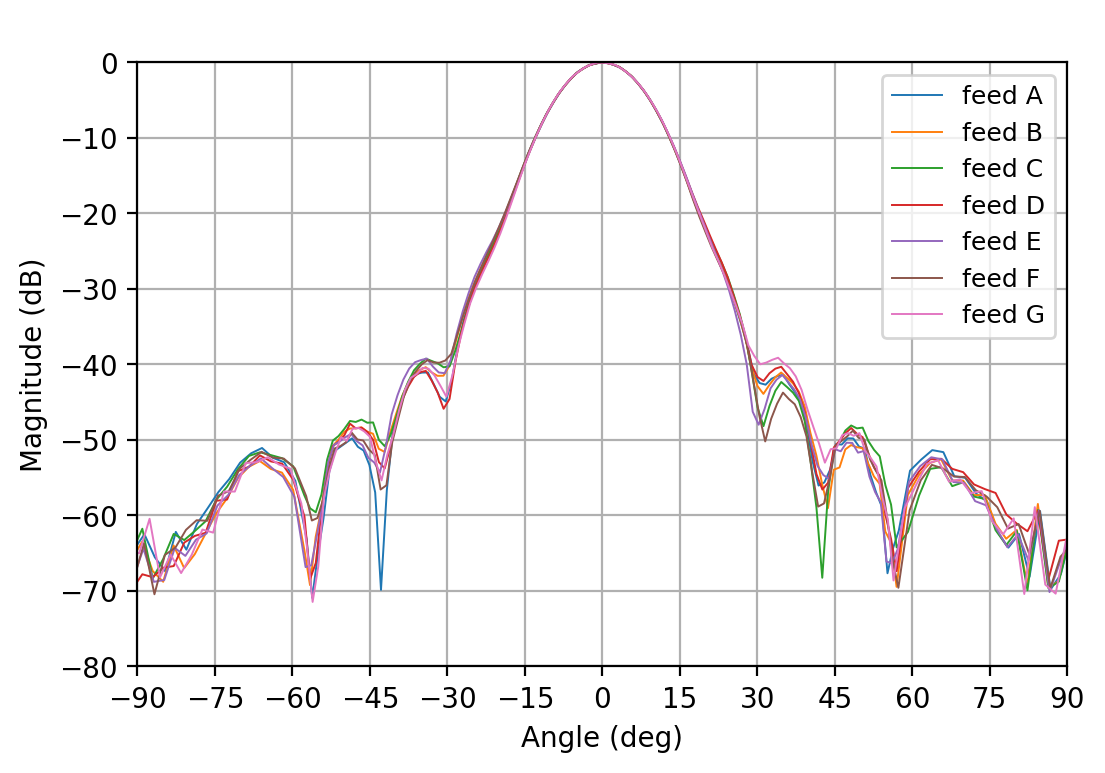}\\
   \includegraphics[width=.5\textwidth]{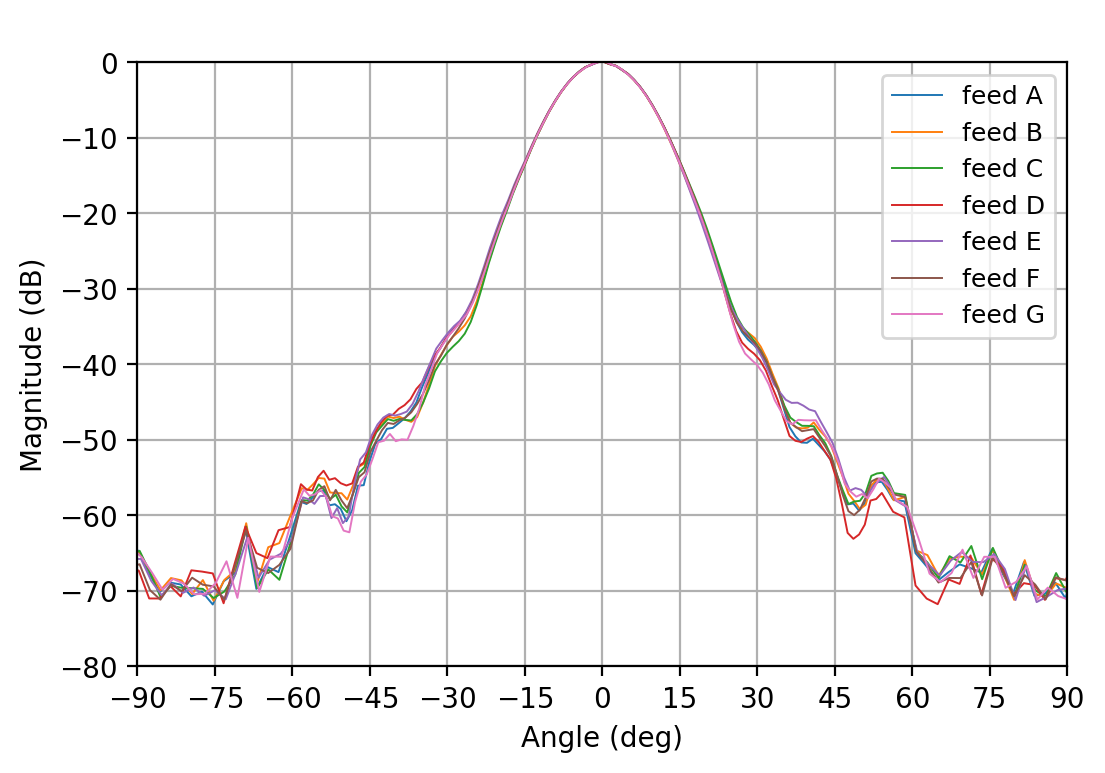}\\ %\vspace{20 pt}
   \includegraphics[width=.5\textwidth]{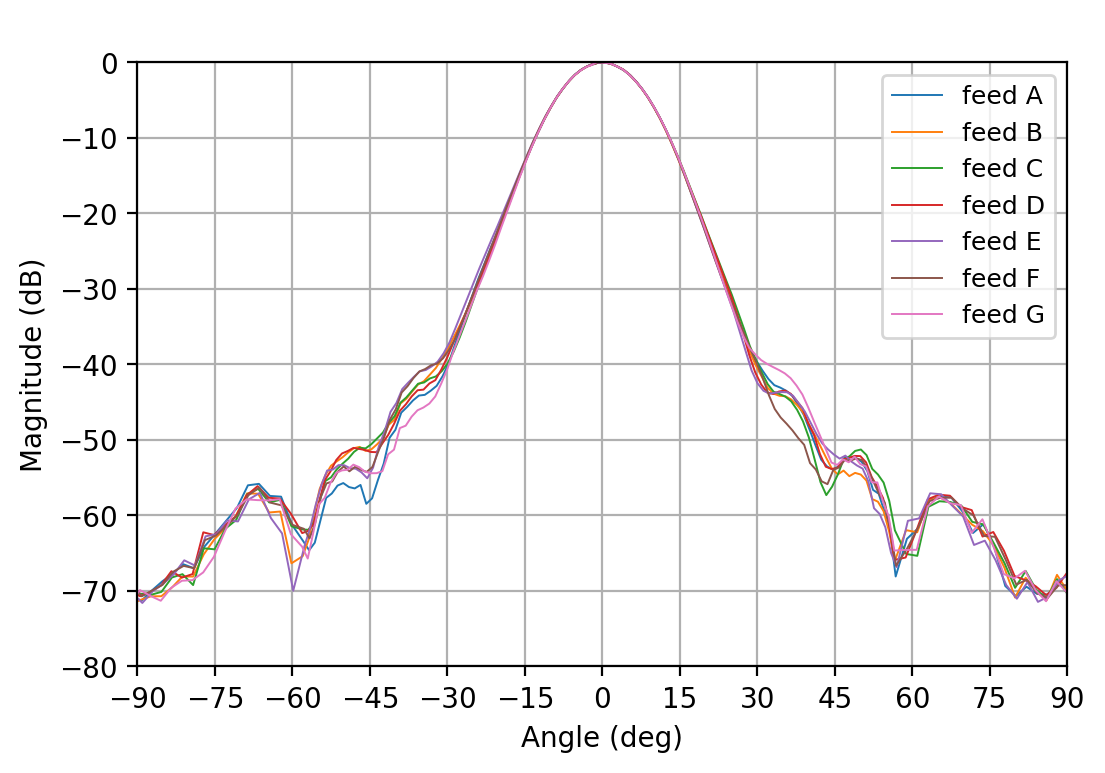}\\ %\vspace{20 pt}
   \caption{Measured radiation patterns of all six feed horns (plus a seventh prototype, named $G$) at 85~GHz ($f_{0}-10\%$). \textit{Top}: co-polar E-plane. \textit{Middle}: co-polar H-plane. \textit{Bottom}: co-polar $45\degr$ plane.}
   \label{STRIP_W_f01}
\end{figure}

\begin{figure}[htbp]
   \centering
   \includegraphics[width=.5\textwidth]{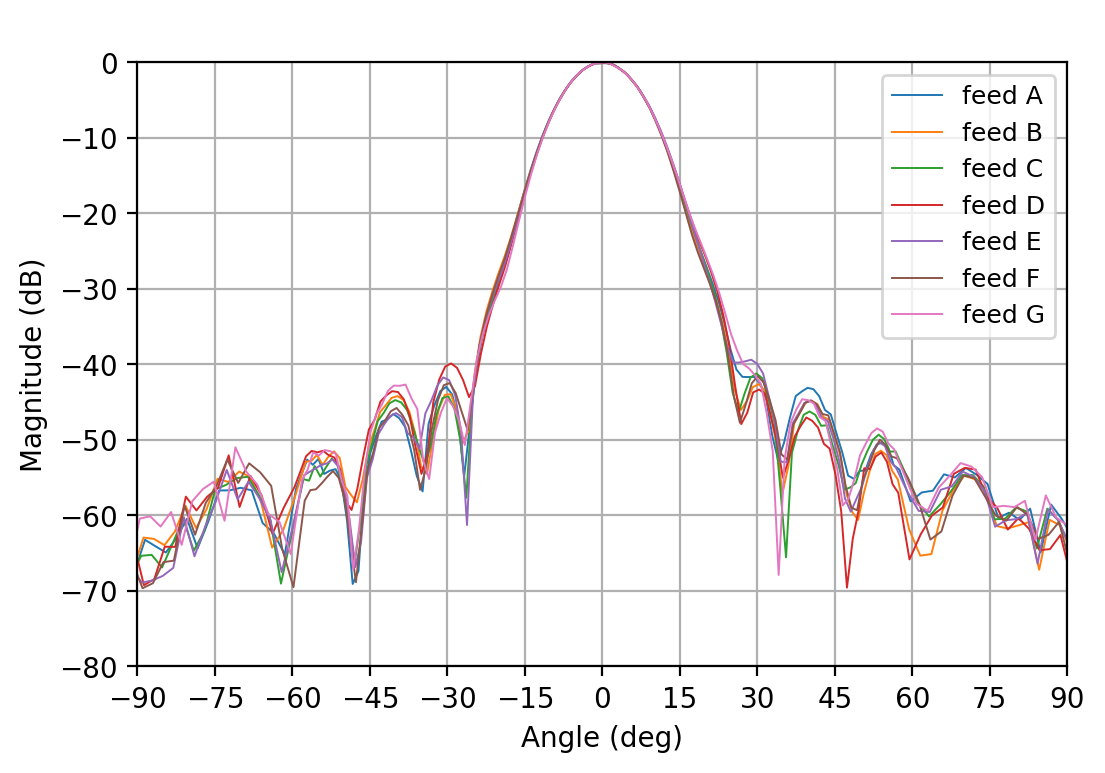}\\
   \includegraphics[width=.5\textwidth]{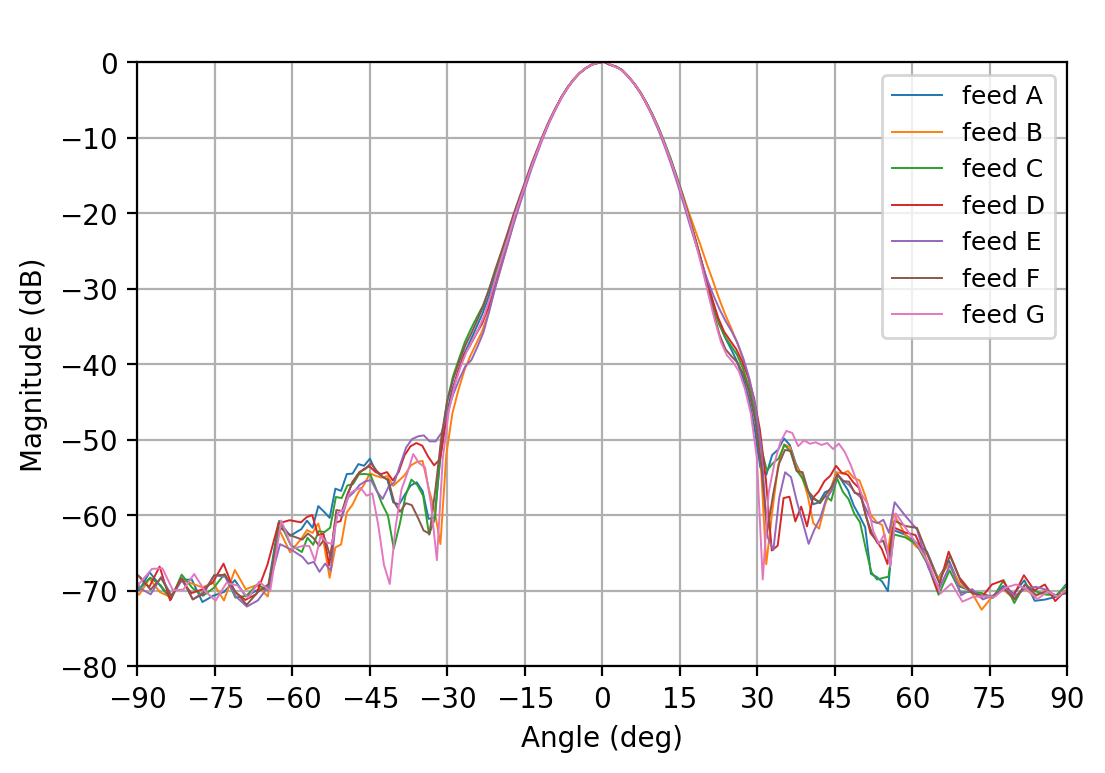}\\ %\vspace{20 pt}
   \includegraphics[width=.5\textwidth]{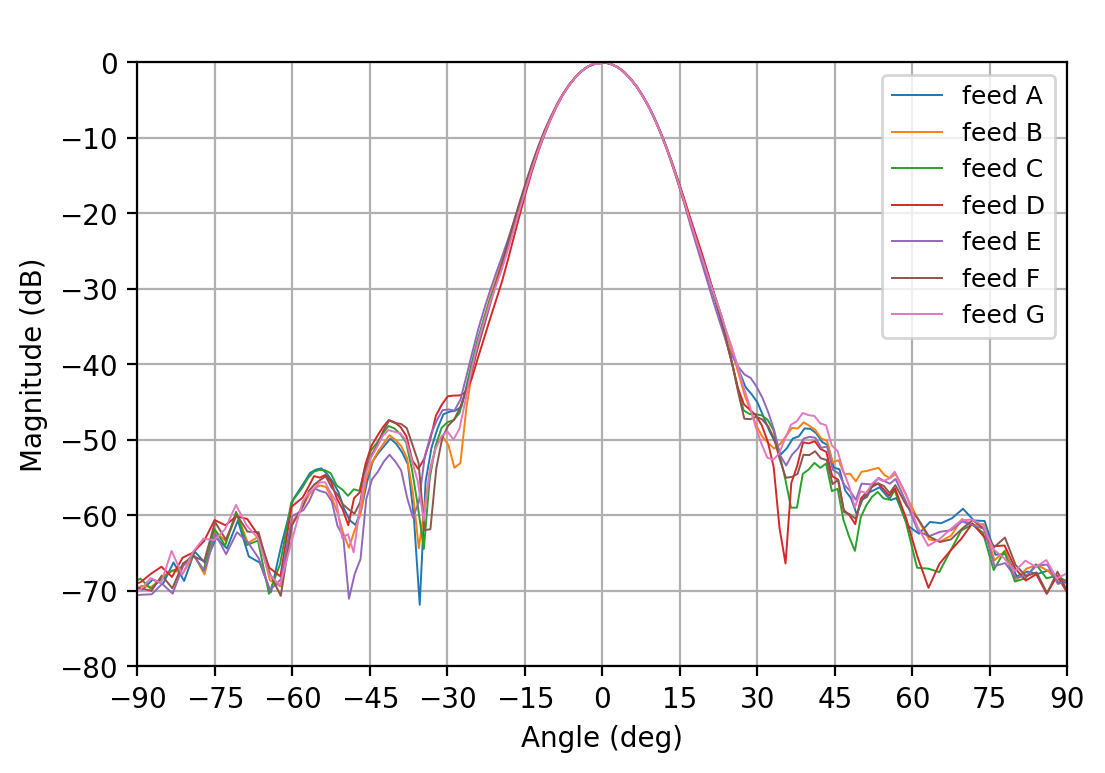}\\ %\vspace{20 pt}
   \caption{Measured radiation patterns of all six feed horns (plus a seventh prototype, named $G$) at 95~GHz ($f_{0}-10\%$). \textit{Top}: co-polar E-plane. \textit{Middle}: co-polar H-plane. \textit{Bottom}: co-polar $45\degr$ plane.}
   \label{STRIP_W_f03}
\end{figure}

\begin{figure}[htbp]
   \centering
   \includegraphics[width=.5\textwidth]{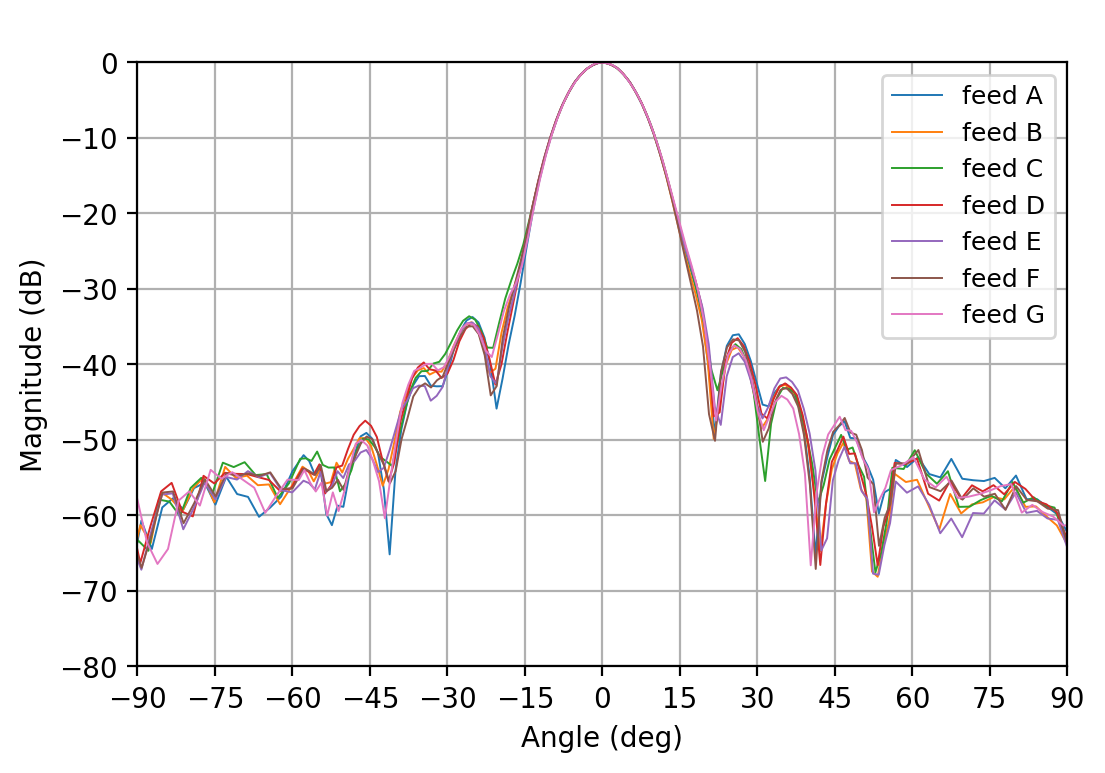}\\
   \includegraphics[width=.5\textwidth]{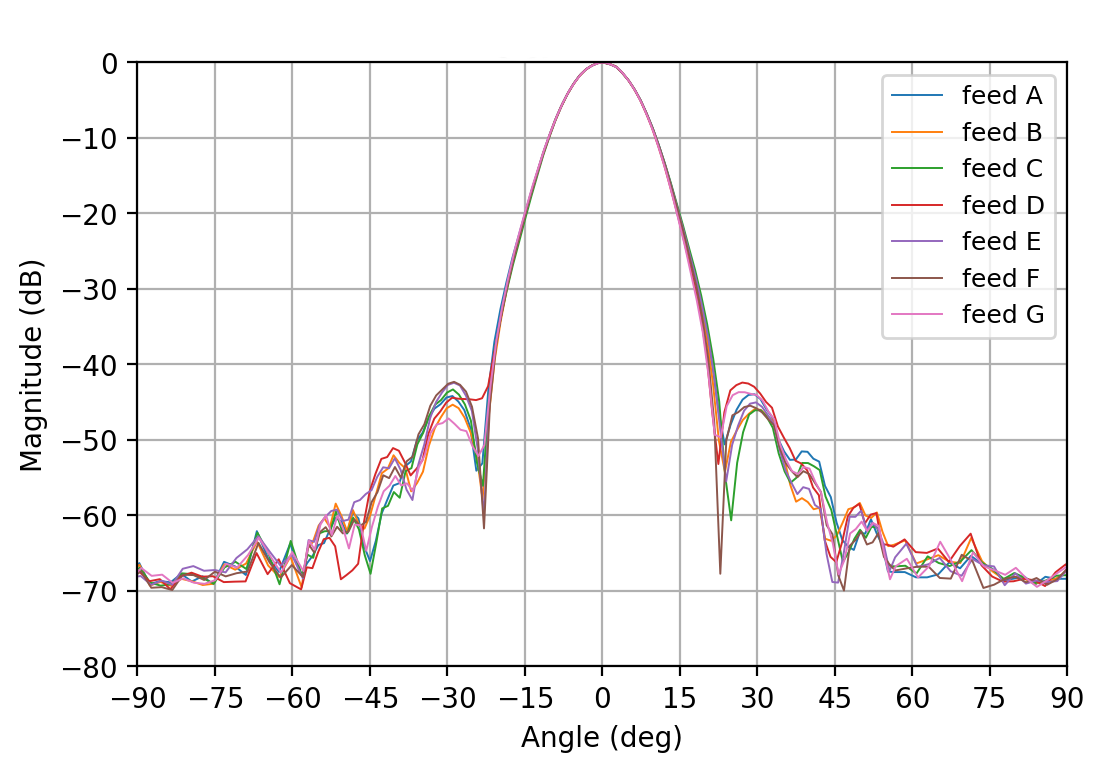}\\ %\vspace{20 pt}
   \includegraphics[width=.5\textwidth]{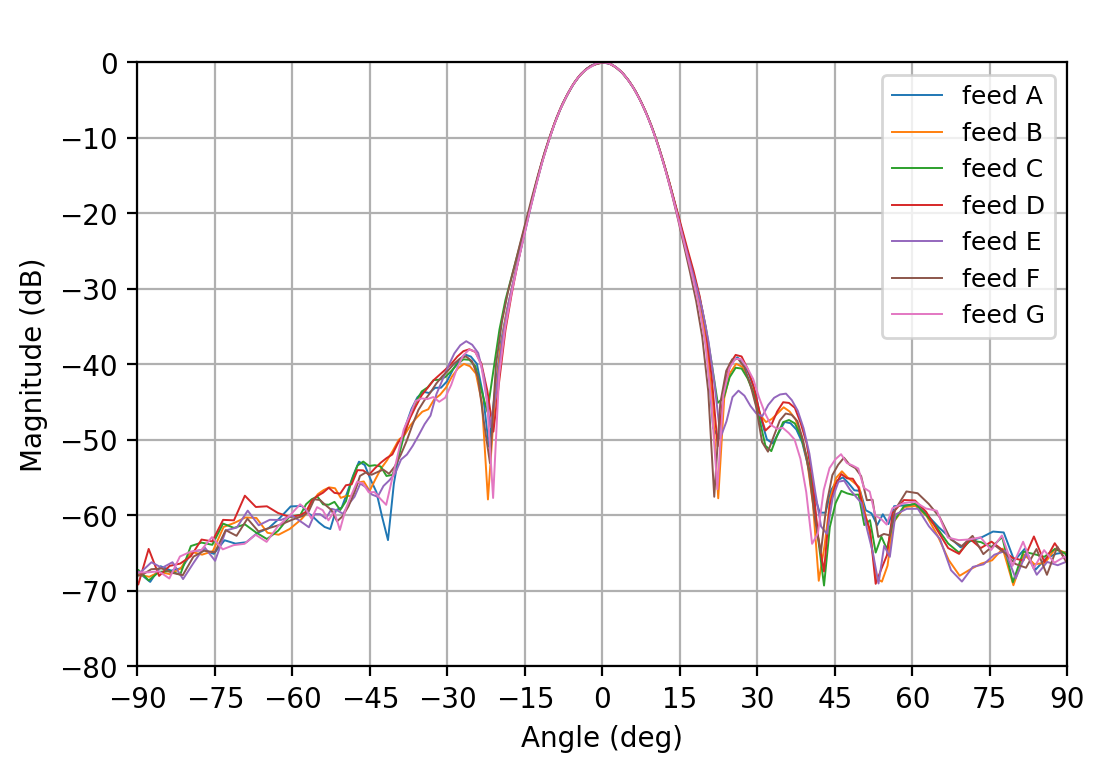}\\ %\vspace{20 pt}
   \caption{Measured radiation patterns of all six feed horns (plus a seventh prototype, named $G$) at 105~GHz ($f_{0}-10\%$). \textit{Top}: co-polar E-plane. \textit{Middle}: co-polar H-plane. \textit{Bottom}: co-polar $45\degr$ plane.}
   \label{STRIP_W_f05}
\end{figure}

We collected a total of 210 radiation patterns. As a reference, Figure~\ref{STRIP_WC} shows a comparison between measured and simulated radiation patterns at $f_{0} = 95$~GHz relevant to one of the six feed horns. Figures~\ref{STRIP_W_f01} to \ref{STRIP_W_f05} show the co-polar
%and cross-polar
radiation patterns on the principal planes of all six W-band horns (plus a seventh prototype) at 85, 95 and 150~GHz, respectively. The simulated patterns are superimposed on all plots.

Differently from what was discussed for the Q-band radiation patterns, the agreement between measurement and simulation is not fully satisfactory for the W-band horns. Two main non-conformities affect the W-band feed horns: (a) the co-polar radiation patterns differ from simulations in the sidelobes angular region $|\theta|>30$\degr; (b) the cross-polar radiation patterns on the 45\degr\ plane show a relatively strong (approximately $-20$~dB) co-polar component.

We took several corrective actions to verify the cause of the observed discrepancies. In particular, we replaced the homemade circular-to-rectangular WR10 waveguide transition with a new commercial transition, whose cross-polar contribution was verified to be better than $-40$~dB over the entire operating band. This did not lead to any improvement in terms of the agreement between co-polar measurements and simulations and in the suppression of the co-polar residual in the cross-polar diagrams. Another possible source of the spurious cross-polar component could come from non-optimal alignment between the source antenna and the feed horn under test. However, even after extensive re-alignment tests, it was not possible to minimize the cross-polar contribution below the observed level.

Measurements with a metrology machine showed that the horns were $\sim$2.25~mm shorter than designed, due to thinner plates. We repeated simulations of the "shortened" configuration to verify a possible better agreement with measurements, but no significant improvement was achieved: the sidelobes have somewhat different shape but they do not improve overall the match with the measurements.

Our most likely guess is that feed horn plates misalignment is playing a key role in the measured horn performance. In fact, we observed that the direction corresponding to the maximum directivity of co-polar radiation patterns does not coincide with the nominal axis of the horns. The maximum discrepancy is $\sim$16\arcmin\ and $\sim$12\arcmin\ for the E- and H-plane, respectively. Moreover, the mechanical axis of the horns was misaligned with respect to the nominal axis by the same amount. We conclude that the alignment of the ring plates with internal dowel pins seems to be not ideal for such a long structure.
%--- begin rev. #10
As discussed for the Q-band feed horns, no tolerance stack-up analysis was possible due to the restrictions of our simulation tools to axisymmetric structures. Therefore it was not possible to investigate the impact on cross-polarization and on the return loss of such a Torre di Pisa like structure.
%--- end rev. #10
Given the agreement between measurements and simulations in the main beam region, we decided to proceed with the integration of W-band horns into the focal plane of the telescope.

%----------------------------------------
\subsubsection{Return loss}
\label{sec:W-meas-rl}
\begin{figure}[htbp]
   \centering
   \includegraphics[width=10 cm]{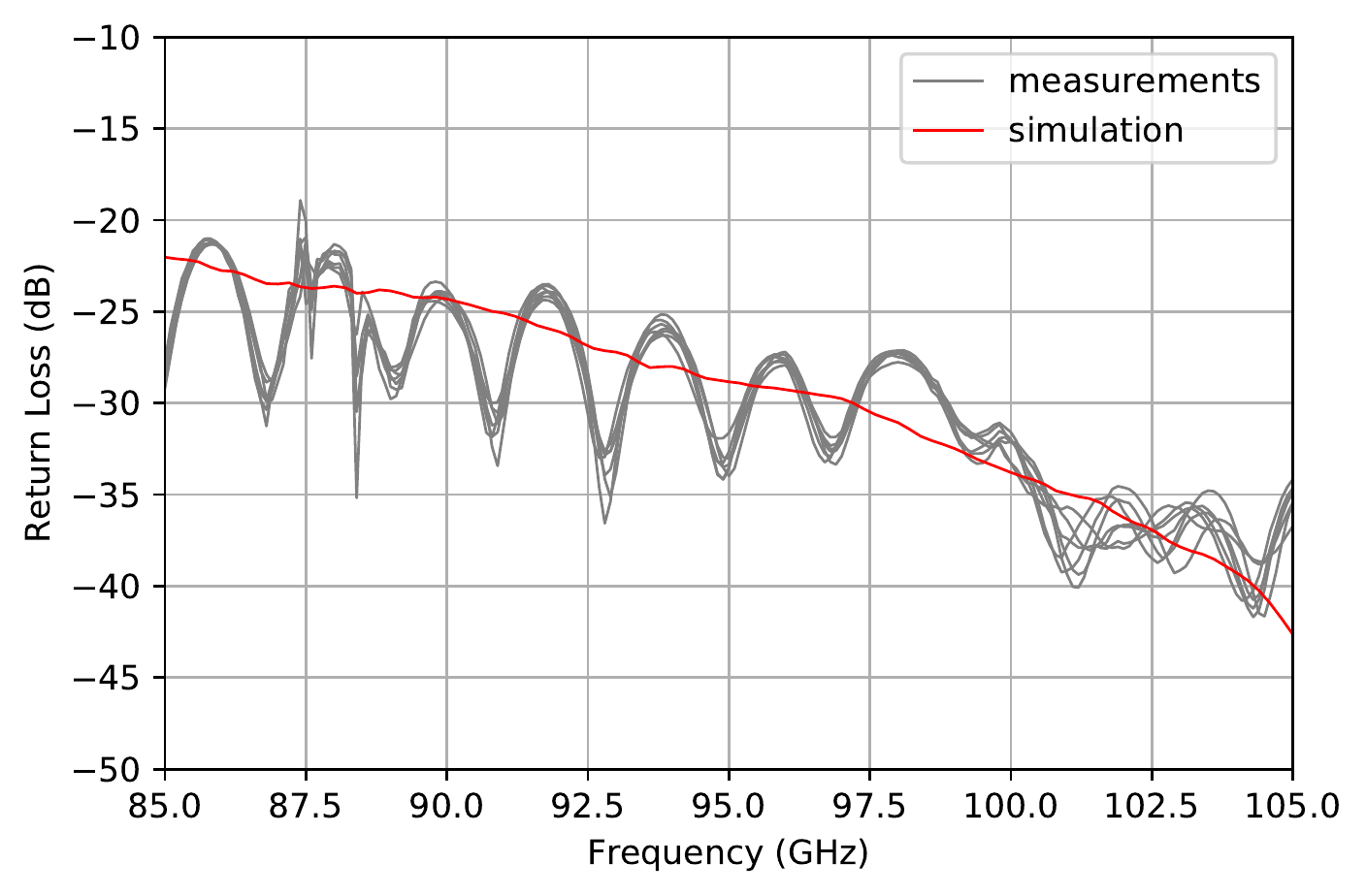}
   \caption{Measured return loss of the six W-band feed horns. Return loss levels are better than $-20$~dB on the whole 20\% bandwidth.}
   \label{STRIP_W_RLmeas}
\end{figure}

The measured return loss of the W-band horns was compared to the expected value in the operative frequency band. Figure~\ref{STRIP_W_RLmeas} shows that the level of return loss is better than $-20$~dB on the whole 20\% bandwidth.

%--- begin rev. #10
%--- end rev. #10

%----------------------------------------
\section{Conclusion} %--- all
\label{sec:conclusion}
In this paper, we discussed the development of the Q- and W-band corrugated feed horn arrays populating the focal plane of the Strip telescope of the Large Scale Polarization Explorer experiment. Their electromagnetic design is based on a dual-profiled configuration, which satisfies the instrumental requirements in terms of telescope illumination in the relevant frequency band, while keeping the feed horn size compact.

We discussed the mechanical engineering of the feed horn profiles with the platelet technique. The forty-nine feed horns in the Q-band are arranged into seven hexagonal modules of seven elements each, while the W-band feed horns are stand-alone assemblies placed at the edges of the focal plane extent.

We measured the radiation pattern of all the Strip feed horns in the anechoic chamber and characterized their impedance matching in their operative frequency band. The measured radiation patterns of the Q-band feed horns resulted in a very good agreement with simulations, at the level of a fraction of a dB down to the first sidelobes. Moreover, we observed an excellent repeatability in their angular response, confirming both the quality of the mechanical manufacturing and the reliability of the measurements, as well as the simulations. The measured return loss also confirms the expected impedance matching for all forty-nine elements, below $-40$~dB over the whole operative bandwidth in the Q-band.

The W–band horns showed a larger discrepancy in angular response between measurements and simulations, particularly in the sidelobe region at angles $|\theta|>30$\degr. Furthermore, the cross-polar measurements show a residual co-polar component, approximately at the $-20$~dB level. After an in-depth assessment of the possible causes, and after excluding possible systematic effects in the measurement setup, the measured behaviour has been attributed to the non-optimal alignment of the ring plates making up the feed horn structure. In fact, the use of internal pins for the alignment is most likely not an optimal choice for such a long and densely stacked structure (197 plates in $\sim$102~mm). Overall, however, the agreement between measurements and simulations in the main beam region, well supported by our return loss measurements, is considered acceptable for the purpose of atmospheric monitoring of the W-band channel.

%----------------------------------------
%\appendix
%\section{TBW}
%Please always give a title also for appendices.

%----------------------------------------
% \begin{acknowledgements} 
\acknowledgments
The LSPE--Strip instrument has been developed thanks to the support of ASI contract I/022/11/1 and Agreement 2018-21-HH.0 and by funding from INFN (Italy).

The authors wish to thank Lorenzo Curioni at Cloema S.c.r.l. (Florence, Italy) for the fruitful discussions on the mechanical processing of almost a thousand Aluminum sheets, which later on became antennas.

%----------------------------------------
\clearpage
\bibliographystyle{JHEP}
\bibliography{C.Franceschet_The_LSPE_Strip_feed_horn_array}

\end{document}